\documentclass[twocolumn, twocolappendix]{aastex631}

\newcommand{\OI}{O~{\sc i}}

\newcommand{\CII}{C~{\sc ii}}
\newcommand{\CI}{C~{\sc i}}
\newcommand{\NaI}{Na~{\sc i}}

\newcommand{\SiII}{Si~{\sc ii}}

\newcommand{\CaII}{Ca~{\sc ii}}
\newcommand{\TiII}{Ti~{\sc ii}}

\newcommand{\Nifs}{$^{56}$Ni}

\newcommand{\DmB}{$\Delta \mathrm{m}_{15}(B)$}

\newcommand{\Ia}{SNe Ia}

\newcommand{\timeB}{$t_{B}^{max}$}

\shorttitle{Early-Time UV Magnitudes and Colors of SNe Ia}
\shortauthors{Hoogendam et al.}

\begin{document}

\title{From Out of the Blue: {\it Swift} Links 2002es-like, 2003fg-like, and Early-Time Bump Type Ia Supernovae}

\correspondingauthor{W. B. Hoogendam}
\email{willemh@hawaii.edu}

\author[0000-0003-3953-9532]{W. B. Hoogendam}
\altaffiliation{NSF Graduate Research Fellow}
\affiliation{Institute for Astronomy, University of Hawaii, 2680 Woodlawn Drive, Honolulu, HI 96822, USA}

\author[0000-0003-4631-1149]{B. J. Shappee}
\affiliation{Institute for Astronomy, University of Hawaii, 2680 Woodlawn Drive, Honolulu, HI 96822, USA}

\author[0000-0001-6272-5507]{P. J. Brown}
\affiliation{George P. and Cynthia Woods Mitchell Institute for Fundamental Physics and Astronomy, Department of Physics and Astronomy, Texas A\&M University, College Station, TX 77843, USA}

\author[0000-0002-2471-8442]{M. A. Tucker}
\altaffiliation{CCAPP Fellow}
\affiliation{Center for Cosmology and Astroparticle Physics,
The Ohio State University, 
191 West Woodruff Ave,
Columbus, OH, USA}
\affiliation{Department of Astronomy, 
The Ohio State University, 
140 West 18th Avenue,
Columbus, OH, USA}
\affiliation{Department of Physics,
The Ohio State University,
191 West Woodruff Ave,
Columbus, OH, USA}

\author[0000-0002-5221-7557]{C. Ashall}
\affiliation{Department of Physics, Virginia Tech, Blacksburg, VA 24061, USA}

\author[0000-0001-6806-0673]{A. L. Piro}
\affiliation{The Observatories of the Carnegie Institution for Science, 813 Santa Barbara St., Pasadena, CA 91101, USA}

\begin{abstract}
We collect a sample of 42 SNe Ia with \emph{Swift} UV photometry and well-measured early-time light curve rises and find that 2002es-like and 2003fg-like SNe Ia have different pre-peak UV color evolutions compared to normal SNe Ia and other spectroscopic subtypes. 
Specifically, 2002es-like and 2003fg-like SNe Ia are cleanly separated from other SNe Ia subtypes by $UVM2-UVW1\gtrsim1.0$~mag at $t=-10$~days relative to $B$-band maximum.
Furthermore, the SNe Ia that exhibit non-monotonic bumps in their rising light curves, to date, consist solely of 2002es-like and 2003fg-like SNe Ia.
We also find that SNe Ia with two-component power-law rises are more luminous than SNe Ia with single-component power-law rises at pre-peak epochs.
Given the similar UV colors, along with other observational similarities, we discuss a possible progenitor scenario that places 2002es-like and 2003fg-like SNe Ia along a continuum and may explain the unique UV colors, early-time bumps, and other observational similarities between these objects.
Ultimately, further observations of both subtypes, especially in the near-infrared, are critical for constraining models of these peculiar thermonuclear explosions.
\end{abstract}

\section{Introduction} \label{sec:intro}

Type Ia Supernovae (SNe Ia) are important astrophysical explosions that drive chemical enrichment \citep{Matteucci01}, produce heavy elements \citep{Raiteri96}, and enable precise distance determinations \citep{Phillips93, riess98, Perlmutter99, Phillips99, Burns18}. Despite many large \Ia\ data sets (recent examples include \citealp{Holoien17, Holoien17B, Holoien17C, Holoien19, Jones19, Phillips19, Fremling20, POISE_ATel, Jones21, Neumann23, Peterson23} and Do et al. 2024 in prep.)
and numerous theoretical models (e.g., \citealt{Nomoto82, Khokhlov91, Woosley94, Ropke07, Kashi11, Woosley11, thom11, Pakmor12, Hoeflich17}), the progenitor systems of SNe Ia are not yet comprehensively connected to observations (reviews include \citealt{Maoz14}, \citealt{Livio18}, \citealt{Jha19}, and \citealt{Liu23}). While there is broad consensus that \Ia\ originate from carbon-oxygen white dwarf stars (CO WDs; \citealt{hoy60}), theoretical models struggle to replicate all of the observed diversity of \Ia\ \citep{Maoz14, Livio18}.

Several progenitor scenarios may explain the origin of \Ia, including the single-degenerate (SD), double-degenerate (DD), and core-degenerate (CD) scenarios. The SD scenario consists of a CO WD with a non-degenerate companion such as a main sequence or red giant star (e.g., \citealp{Whelan73}), the DD scenario consists of two CO WDs or a CO WD and a He WD (e.g., \citealp{Nomoto80}), and the CD scenario consists of a CO WD and a degenerate CO core of an AGB star (e.g., \citealp{Hoeflich:Khokhlov:96}).

A variety of explosion mechanisms for each progenitor scenario may produce \Ia\ below, at, or above the Chandrasekhar mass. In either the SD or DD progenitor scenario, material from a companion can accrete onto the CO WD. It can trigger an explosion through central carbon ignition as the CO WD approaches the Chandrasekhar mass \citep{hoy60, Whelan73, Nomoto82, Piersanti03} or through a He detonation on the surface of a CO WD below the Chandrasekhar mass \citep{Nomoto80, Livne90, Woosley94, Hoeflich:Khokhlov:96, Hoeflich17, Maeda18, Polin19}. In addition to the aforementioned explosion mechanisms, the DD scenario has additional explosion mechanisms which include mergers below, at, or above $M_{ch}$ \citep{iben84, Webbink84, vanKerkwijk10, Scalzo10, Pakmor10, Pakmor13, Kromer13, Kromer16} and third- or fourth-body induced collisions \citep{thom11, Shappee13c, Pejcha13}. Finally, the CD scenario may explode via the merger of a degenerate CO core of an asymptotic giant branch (AGB) star and a CO  WD \citep{Hoeflich:Khokhlov:96, Noebauer16, Maeda23}.

Despite differences in the explosion mechanism and potentially the progenitor system(s) for \Ia, \citet{Phillips93} find an empirical relationship between the decline rate of \Ia\ light curves and their absolute magnitudes, which holds for a majority of \Ia. Conversely, UV observations of \Ia\ show greater spectral diversity than in the optical \citep{Ellis08, Foley08, Walker12}, and photometrically, \Ia\ can be grouped into NUV-red and NUV-blue classes based on the UV$-$optical color curves \citep{Milne13}. Furthermore, the UV colors have an intrinsic scatter that is incompletely explained by extinction and redshift \citep{Brown17} and are redder than the \citet{Kasen07} predictions for asymmetric explosions \citep{Brown18}. \citet{Pan20} claim a correlation between UV flux and host-galaxy metallicity; however, \citet{Brown20} were unable to confirm this correlation.

In addition to UV differences between spectroscopically normal \Ia, some \Ia\ show significant spectroscopic differences from normal \Ia\ yet remain on the \citet{Phillips93} relation (e.g., SN~1991T \citealt{Phillips92,fil92b}, and SN~1991bg \citealt{fil92a}), and other spectroscopically different \Ia\ deviate from the \citet{Phillips93} relationship (e.g., SN~2002es \citealt{Ganeshalingam12}, SN~2003fg \citealt{Howell06}, and SN~2006bt \citealt{Foley10}). The growing number of these extreme \Ia\ offers a unique chance to probe the generally homogeneous nature of \Ia.

Two particularly interesting \Ia\ subtypes are 2002es-like and 2003fg-like \Ia. 2002es-like \Ia\ are subluminous and have spectra that are similar to 1991bg-like \Ia\ with \SiII\ $\lambda 5972$, \OI, and \TiII\ features near \timeB. They lack a secondary $i$-band rebrightening, similar to other subluminous \Ia. However, unlike other fast-declining, subluminous \Ia, 2002es-like \Ia\ decline at a rate similar to normal \Ia\ \citep{Taubenberger17}. One model for these objects is in the DD scenario with CO WDs whose masses sum greater than $M_{ch}$ and violently merge \citep{Pakmor10, Kromer16}. Unfortunately, this class of objects is small with $\sim$10 objects, and a homogeneous data set does not yet exist. 
Conversely, 2003fg-like \Ia\ have overluminous and broader optical light curves than normal \Ia\ yet also lack the secondary $i$-band peak \citep{Taubenberger17, Ashall21}. In the UV, 2003fg-like \Ia\ have different colors than normal and 1991T-like \Ia\ \citep{Brown14a}, and spectroscopically, they have weaker \CaII\ and stronger \OI\ and \CII\ features. For the first members discovered in this class, the \citet{Arnett82} relationship yields \Nifs\ masses for 2003fg-like \Ia\ exceeding $M_{ch}$ (e.g., \citealt{Howell06, Tanaka10, Scalzo10, Taubenberger11}), leading to the colloquial ``Super-Chandresekhar'' designation (e.g., \citealp{Howell06, Chen09, Scalzo10, Silverman11, Scalzo12, Das13, Taubenberger13a, Hsiao20}). However, lower luminosity 2003fg-like \Ia\ with derived \Nifs\ masses below $M_{ch}$ have also been discovered, suggesting that not all members of this class are ``super-Chandrasekhar'' explosions
(e.g., \citealt{Hicken07,Chakradhari14,Chen19,Lu21}). Potential models for 2003fg-like \Ia\ included a rapidly spinning CO WD \citep{Langer00,Yoon05,Hachisu12}, the merger of two WDS \citep{Bulla16}, and the CD scenario (e.g., \citealt{Hoeflich:Khokhlov:96}, \citealt{Ashall21}, \citealt{Maeda23}).

Despite their spectroscopic and photometric differences, both 2002es-like and 2003fg-like have had observations of rising light curve bumps\footnote{Here we define a SN Ia with a non-monotonic rising light curve to have a bump (i.e., a light curve that has a decrease in flux at early times). Section \ref{sec:bumps} explains this definition more thoroughly.} \citep{Cao15, Miller20, Jiang21, Dimitriadis23, Srivastav23a, Srivastav23b, Xi23}; conversely, other spectroscopic subtypes and normal \Ia\ show deviations from a single-component power-law rise, and it remains unclear if these events are connected to the rising light curve behavior of 2002es-like and 2003fg-like \Ia. 
While theoretical models of normal SNe Ia are similar at peak $B$-band magnitude, at early times (0-5 days after explosion), various models make different predictions about the light curve shape (e.g., \citealt{Kasen10, Piro13, Piro14, Piro16, Maeda18, Polin19, Magee20b, Maeda23}). Thus, probing the earliest stages of \Ia\ explosions can provide otherwise unavailable information about the underlying physics.

In this paper, we analyze the UV absolute luminosities and colors of \Ia\ with different rising light curve behavior. §\ref{sec:data} details the observations, data reduction process, and sample selection. §\ref{sec:abs_mags} presents the absolute magnitudes of our sample, and §\ref{sec:colors} presents the UV and optical color curves. Discussion is in §\ref{sec:discussion}, and conclusions are presented in §\ref{sec:conclusions}. Throughout this work, we adopt $H_0 = 73$ km sec$^{-1}$ Mpc$^{-1}$, $\Omega_m = 0.30$, and $\Omega_{vac} = 0.70$.

\section{Data}\label{sec:data}

We perform an exhaustive literature search and identify 42 \Ia\ with pre-peak ( $t<-5$~days) \emph{Swift} UV photometry and early-time ($t<-14$~days) optical observations which can be used to constrain the shape of the rising light curve. We relax the optical phase constraint if the SN Ia rises over four magnitudes after discovery (and thus must be young; iPTF13ebh \citealp{Hsiao15} and 2019ein \citealp{Pellegrino20, Kawabata20}) or if there is a non-detection within two days of the first detection (this is only for some \Ia\ in \citealt{Burke22} who use DLT40 discoveries). Of the 42 \Ia\ with pre-peak \emph{Swift} observations, we exclude five due to issues with host-galaxy contamination or low signal-to-noise. Details for all \Ia\ in our sample, including reasons for exclusion from the subsequent analyses, are provided in Appendix \ref{sec:individual_comments} along with values for \timeB\ and extinction. Table \ref{tab:sample} lists the \Ia\ in the final sample along with their redshift, distance, early-time light curve category, and host galaxy name and morphology. 

\emph{Swift} photometry is taken from the Swift Optical/UV Supernova Archive (SOUSA; \citealt{Brown14b})\footnote{Accessed via\\ \href{https://github.com/pbrown801/SOUSA/tree/master/data}{https://github.com/pbrown801/SOUSA/tree/master/data}}, which uses the \citet{Breeveld11} Vega magnitude system zero points that update the original \citep{Poole08} zero points. Either a 3- or 5-arcsecond aperture is used to perform photometry with aperture size chosen to maximize S/N. The host galaxy counts from a post-SN template image are subtracted to produce the final photometry. Three exceptions for host-galaxy subtraction are SN~2021hpr and SN~2021aefx, which had galaxy flux from pre-explosion \emph{Swift} observations subtracted, and SN~2022ilv, which does not have a clear host galaxy. Observations obtained within 0.75d are combined with a weighted average to increase the S/N in the final photometry. 

Two Swift filters, $UVW2$ and $UVW1$, have transmission functions extending into the optical wavelengths (i.e., a red leak). This creates a broader distribution of photons from the UV to the optical \citep{Brown10}. The relative similarity of \Ia\ in the optical means that any peculiar behavior comes from the UV regime. Still, the effect of UV spectral variations is diluted in the $UVW2$ filter compared to the neighboring $UVM2$ filter. We use all filters in this work.

To calculate absolute magnitudes in each of the six \emph{Swift} filters, \textcolor{black}{we correct for Milky Way and host-galaxy extinction using literature values and SN-independent distances also from the literature. We do not perform $K$-corrections since our sample is low-redshift}. We adopt a \citet{CCM89} extinction law to convert $A_V$ into a filter-specific extinction estimate for each \emph{Swift} filter. \textcolor{black}{We note that the 2002es-like and 2003fg-like SNe Ia in our sample all have extinction estimates based on \NaI\ D, not SN color, and the estimates are all $\le 0.05 $~mag. Thus, extinction is not a significant factor in the main results of the paper discussed in Section \ref{sec:discussion}.}

We also include a comparison sample of 2003fg-like \Ia\ that lack early-time optical observations yet still have \emph{Swift} photometry. The comparison sample consists of SN2009dc \citep{Yamanaka09,Silverman11,Taubenberger11}, SN2012dn \citep{Chakradhari14,Taubenberger19}, SN2015M \citep{Ashall21}, ASASSN-15hy \citep{Lu21}, ASASSN-15pz \citep{Chen19} with extinction and \timeB\ values from \citet{Ashall21}. A comparison sample of 2002es-like \Ia\ with \emph{Swift} observations would also be useful in this work; however, such a sample does not yet exist.

\begin{deluxetable*}{ccccccc}
\tablenum{1}
\tablecaption{
Object information for the \Ia\ sample. SNe Ia subtypes are abbreviated as follows: 
Norm $\Rightarrow$ Normal SN Ia; 
Fast $\Rightarrow$ subluminous, transitional, 1991bg-like, etc.; 
99aa $\Rightarrow$ 1999aa-like SN Ia (similar to 1991T-like SN Ia); 
03fg $\Rightarrow$ 2003fg-like SN Ia; 
02es $\Rightarrow$ 2002es-like SN Ia. 
Host galaxy types are taken from \citet{deVaucouleurs91}.
The full table is available in the electronic publication.
}\label{tab:sample}
\tablewidth{0pt}
\tablehead{
\colhead{SN Name} & \colhead{Subtype} & \colhead{$z$ \tablenotemark{a}} & \colhead{$\mu$\tablenotemark{b} [mag] }& \colhead{Rise} 
& \colhead{Host Name} & \colhead{Host Type}}
\startdata
SN~2009ig   & Norm        & 0.0088 (1)      & $32.56\pm0.07$ (a) & Single   &   NGC 1015    & SB(r)a        \\
SN~2011fe   & Norm        & 0.0008 (2)      & $29.04\pm0.19$ (b) & Single   &   M 101       & SAB(rs)cd     \\
SN~2012cg   & Norm        & 0.001458 (3)    & $30.84\pm0.13$ (a) & Double   &   NGC 4424    & SB(s)a        \\
SN~2012fr   & Norm        & 0.005457 (4)    & $31.38\pm0.06$ (a) & Double   &   NGC 1365    & SB(s)b        \\
SN~2012ht   & Fast        & 0.00356 (5)     & $31.94\pm0.03$ (a) & Single   &   NGC 3447    & SAB(s)m pec   \\
LSQ12gdj    & 91T         & 0.030324 (7)    & $35.46\pm0.15$ (f) & Single   &   ESO 472-007 & Unclassified  \\
SN~2013dy   & Norm        & 0.003889 (6)    & $31.63\pm0.13$ (a) & Double   &   NGC 7250    & Sdm?          \\
SN~2013gy   & Norm        & 0.014023 (8)    & $33.25\pm0.20$ (d) & Single   &   NGC 1418    & SB(s)b:       \\
iPTF13dge   & Norm        & 0.01586 (9)     & $34.03\pm0.47$ (e) & Single   &   NGC 1762    & SA(rs)c:      \\ 
iPTF13ebh   & Fast        & 0.01316 (10)    & $33.30\pm0.08$ (g) & Double   &   NGC 0890    & SAB0          \\
ASASSN-14lp & Norm        & 0.0051 (1)      & $30.73\pm0.45$ (c) & Single   &   NGC 4666    & SABc:         \\
iPTF14atg   & 02es        & 0.02129 (11)    & $35.32\pm0.33$ (j) & Bump     &   IC 831      & Unclassified  \\
\vdots & \vdots & \vdots & \vdots & \vdots & \vdots & \vdots \\
\vdots & \vdots & \vdots & \vdots & \vdots & \vdots & \vdots \\
SN~2019ein  & Norm        & 0.00775 (24)    & $32.71\pm0.08$ (g) & Single   &   NGC 5353    & S0 edge-on    \\
SN~2019yvq  & 02es        & 0.0094 (25)     & $33.14\pm0.11$ (m) & Bump     &   NGC 4441    & SAB0-pec      \\
SN~2020hvf  & 03fg        & 0.00581 (26)    & $32.34\pm0.15$ (f) & Bump     &   NGC 3643    & SB0+(r)       \\
SN~2020nlb  & Norm        & 0.00243 (27)    & $31.00\pm0.12$ (d) & Single   &   NGC 4382    & SA0 pec       \\
SN~2020tld  & Fast        & 0.011201 (28)   & $33.69\pm0.48$ (h) & Single   &   ESO 194-021 & SA0           \\
SN~2020udy  & 02cx        & 0.01722 (9)     & $33.68\pm0.45$ (c) & Single   &   NGC 0812    & S pec         \\
SN~2021fxy  & Norm        & 0.0094 (29)     & $32.57\pm0.40$ (i) & Single   &   NGC 5018    & E3            \\ 
SN~2021hpr  & Norm        & 0.009346 (30)   & $33.01\pm0.17$ (a) & Double   &   NGC 3147    & SA(rs)bc      \\
SN~2021zny  & 03fg        & 0.026602 (7)    & $35.19\pm0.20$ (d) & Bump     &  CGCG 438-018 & Unclassified  \\
SN~2021aefx & Norm        & 0.005017 (31)   & $31.27\pm0.49$ (n) & Double   &   NGC 1566    & SAB(s)bc      \\
SN~2022eyw  & 02cx        & 0.0087 (12)     & $33.12\pm0.15$ (f) & Single   & MCG +11-16-003& Unclassified  \\
SN~2022ilv  & 03fg        & 0.0310 (32)     & $35.28\pm0.47$ (o) & Bump     & Hostless      & Hostless      \\
SN~2023bee  & Norm        & 0.0067 (9)      & $33.04\pm0.20$ (d) & Double   &   NGC 2708    & SAB(s)b       \\
\enddata
\tablenotemark{a}{(1) \citet{Meyer04}; (2) \citet{deVaucouleurs91}; (3) \citet{Kent08}; (4) \citet{Bureau96}; (5) \citet{Kerr86}; (6) \citet{Schneider92}; (7) \citet{Springob05}; (8) \citet{Catinella05}; (9) \citet{Falco99}; (10) \citet{Bosch_v/d15}; (11) \citet{Rines16}; (12) \citet{Albareti17};  (13) \citet{Tak_v/d08}; (14) \citet{Beers95}; (15) \citet{Koribalski04}; (16) \citet{Theureau05}; (17) \citet{Cappellari11}; (18) \citet{Schneider90}; (19) \citet{Smith2000}; (20) \citet{Jones09}; (21) \citet{Norris11}; (22) \citet{Bilicki14};  (23) \citet{Rhee96}; (24) \citet{Dried_van01}; (25) \citet{Miller20}; (26) \citet{van_Driel16}; (27) \citet{Smith2000}; (28) \citet{Loveday96}; (29) \citet{Rothberg06}; (30) \citet{Epinat08}; (31) \citet{Allison14}; (32) \citet{22ilv_z}, SN~2022ilv is hostless, so the redshift is determined using SNID \citep{Blondin07}.}\\
\tablenotemark{b}{(a) Cepheids; \citet{Riess22}; (b) Cepheids; \citet{Shappee11}; (c) Tully-Fisher \citet{Tully16}; (d) Tully-Fisher \citet{Tully13}; (e) Tully-Fisher \citet{Theureau07}; (f) Hubble flow using $H_0 = 73$ km s$^{-1}$ Mpc$^{-1}$ and correcting for Virgo + GA + Shapley. (g) Tully-Fisher \citet{Jensen21}; (h) Fundamental Plane \citet{Springob14}; (i) Tully-Fisher \citet{Courtois12}; (j) Fundamental Plance \citet{Saulder16}; (k) TRGB \citet{Tully13}; (l) TRGB \citet{Hoyt21}; (m) Peculiar Velocity Modelling \citet{Carrick15}; (n) TRGB \citet{Sabbi18}; (o) Hostless; $z$ from SN spectrum distance \citet{Srivastav23a}.
}\\
\end{deluxetable*}

\section{Optical and UV Absolute Magnitudes}\label{sec:abs_mags}

\begin{figure*}
    \includegraphics[width=\textwidth]{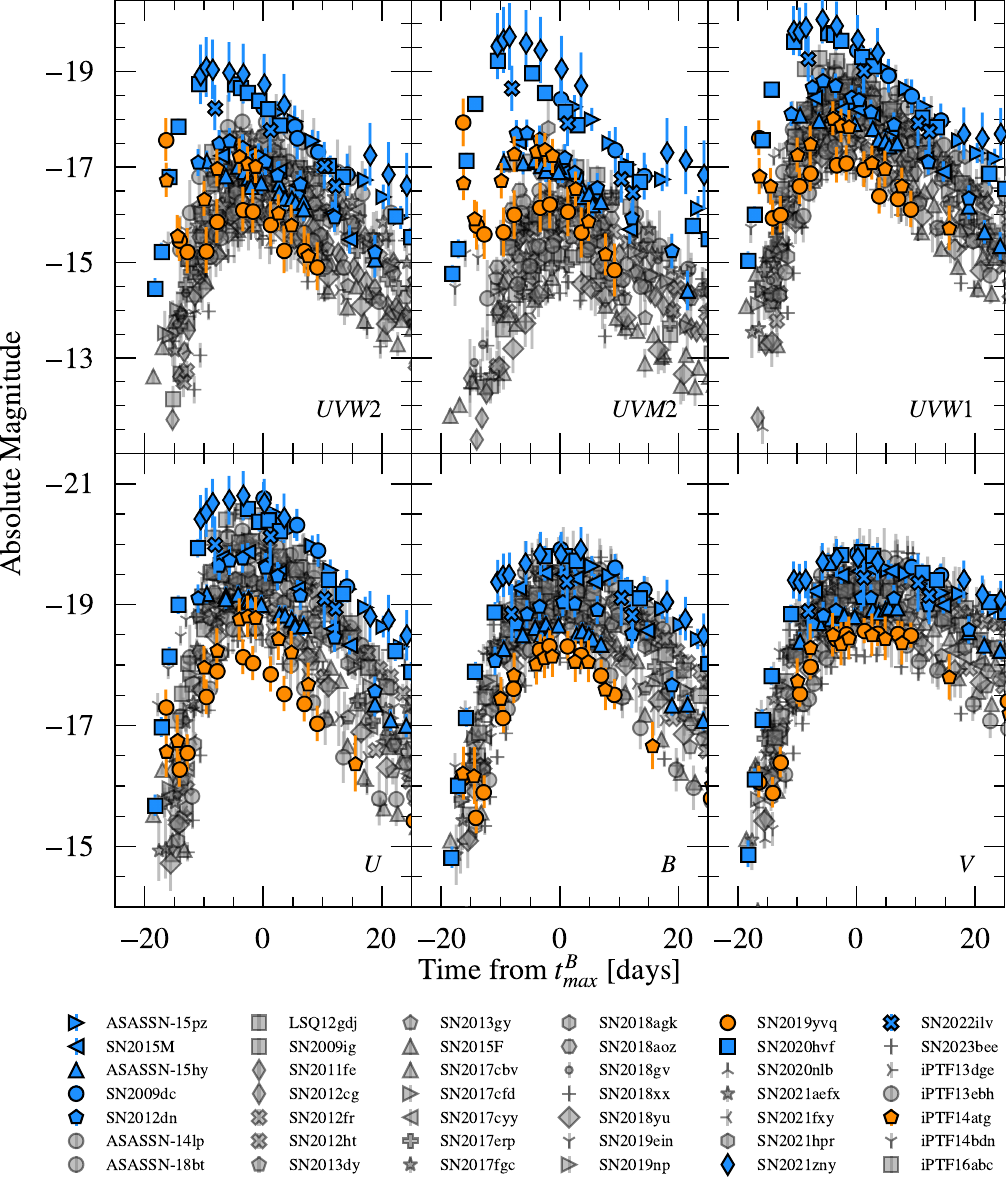} \\
    \caption{
    Absolute \textcolor{black}{\emph{Swift}} Vega magnitudes of our sample. 2002es-like \Ia\ are orange, 2003fg-like \Ia\ are blue, and other \Ia\ spectral sub-types are gray. Note that errors within an individual SN light curve are correlated because the distance and extinction uncertainties are added in quadrature with photometric uncertainties. 
    }
    \label{fig:all_mags}
\end{figure*}

The 2002es-like and 2003fg-like \Ia\ in our sample have \textcolor{black}{light curves which deviate from the average SN Ia light curve,} as shown in Figure \ref{fig:all_mags}. First, the two 2002es-like \Ia\ (iPTF14atg and SN~2019yvq) are underluminous at optical wavelengths yet \textcolor{black}{more UV luminous than their optical luminosity would suggest}, especially in the $UVM2$ band. Second, the 2003fg-like SNe~2020hvf, 2021zny, and 2022ilv are overluminous in both optical and UV wavelengths. This behavior conflicts with the paradigm that more luminous \Ia\ should be powered by more \Nifs, which in turn increases the opacity, reducing the ratio of UV to optical emission \citep{Lentz00, Walker12, DerKacy20}. The observed UV brightness may be from the shock heating of an envelope of H/He-devoid material around the SN \citep{Piro16, Maeda23}, which is consistent with the findings of \citet{Ashall21}, who found the most likely progenitor system was one within an envelope.

Finally, SNe~2020hvf and 2021zny peak\footnote{The main light curve peak from \Nifs\ decay.} in the UV much earlier than the other \Ia. While iPTF14atg and SN~2019yvq both have UV peaks that are approximately concurrent with their optical maxima, the $UVM2$ peaks of SN~2020hvf and SN~2021zny are at least $\sim$10~days earlier than the optical peak. We fit the $UVM2$- and $B$-band light curves with the template-independent Gaussian process method in \verb|SNooPy| \citep{Burns11,Burns14} to find the respective peak times, $t_{max}^{UVM2}$ and $t_{max}^B$. For iPTF14atg and SN~2019yvq, $t_{max}^{UVM2}$ is $\sim$3.6~days and $\sim$0.9~days prior to \timeB, respectively. For SN~2020hvf and SN~2021zny, $t_{max}^{UVM2}$ is 8.9~days and 10.8~days prior to \timeB, respectively. 

\section{Optical and UV Colors}\label{sec:colors}

Figure \ref{fig:all_colors} shows each unique color permutation of the \emph{Swift} filters using the \textcolor{black}{apparent magnitudes to compute the color}. We find two disparate UV color evolutions in Figure \ref{fig:all_colors}, so we split our sample into two groups based on the UV color evolution. To avoid confusion with the existing terms of UV-blue and UV-red from \citet{Milne13}, we adopt different terms to refer to these groups. We define Group 1 as \Ia\ with $UVM2-UVW1 > 1.5$~mag at $t=-10$~days, and the inverse is true for Group 2. Interestingly, these groupings also correspond to differentiation by spectral classification, with Group 2 consisting solely of 2002es-like and 2003fg-like \Ia, whereas Group 1 consists of other spectral subtypes (e.g., normal \Ia, 1999aa-like \Ia, fast-declining/transitional \Ia). The majority of \Ia\ are in Group 1. Group 2 consists of the 2003fg-like \Ia\ reference sample and iPTF14atg, SN~2019yvq, SN~2020hvf, SN~2021zny, and SN~2022ilv. 

\begin{figure*}
    \includegraphics[width=\textwidth]{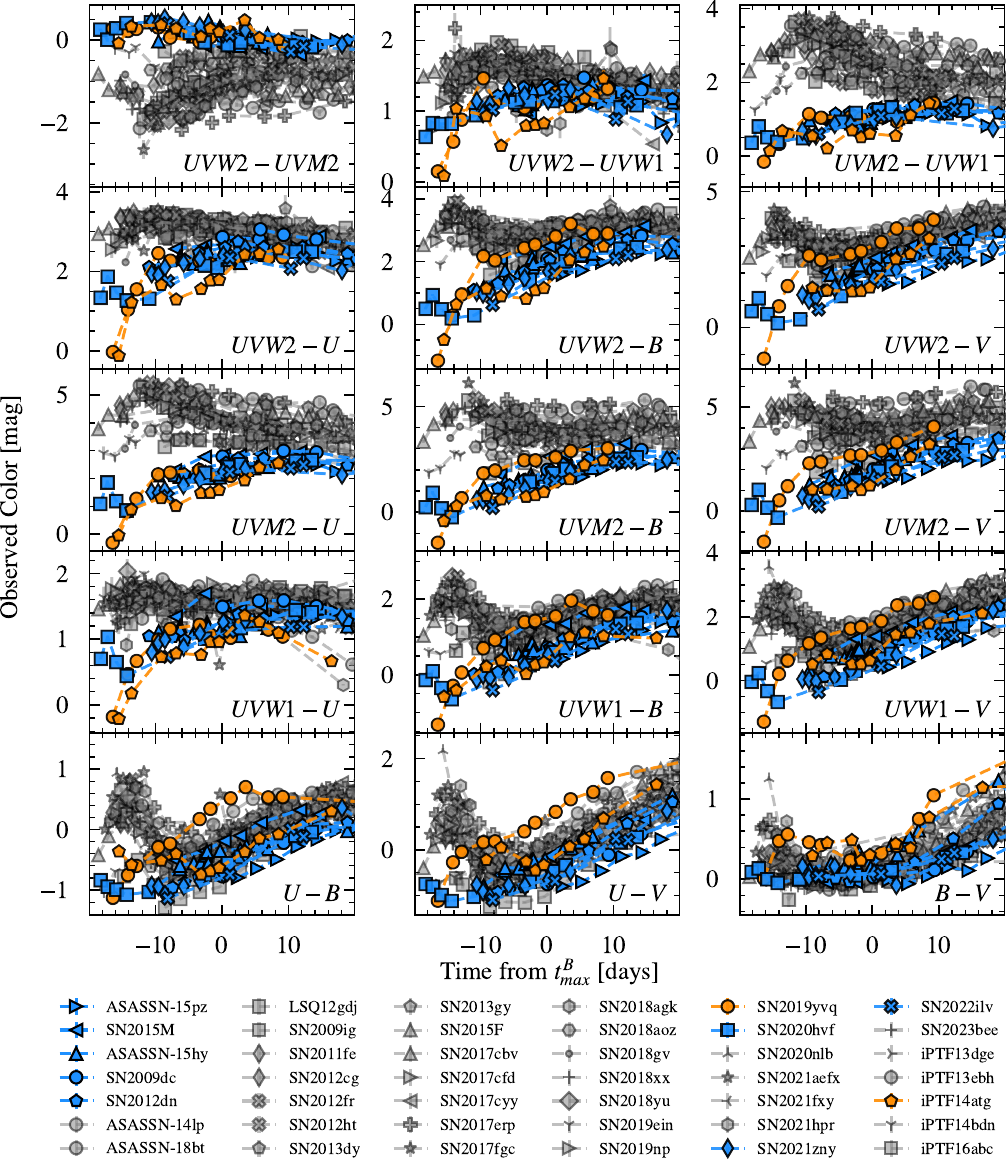} \\
    \caption{Each unique \textcolor{black}{\emph{Swift}} observed color curve of our sample. The color scheme is the same as Figure \ref{fig:all_mags}. 
    \emph{Row 1:} Observed UV colors of our sample. 
    \emph{Row 2:} Observed $UVW2-U/B/V$ colors of our sample.
    \emph{Row 3:} Observed $UVM2-U/B/V$ colors of our sample.
    \emph{Row 4:} Observed $UVW1-U/B/V$ colors of our sample.
    \emph{Row 5:} Observed optical colors of our sample.
    }
    \label{fig:all_colors}
\end{figure*}

\subsection{UV Colors}
In the top row of Figure \ref{fig:all_colors}, Group 2 \Ia\ are bluer than Group 1 \Ia\ in the $UVW2-UVW1$ and $UVM2-UVW1$ color curves, whereas in the $UVW2-UVM2$ Group 2 \Ia\ are redder. This is due to the dilution of the UV excess due to the \emph{Swift} $UVW2$ filter transmission, i.e., the same amount of UV flux in the $UVW2$ filter will have a small effect on the UV+optical flux, whereas the excess flux is a larger percentage in the $UVM2$ filter. In the $UVW2-UVM2$ color curve, Group 2 \Ia\ have $(UVW2-UVM2) > -0.2$ mag from explosion to \timeB. After \timeB, these same \Ia\ remain redder than other \Ia. 

While Group 2 \Ia\ are slightly separated from the rest of the sample in the $UVW2-UVW1$ color curve, they are closer to the rest of the sample in $UVW2-UVW1$ than $UVW2-UVM2$ and $UVM2-UVW1$. 
   
The increased similarity between Group 1 and Group 2 \Ia\ in the $UVW2-UVW1$ color curve may originate from the red leak. In extreme cases, the optical component from the red leak provides over half of the total flux (e.g., \citealp{Brown10}). Thus, contamination from optical light dilutes the observed difference from additional UV flux. The difference is still observed with filters with the red leak, demonstrating that abnormal behavior does not arise from the red leak.

\subsection{UV$-$Optical Colors}
The $UVW2-U/B/V$ color curves are all characterized by the same rapid redward ascent of Group 2 \Ia. Prior to $t_{B}^{max}$, Group 2 \Ia\ are all  $\lesssim$2.6 mag, whereas Group 1 \Ia\ have $2.5$ mag $\le (UVW2-U) \le 3.5$ mag. After \timeB, Group 2 \Ia\ are on the blue edge of the Group 1 distribution. Like the $UVW2-U/B/V$ color curves, Group 2 \Ia\ become similar to the rest of our sample in the $UVW1-U/B/V$ shortly after explosion. Overall, the $UVW1$ colors evolve in the same manner as the $UVW2$ filter.

Because the $UVM2$ filter does not have a red leak, this filter is the best UV probe, so the $UVM2-U/B/V$ color curves are the most important to consider. Like the other UV$-$optical colors, Group 2 \Ia\ quickly rise redward in the $UVM2-U/B/V$ color curves, initially dominated by the UV. The $UVM2-U$ color curve shows that even at $t_{B}^{max}$, Group 2 \Ia\ have different colors, and this difference persists until $t\approx+$15 days. 
The $UVM2-B$ and $UVM2-V$ color curves are generally similar to the $UVM2-U$ color curve. 

\begin{deluxetable}{ccc}
\tablenum{2}
\tablecaption{
Color differences between 2002es-like, 2003fg-like, and other SNe Ia. The colors are calculated by taking the median color for all SNe Ia in each category in the epochal range of $-12 \le t \le -8$~days. Uncertainties are $1\sigma$.
}\label{tab:color_differences}
\tablewidth{0pt}
\tablehead{
\colhead{Color} & \colhead{2002es-like and 2003fg-like} & \colhead{Other SNe Ia}}
\startdata
$UVW2-UVM2$	&	$0.41\pm0.16$	&	$-1.51\pm0.92$	\\
$UVW2-UVW1$	&	$1.03\pm0.16$	&	$1.62\pm0.24$	\\
$UVM2-UVW1$	&	$0.60\pm0.22$	&	$3.03\pm1.11$	\\
$UVW2-U$	&	$1.95\pm0.31$	&	$3.20\pm0.53$	\\
$UVW2-B$	&	$1.05\pm0.48$	&	$2.87\pm0.81$	\\
$UVW2-V$	&	$1.15\pm0.64$	&	$2.97\pm0.82$	\\
$UVM2-U$	&	$1.51\pm0.32$	&	$4.79\pm1.42$	\\
$UVM2-B$	&	$0.60\pm0.51$	&	$4.33\pm1.72$	\\
$UVM2-V$	&	$0.76\pm0.64$	&	$4.39\pm1.67$	\\
$UVW1-U$	&	$0.97\pm0.14$	&	$1.61\pm0.30$	\\
$UVW1-B$	&	$0.03\pm0.30$	&	$1.31\pm0.62$	\\
$UVW1-V$	&	$0.06\pm0.45$	&	$1.39\pm0.66$	\\
$U-B$	    &	$-0.93\pm0.26$	&	$-0.35\pm0.46$	\\
$U-V$	    &	$-0.84\pm0.42$	&	$-0.24\pm0.52$	\\
$B-V$	    &	$0.09\pm0.15$	&	$0.06\pm0.13$	\\
\enddata
\end{deluxetable}

\begin{figure*}
    \includegraphics[width=\textwidth]{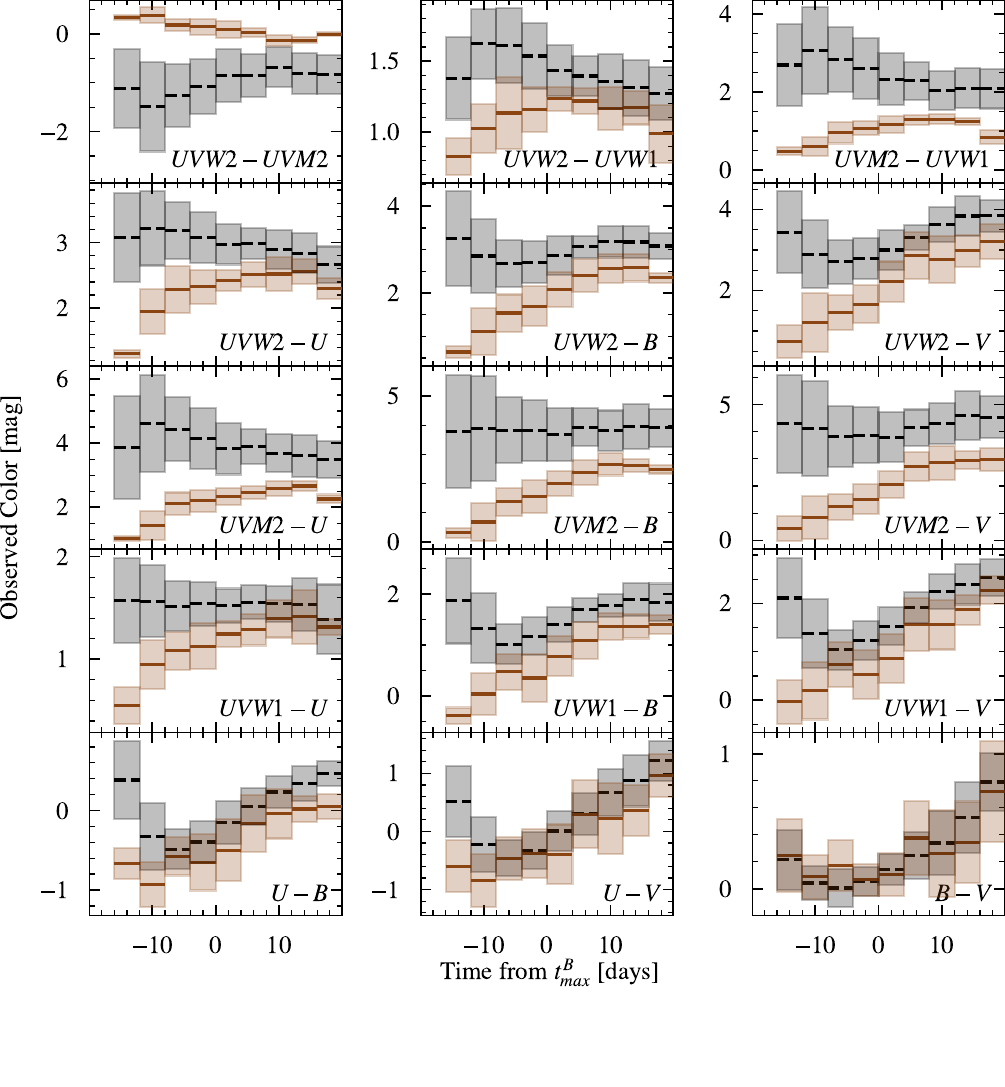}
    \caption{\textcolor{black}{The median (line) and $1\sigma$ uncertainty (shaded region) for the 2002es-like and 2003fg-like (brown) and other (grey) SNe Ia for every \emph{Swift} color combination. The 2002es-like SNe Ia and 2003fg-like SNe Ia have different UV colors throughout their evolution. The intrinsic scatter is larger than the photometric uncertainty, so we take a single averaged color for each SN Ia in each bin to avoid bias in the scatter determination.}
    }
    \label{fig:all_colors}
\end{figure*}

\textcolor{black}{For the $UVM2-U$ color, we compute the significance of observing a color difference between 2002es-like and 2003fg-like SNe Ia and the other spectral subtypes in the $-12 \le t \le -8$~day range. The color range for the rest of our sample is $4.79\pm1.42$~mag. For each of the 2002es-like and 2003fg-like SNe Ia, we compute the significance of the difference between the rest of the sample and the individual SN Ia. For example, the $UVM2-U$ color for iPTF14atg is $1.30\pm0.17$~mag, which means there is a $2.44\sigma$ tension between the average of the rest of the sample and iPTF14atg. This calculation is repeated for each SN Ia. The sum of the significance values is the composite significance (see \citealt{Fisher70}), which is $\sim$$13.9\sigma$. Finally, using the same method to compute the average $UVM2-U$ color for the 2002es-like and 2003fg-like SNe Ia gives values of $1.51\pm0.33$~mag and $1.73\pm0.43$~mag, respectively. These values are consistent with each other, confirming the visual evidence that these subtypes have matching UV colors.}

\subsection{Optical Colors}
There are two phenomena in the optical color curves shown in the bottom row of Figure \ref{fig:all_colors}. First, Group 2 \Ia\ are not as uniform as in the UV, and second, Group 2 \Ia\ are not wholly different than Group 1 \Ia\ near and after \timeB.

There are several interesting features in the optical colors. First, in the $U-B$ and $U-V$ colors, SN~2019yvq is redder than other Group 2 \Ia. \textcolor{black}{As the ejecta slow, they are less blueshifted, so absorption features in the $B$ and $V$ bands become split between the bands, causing SN~2019yvq to appear redder than the other Group 2 \Ia.} Second, iPTF14atg evolves similarly to SN~2019yvq when $-15 \leq t \leq -7$~days, but then evolves similarly to the other Group 1 members when $t \geq -7$. Either SN~2019yvq is evolving on an earlier time scale than the other bump \Ia\ and other early-time \Ia, similar to the $V-r$ color evolution in subluminous \Ia\ (see Figure 8 in \citealt{Hoogendam22}), or the mechanism driving the first inflection point in the $U-B$ and $U-V$ colors for iPTF14atg may be absent from SN~2019yvq. Finally, in the $B-V$ color curve, it is difficult to differentiate Group 2 \Ia\ and the 2003fg-like \Ia\ from Group 1 \Ia\ in our sample.

\section{Discussion}\label{sec:discussion} 
\textcolor{black}{\citet{Milne13} previously studied the UV light curves of SNe Ia. Our color plots are similar to Figure 3 in \citet{Milne13}, albeit with a larger plotted range that obscures the similarity of the figures (the increased range is due to the inclusion of 2002es-like and 2003fg-like SNe Ia). Direct comparison with \citet{Milne13} is difficult since only two SNe Ia (SNe 2009ig and 2011fe) are shared between the samples. SN 2009ig is ``NUV-red,'' and SN 2011fe is ``NUV-blue.'' In our analysis, both SNe 2009ig and 2011fe are categorized as ``single'' SN Ia. When comparing the colors in Figure \ref{fig:all_colors}, both the ``single'' and ``double'' SNe Ia are located in both the ``NUV-red'' and ``NUV-blue'' regions. This suggests no strong link exists between the \citet{Milne13} classification and the rising light curve.}

\subsection{The Interesting Bump Cases}\label{sec:bumps}
\begin{figure}
    \includegraphics[width=\linewidth]{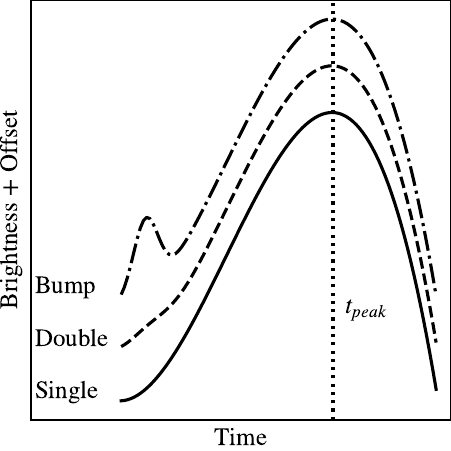}
    \caption{
    Idealized light curve categorization examples for bumps (dot-dash), double power law (dashed), and single power law (solid). The peak brightness for each category has been vertically offset for visual clarity.
    }
    \label{fig:cats}
\end{figure}

\begin{figure*}
    \centering{
    \includegraphics[width=\textwidth]{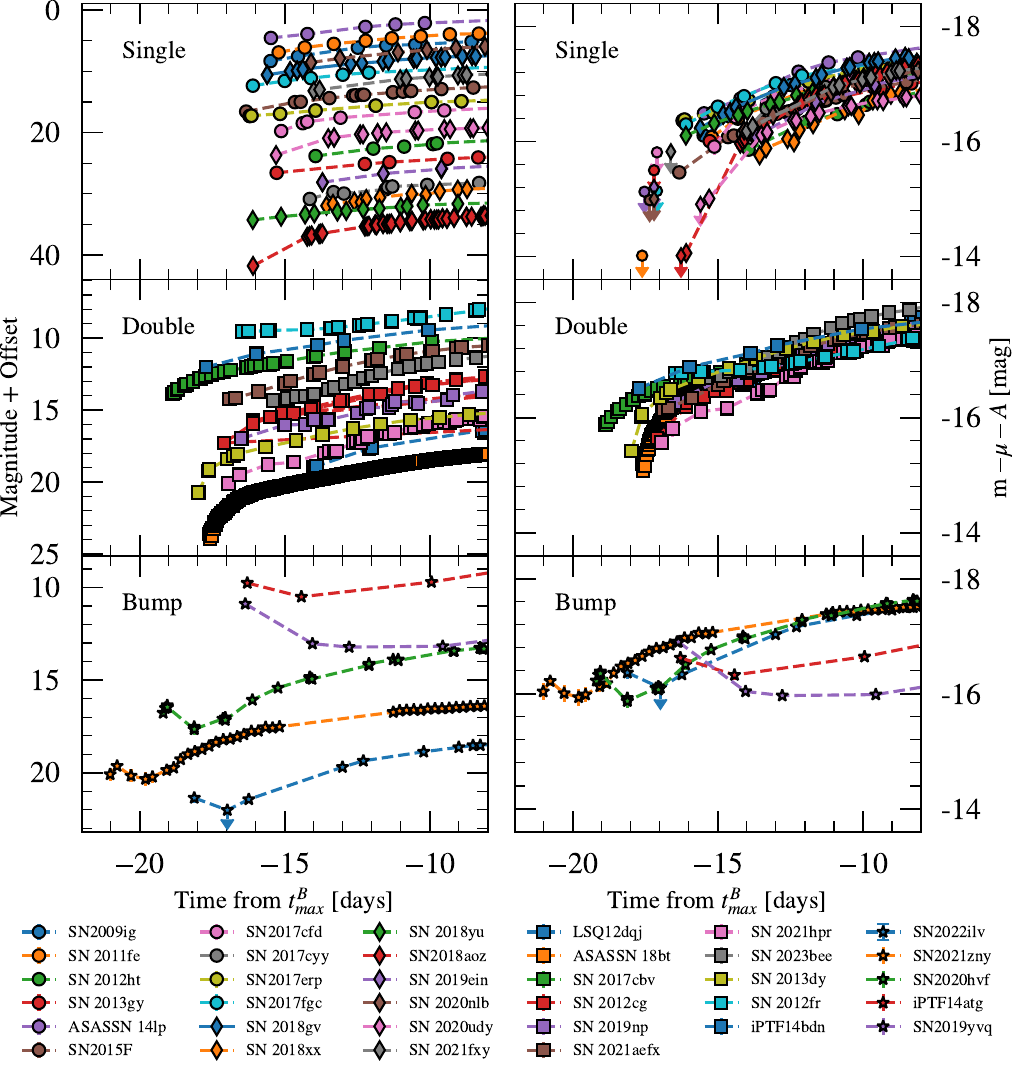}}
    \caption{
    \textcolor{black}{Non-\emph{Swift}} light curves grouped by category. For several \Ia, data were not made available after publication. In these cases, the figures containing the data were digitalized. \textcolor{black}{Categorization information for each SN Ia is in the appendix.}
    \emph{Left:} Apparent magnitudes with offset. The top and middle panels use $B$/$V$-band data, whereas the bottom panel uses $UVM2$ data for iPTF14atg, SN~2019yvq, ATLAS data for SN~2020hvf and SN~2022ilv, and TESS data for SN~2021zny. The left panel demonstrates the differences between single, double, and bump \Ia. If not observed early enough, bump \Ia\ may look like double \Ia. Likewise, double \Ia\ not observed early enough would look like single \Ia. Our sample selection should prevent this issue from influencing our results because we select \Ia\ with data before $-15$~d. \emph{Right:} Extinction-corrected absolute magnitudes. For the single and double panels, $B$-band data is used where available. Otherwise, $g$- or $V$-band data is used. For the bump panel, the data is the same as the left-hand side. The upper limits are taken from the latest available 3$\sigma$ upper limit or detections with $S/N < 5$, regardless of the bandpass. Interestingly, at concurrent times the \Ia\ with two-component rises are more luminous than \Ia\ with single-component rises, and additionally, the bump 2003fg-like \Ia\ appear to have longer rise times than the bump 2002es-like \Ia.
    }
    \label{fig:cats_obs}
\end{figure*}

We separate the rising light curves of \Ia\ into three categories as shown in Figure \ref{fig:cats}: ``single'', ``double'', and ``bump''. Single \Ia\ have rising light curves well fit by a single power law, whereas double \Ia\ have rising light curves better fit by a broken or two-component power law. Bump \Ia\ have non-monotonic light curve bumps in the UV and/or the optical (i.e., the flux decreases by at least 1$\sigma$ between any two epochs during the light curve rise). Figure \ref{fig:cats} and the left-hand panel of Figure \ref{fig:cats_obs} elucidate these different rising light curve behaviors using idealized and observed light curves, respectively. The \Ia\ with bluer UV colors, iPTF14atg, SN~2019yvq, SN~2020hvf, SN~2021zny, and SN~2022ilv, are the only bump \Ia, and no spectroscopically normal \Ia\ display bumps by our definition in their rising light curve, despite composing the majority of our sample and the majority of SN Ia volumetrically \citep{Desai23}. Thus, the color scheme in Figures \ref{fig:all_mags} and \ref{fig:all_colors} also correlates with rising light curve morphology. The blue and orange points correspond to \Ia\ with a rising light curve bump and the grey points correspond to \Ia\ without a rising light curve bump.
However, we note that simultaneous high-cadence, high-signal-to-noise, early-time optical and UV data do not exist for most of the supernovae in our sample since the current observations come from heterogeneous survey and follow-up efforts. For example, there are no simultaneous UV observations of 2003fg-like \Ia\ since all the bumps are observed in the optical.  Conversely, both 2002es-like \Ia\ in our sample have observed UV bumps but iPTF14atg does not show a clear bump in its optical light curves due to a gap in ground-based optical coverage at these epochs and low signal-to-noise in in the \emph{Swift} optical observations. 
Ultimately, simultaneous high-cadence, high-signal-to-noise optical and UV data are needed for further study of \Ia\ rising light curves.

When considering the colors of the bump \Ia, the observed dichotomy of UV colors is inconsistent with the excess luminosity originating from interaction with a companion, which is viewing-angle dependent \citep{Kasen10, Brown12a}. This implies companion interaction and other viewing-angle dependent models are insufficient to explain \emph{both} two-component rising light curves and rising light curve bumps. Thus, monotonically and non-monotonically rising light curves may have different physical origins. Despite our small sample, we can disfavor companion interaction as the cause of the early-time light curve bumps. Using the same viewing-angle argument as \citet{Burke21}, the probability of observing bumps in five out of five 2002es-like and 2003fg-like \Ia\ is 1 in $10^5$. Underlying this claim is the assumption that the viewing angle is independent of the spectral subtype, which seems likely to be true but observations do not guarantee this is reality, and that the companion interaction results in the observables predicted by \citet{Kasen10}. Thus, at the population level, the observed rate of \Ia\ with rising light curve bumps is too high to be fully explained by companion interaction.

Intriguingly, SN~2020hvf, SN~2021zny, and SN~2022ilv all display rising light curve bumps \citep{Jiang21, Dimitriadis23, Srivastav23a}, and Figure \ref{fig:all_colors} shows that the color evolutions of SN~2020hvf, SN~2021zny, and SN~2022ilv are consistent with the regular 2003fg-like \Ia\ in our comparison sample. This would naively suggest that other 2003fg-like \Ia\ have unobserved early-time light curve bumps. Such a conclusion remains speculative due to the small sample size. Initial analysis of SN~2022pul suggests there may not be a rising light curve bump \citealp{Kwok23, Siebert23b}, but an in-depth study of the rising light curve is yet to be published. Similarly, iPTF14atg, SN~2019yvq are the only 2002es-like \Ia\ with UV data observed early enough to detect a bump. Three other 2002es-like \Ia, iPTF14dpk \citep{Cao16, Burke21}, SN~2022ywc \citep{Srivastav23b}, and SN~2022vqz \citep{Xi23}, have reported rising light curve bumps but no UV data. While the current sample is small, we can obtain preliminary statistical conclusions despite the small-number statistics.

Assuming the \Ia\ in our sample are representative, we can compute the \textcolor{black}{minimum} fraction of 2002es-like and 2003fg-like SNe Ia that display a rising light curve bump at a $90\%$ confidence level \textcolor{black}{using binomial statistics. Given $b$ detected bumps from $n$ \Ia, what is the minimum percent of the population that has detectable bumps such that there is only a 10\% chance of the observations randomly occurring?} For 2002es-like \Ia, observing 2 rising light curve bumps from our 2-object sample implies that at least $32\%$ of early-time 2002es-like SN Ia light curves display a rising light curve bump. Similarly, for our sample of 2003fg-like \Ia, at least $47\%$ of 2003fg-like \Ia\ display a rising light curve bump \textcolor{black}{($b=n=3$)}. Combining our sample of 2002es-like and 2003fg-like \Ia\ and assuming 2002es-like and 2003fg-like \Ia\ arise from the same distribution, then at least 63\% of the \Ia\ display a rising light curve bump \textcolor{black}{($b=n=5$)}. Finally, if we include 2002es-like and 2003fg-like SNe Ia that lack UV data in our statistical analysis, then 8 out of 9 SNe display a rising light curve bump, and thus, at least 65\% of \Ia\ in these subtypes display a rising light curve bump.

\subsection{A Link Between 2002es-like \Ia\ and 2003fg-like \Ia}

\begin{figure*}
    \includegraphics{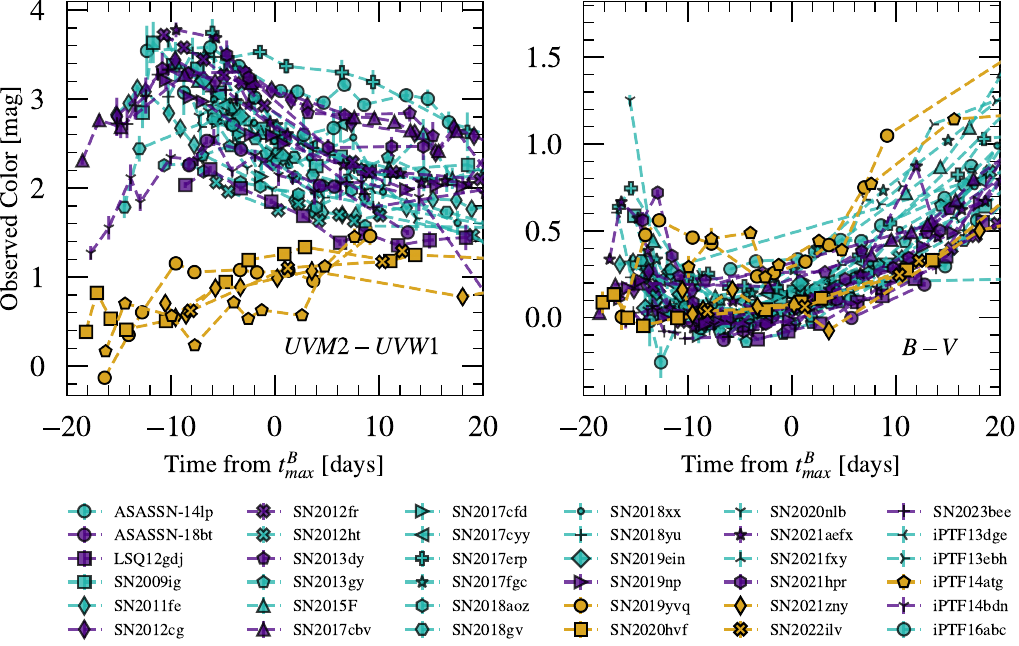}
    \caption{Enlarged \textcolor{black}{\emph{Swift}} color curves of the sample colored by the shape of the rising light curve: single power law (cyan), double power law (purple), and bump (khaki).
    \emph{Left:} The $UVM2-UVW1$ color curve; all bump \Ia\ are below $\sim$1.5 mag until $t\approx10$~days after \timeB. Single and double \Ia\ are above $\sim$1.5~mag until $t\approx10$~days. We expect future SN with $UVM2-UVW1$ colors below this region to be 2002es-like or 2003fg-like.
    \emph{Right:} The $B-V$ color curve shows minimal distinction between the different groups if any at all. 
    }
    \label{fig:Super_C_comp}
\end{figure*}

\citet{Brown14a} show that 2003fg-like \Ia\ are different from normal as well as 1991T-like \Ia\ in the $UVW1-V$ and $UVM2-UVW1$ colors. This work expands that analysis with color evolution data at $t\le-10$~days for more \Ia\ and the inclusion of three additional 2003fg-like \Ia\ and two 2002es-like \Ia.

There are several observational similarities between 2002es-like and 2003fg-like \Ia. Both classes lack a secondary $i$-band maximum \citep{Ashall21, Burke21}, have some members with nebular [\OI] emission \citep{Taubenberger13a, Taubenberger13b, Kromer16, Taubenberger19, Dimitriadis23, Siebert23b}, and have members with \CII\ absorption of varying strengths in their near-peak optical spectra \citep{Kromer13, Cao15, Ashall21, Li23, Siebert23b}.  Lastly, 2003fg-like \Ia\ prefer metal-poor, young environments (e.g., \citealt{ Lu21}, Galbany et al. in prep), whereas \citet{White15} suggest 2002es-like \Ia\ prefer older, elliptical galaxies, but an in-depth study has not yet been performed for the local environments of 2002es-like \Ia.

Of the observational similarities between the two subtypes, the shared carbon absorption is perhaps the most intriguing. 
One proposed source for the \CII\ feature is a C-rich envelope either from the merger of two CO WDs in the DD scenario \citep[e.g.,][]{Moll14, Raskin13, Raskin14} or a CO WD and an AGB core in the CD scenario where the AGB star has lost its H and/or He envelope \citep[e.g.,][]{Lu21, Hsiao20, Ashall21}. \citet{Ashall21} find a weak linear relationship between the pEW of \CII\ $\lambda$6580\AA\ and \DmB\ where faster-declining 2003fg-like \Ia\ have a smaller pEW value. Therefore, if there is a link between the progenitor scenario and/or explosion mechanism of 2002es-like and 2003fg-like \Ia, one may expect 2002es-like \Ia\ to follow this relationship with 2002es-like \Ia\ showing weak \CII\ features. 

Observationally, only some 2002es-like \Ia\ display \CII\ absorption (e.g, SN~2010lp, \citealp{Kromer13}; iPTF14atg, \citealp{Cao15}; and SN~2016ije, \citealp{Li23}), while others completely lack \CII\ (e.g., PTF10ops, \citealp{Maguire11}; PTF10ujn, \citealp{White15}; PTF10acdh, \citealp{White15}; and SN~2019yvq, \citealp{Miller20}). However, 2002es-like \Ia\ without \CII\ may still have an envelope which could be observed in a lower ionization state of carbon. \CI\ $\lambda$10693\AA\ is the strongest \CI\ feature, and visual inspection of Figure 2 in \citet{Burke21} suggests the presence of \CI\ in SN~2019yvq. Unfortunately, no other 2002es-like \Ia\ have published NIR spectra near peak light.

If we assume 2002es-like and 2003fg-like \Ia\ are linked, different scenarios can be constructed to explain a potential continuum in either the DD scenario or the CD scenario. The first-order explosion parameters would be the mass of the carbon envelope and the mass of the primary degenerate object (CO WD or CO AGB core, depending on the progenitor scenario). We do not consider potential higher-order effects such as flame speed or burning efficiency in our brief qualitative comparisons.

A larger core mass will produce a more optically luminous explosion with higher ionization features. In this picture,  2002es-like SNe Ia will have smaller core masses because they have lower optical luminosities (cf. Figure 1) and show lower ionization lines (e.g., \citealt{Taubenberger17}), and conversely, 2003fg-like SNe Ia will have larger core masses due to their larger optical luminosities (cf. Figure 1) and higher-ionization spectral features (e.g., \citealt{Taubenberger17}). Less massive circumstellar material results in weaker C features (i.e., smaller pEWs) since these features trace envelope mass \citep{Ashall21}, whereas a larger envelope should produce stronger \CII\ and \CI\ absorption, more luminous peak magnitudes, and lower \SiII\ velocities.

In addition to a nearby envelope, both a violent merger in the DD scenario and the CD scenario may have material extending out to larger distances \citep{Kashi11, Raskin13, Hsiao20}. If that material is either launched in a wind or dynamically ejected at a constant velocity, then the extended material will have a r$^{-2}$ density profile (e.g., \citealp{Moriya23}). That can be compared to the expected r$^{-3}$ density profile of the nearby envelope \citep{Piro16}. Studies have investigated the effects of shock heating the envelope and wind separately, but to our knowledge, no study has simultaneously simulated both. \citet{Piro16} and \citet{Maeda23} show that a nearby envelope with an r$^{-3}$ density profile is expected to produce a distinct bump in the early-time light curves whereas \citet{Moriya23} show that a diffuse, r$^{-2}$ wind may provide persistent additional UV luminosity through maximum light. Qualitatively, the combination of these two effects may reproduce the observed 2003fg-like and 2002es-like light curves with their observed rising light curve bumps and persistent blue UV colors through maximum light. A combined envelope and wind-driven circumstellar medium may produce a correlation between the size of the $\sim 2$-day rising light curve bump or the UV luminosity and the combined pEW of the \CI\ and \CII\ features at early times.  Future theoretical and observational work is needed to test these qualitative considerations.

Finally, determining if 2003fg-like \Ia\ originate from the DD or CD scenario is another important open question. There are several predicted observational differences between the DD and CD scenarios. First, the DD scenario is predicted to show [\OI] emission during the nebular phase \citep{Pakmor12} which may arise from low-velocity, asymmetric oxygen distributions caused by incomplete burning during a violent merger \citep{Mazzali22}, whereas the CD scenario does not predict [\OI] emission. Second, a violent merger in the DD scenario is predicted to be aspherical \citep{Bulla16}; currently, no CD models offer predictions about polarization but intuitively one could expect these explosions to be spherical. Third and finally, the CD model of \citet{Lu21} predicts a high X-ray flux due to a fast-receding photosphere and low opacity.
This X-ray flux is as of yet undetected in any SN Ia to date \citep{Lu21}, and future X-ray studies of 2002fg-like and 2003fg-like \Ia\ are also needed.

\section{Conclusion}\label{sec:conclusions}

We present UV and optical photometry from \emph{Swift} compiled in the SOUSA catalog \citep{Brown14b} for \Ia\  with early-time optical observations and pre-\timeB\ UV photometry. The data can be categorized by either the UV color evolution or the rising light curve morphology, and we find that both categorization criteria separate 2002es-like and 2003fg-like \Ia\ from the other spectroscopic subtypes of \Ia. Observationally,
\begin{enumerate}
    \item 2002es-like and 2003fg-like \Ia\ are, on average, UV brighter than other \Ia\ (Figure \ref{fig:all_mags}),
    \item 2002es-like and 2003fg-like \Ia\ have extreme blue UV colors through 10 days after maximum (Figure \ref{fig:all_colors}), and
    \item 2002es-like and 2003fg-like \Ia\ are the only spectroscopic subtypes that exhibit rising light curve bumps (Figure \ref{fig:cats_obs}).
\end{enumerate}

Because 2002es-like and 2003fg-like \Ia\ show similar UV colors and also are the only spectroscopic subtypes to exhibit rising light curve bumps, we examine a potential relationship between these two subtypes. We propose a potential continuum between 2002es-like and 2003fg-like \Ia\ with the following progenitor properties. These \Ia\ may originate in low-metallicity DD or CD scenarios enshrouded by a carbon-rich circumstellar medium \citep{Kromer16, Lu21, Ashall21, Siebert23b, Kwok23}. 2003fg-like \Ia\ should have higher luminosities and 2002es-like \Ia\ are the lower-luminosity members of the continuum; the rising light curve bumps may arise from shock heating of an inner carbon envelope (e.g., \citealt{Piro16, Maeda23}) and wind-originated outer carbon material may cause the blue UV color evolution (e.g., \citealt{Moriya23}). Future theoretical modeling should include two-layer circumstellar medium initial conditions. 

With the advent of transient surveys (e.g., All-Sky Automated Survey for SuperNovae (ASAS-SN; \citealt{Shappee14, Kochanek17, Hart23}), Asteroid Terrestrial-impact Last Alert System (ATLAS; \citealt{Tonry18}), Panoramic Survey Telescope And Rapid Response System (Pan-STARRS; \citealt{Chambers16} and Zwicky Transient Facility (ZTF; \citealt{Bellm19}), rapid classification groups (e.g., SCAT; \citealp{Tucker22c}; ePESSTO+ \citealp{Smartt15}) and dedicated follow-up groups (e.g., POISE; \citealp{POISE_ATel}; YSE \citealp{Jones21}), many more \Ia\ will be quickly discovered and rapidly observed after the explosion. This is especially important for 2002es-like \Ia, which may initially be spectroscopically misclassified as 1991bg-like \Ia. Because of this, rapid follow-up programs with \emph{Swift} (e.g., \citealt{Brown23}) are doubly important. First, such programs increase the sample of well-observed \Ia\ in the UV, and more importantly, 2002es-like or 2003fg-like SN Ia can be differentiated from other subtypes with a single \emph{Swift} epoch. If $UVM2-UVW1 < 1.0$~mag, then the SN Ia is either 2002es-like or 2003fg-like, with the difference between 2002es-like and 2003fg-like \Ia\ determined by peak absolute magnitude. Furthermore, the planned ULTRASAT satellite \citep{Sagiv14} will provide many transient discoveries and early-time UV light curves, which are crucial for \Ia\ science and may display early-time light curve bumps. Ultimately, further early-time observations across the electromagnetic spectrum are needed of 2002es-like and 2003fg-like \Ia.

\bibliography{sample631}{}

\begin{thebibliography}{}
\expandafter\ifx\csname natexlab\endcsname\relax\def\natexlab#1{#1}\fi
\providecommand{\url}[1]{\href{#1}{#1}}
\providecommand{\dodoi}[1]{doi:~\href{http://doi.org/#1}{\nolinkurl{#1}}}
\providecommand{\doeprint}[1]{\href{http://ascl.net/#1}{\nolinkurl{http://ascl.net/#1}}}
\providecommand{\doarXiv}[1]{\href{https://arxiv.org/abs/#1}{\nolinkurl{https://arxiv.org/abs/#1}}}

\bibitem[{{Albareti} {et~al.}(2017){Albareti}, {Allende Prieto}, {Almeida},
  {Anders}, {Anderson}, {Andrews}, {Arag{\'o}n-Salamanca},
  {Argudo-Fern{\'a}ndez}, {Armengaud}, {Aubourg}, {Avila-Reese}, {Badenes},
  {Bailey}, {Barbuy}, {Barger}, {Barrera-Ballesteros}, {Bartosz}, {Basu},
  {Bates}, {Battaglia}, {Baumgarten}, {Baur}, {Bautista}, {Beers}, {Belfiore},
  {Bershady}, {Bertran de Lis}, {Bird}, {Bizyaev}, {Blanc}, {Blanton},
  {Blomqvist}, {Bolton}, {Borissova}, {Bovy}, {Brandt}, {Brinkmann},
  {Brownstein}, {Bundy}, {Burtin}, {Busca}, {Camacho Chavez}, {Cano D{\'\i}az},
  {Cappellari}, {Carrera}, {Chen}, {Cherinka}, {Cheung}, {Chiappini},
  {Chojnowski}, {Chuang}, {Chung}, {Cirolini}, {Clerc}, {Cohen}, {Comerford},
  {Comparat}, {Correa do Nascimento}, {Cousinou}, {Covey}, {Crane}, {Croft},
  {Cunha}, {Darling}, {Davidson}, {Dawson}, {Da Costa}, {Da Silva Ilha},
  {Deconto Machado}, {Delubac}, {De Lee}, {De la Macorra}, {De la Torre},
  {Diamond-Stanic}, {Donor}, {Downes}, {Drory}, {Du}, {Du Mas des Bourboux},
  {Dwelly}, {Ebelke}, {Eigenbrot}, {Eisenstein}, {Elsworth}, {Emsellem},
  {Eracleous}, {Escoffier}, {Evans}, {Falc{\'o}n-Barroso}, {Fan}, {Favole},
  {Fernandez-Alvar}, {Fernandez-Trincado}, {Feuillet}, {Fleming},
  {Font-Ribera}, {Freischlad}, {Frinchaboy}, {Fu}, {Gao}, {Garcia},
  {Garcia-Dias}, {Garcia-Hern{\'a}ndez}, {Garcia P{\'e}rez}, {Gaulme}, {Ge},
  {Geisler}, {Gillespie}, {Gil Marin}, {Girardi}, {Goddard}, {Gomez Maqueo
  Chew}, {Gonzalez-Perez}, {Grabowski}, {Green}, {Grier}, {Grier}, {Guo},
  {Guy}, {Hagen}, {Hall}, {Harding}, {Harley}, {Hasselquist}, {Hawley},
  {Hayes}, {Hearty}, {Hekker}, {Hernandez Toledo}, {Ho}, {Hogg},
  {Holley-Bockelmann}, {Holtzman}, {Holzer}, {Hu}, {Huber}, {Hutchinson},
  {Hwang}, {Ibarra-Medel}, {Ivans}, {Ivory}, {Jaehnig}, {Jensen}, {Johnson},
  {Jones}, {Jullo}, {Kallinger}, {Kinemuchi}, {Kirkby}, {Klaene}, {Kneib},
  {Kollmeier}, {Lacerna}, {Lane}, {Lang}, {Laurent}, {Law}, {Leauthaud}, {Le
  Goff}, {Li}, {Li}, {Li}, {Li}, {Liang}, {Liang}, {Lima}, {Lin}, {Lin}, {Lin},
  {Liu}, {Long}, {Lucatello}, {MacDonald}, {MacLeod}, {Mackereth}, {Mahadevan},
  {Maia}, {Maiolino}, {Majewski}, {Malanushenko}, {Malanushenko}, {Mallmann},
  {Manchado}, {Maraston}, {Marques-Chaves}, {Martinez Valpuesta}, {Masters},
  {Mathur}, {McGreer}, {Merloni}, {Merrifield}, {M{\'e}sz{\'a}ros}, {Meza},
  {Miglio}, {Minchev}, {Molaverdikhani}, {Montero-Dorta}, {Mosser}, {Muna},
  {Myers}, {Nair}, {Nandra}, {Ness}, {Newman}, {Nichol}, {Nidever},
  {Nitschelm}, {O'Connell}, {Oravetz}, {Oravetz}, {Pace}, {Padilla},
  {Palanque-Delabrouille}, {Pan}, {Parejko}, {Paris}, {Park}, {Peacock},
  {Peirani}, {Pellejero-Ibanez}, {Penny}, {Percival}, {Percival},
  {Perez-Fournon}, {Petitjean}, {Pieri}, {Pinsonneault}, {Pisani}, {Prada},
  {Prakash}, {Price-Jones}, {Raddick}, {Rahman}, {Raichoor}, {Barboza Rembold},
  {Reyna}, {Rich}, {Richstein}, {Ridl}, {Riffel}, {Riffel}, {Rix}, {Robin},
  {Rockosi}, {Rodr{\'\i}guez-Torres}, {Rodrigues}, {Roe}, {Roman Lopes},
  {Rom{\'a}n-Z{\'u}{\~n}iga}, {Ross}, {Rossi}, {Ruan}, {Ruggeri}, {Runnoe},
  {Salazar-Albornoz}, {Salvato}, {Sanchez}, {Sanchez}, {Sanchez-Gallego},
  {Santiago}, {Schiavon}, {Schimoia}, {Schlafly}, {Schlegel}, {Schneider},
  {Sch{\"o}nrich}, {Schultheis}, {Schwope}, {Seo}, {Serenelli}, {Sesar},
  {Shao}, {Shetrone}, {Shull}, {Silva Aguirre}, {Skrutskie}, {Slosar}, {Smith},
  {Smith}, {Sobeck}, {Somers}, {Souto}, {Stark}, {Stassun}, {Steinmetz},
  {Stello}, {Storchi Bergmann}, {Strauss}, {Streblyanska}, {Stringfellow},
  {Suarez}, {Sun}, {Taghizadeh-Popp}, {Tang}, {Tao}, {Tayar}, {Tembe},
  {Thomas}, {Tinker}, {Tojeiro}, {Tremonti}, {Troup}, {Trump}, {Unda-Sanzana},
  {Valenzuela}, {Van den Bosch}, {Vargas-Maga{\~n}a}, {Vazquez}, {Villanova},
  {Vivek}, {Vogt}, {Wake}, {Walterbos}, {Wang}, {Wang}, {Weaver}, {Weijmans},
  {Weinberg}, {Westfall}, {Whelan}, {Wilcots}, {Wild}, {Williams}, {Wilson},
  {Wood-Vasey}, {Wylezalek}, {Xiao}, {Yan}, {Yang}, {Ybarra}, {Yeche}, {Yuan},
  {Zakamska}, {Zamora}, {Zasowski}, {Zhang}, {Zhao}, {Zhao}, {Zheng}, {Zheng},
  {Zhou}, {Zhu}, {Zinn}, \& {Zou}}]{Albareti17}
{Albareti}, F.~D., {Allende Prieto}, C., {Almeida}, A., {et~al.} 2017, \apjs,
  233, 25, \dodoi{10.3847/1538-4365/aa8992}

\bibitem[{{Allison} {et~al.}(2014){Allison}, {Sadler}, \& {Meekin}}]{Allison14}
{Allison}, J.~R., {Sadler}, E.~M., \& {Meekin}, A.~M. 2014, \mnras, 440, 696,
  \dodoi{10.1093/mnras/stu289}

\bibitem[{{Amanullah} {et~al.}(2014){Amanullah}, {Goobar}, {Johansson},
  {Banerjee}, {Venkataraman}, {Joshi}, {Ashok}, {Cao}, {Kasliwal}, {Kulkarni},
  {Nugent}, {Petrushevska}, \& {Stanishev}}]{Amanullah14}
{Amanullah}, R., {Goobar}, A., {Johansson}, J., {et~al.} 2014, \apjl, 788, L21,
  \dodoi{10.1088/2041-8205/788/2/L21}

\bibitem[{{Arnett}(1982)}]{Arnett82}
{Arnett}, W.~D. 1982, \apj, 253, 785, \dodoi{10.1086/159681}

\bibitem[{{Ashall} {et~al.}(2014){Ashall}, {Mazzali}, {Bersier}, {Hachinger},
  {Phillips}, {Percival}, {James}, \& {Maguire}}]{Ashall14}
{Ashall}, C., {Mazzali}, P., {Bersier}, D., {et~al.} 2014, \mnras, 445, 4427,
  \dodoi{10.1093/mnras/stu1995}

\bibitem[{{Ashall} {et~al.}(2021){Ashall}, {Lu}, {Hsiao}, {Hoeflich},
  {Phillips}, {Galbany}, {Burns}, {Contreras}, {Krisciunas}, {Morrell},
  {Stritzinger}, {Suntzeff}, {Taddia}, {Anais}, {Baron}, {Brown}, {Busta},
  {Campillay}, {Castell{\'o}n}, {Corco}, {Davis}, {Folatelli}, {F{\"o}rster},
  {Freedman}, {Gonzal{\'e}z}, {Hamuy}, {Holmbo}, {Kirshner}, {Kumar}, {Marion},
  {Mazzali}, {Morokuma}, {Nugent}, {Persson}, {Piro}, {Roth}, {Salgado},
  {Sand}, {Seron}, {Shahbandeh}, \& {Shappee}}]{Ashall21}
{Ashall}, C., {Lu}, J., {Hsiao}, E.~Y., {et~al.} 2021, \apj, 922, 205,
  \dodoi{10.3847/1538-4357/ac19ac}

\bibitem[{{Ashall} {et~al.}(2022){Ashall}, {Lu}, {Shappee}, {Burns}, {Hsiao},
  {Kumar}, {Morrell}, {Phillips}, {Shahbandeh}, {Baron}, {Boutsia}, {Brown},
  {DerKacy}, {Galbany}, {Hoeflich}, {Krisciunas}, {Mazzali}, {Piro},
  {Stritzinger}, \& {Suntzeff}}]{Ashall22}
{Ashall}, C., {Lu}, J., {Shappee}, B.~J., {et~al.} 2022, \apjl, 932, L2,
  \dodoi{10.3847/2041-8213/ac7235}

\bibitem[{{Beers} {et~al.}(1995){Beers}, {Kriessler}, {Bird}, \&
  {Huchra}}]{Beers95}
{Beers}, T.~C., {Kriessler}, J.~R., {Bird}, C.~M., \& {Huchra}, J.~P. 1995,
  \aj, 109, 874, \dodoi{10.1086/117329}

\bibitem[{{Bellm} {et~al.}(2019){Bellm}, {Kulkarni}, {Graham}, {Dekany},
  {Smith}, {Riddle}, {Masci}, {Helou}, {Prince}, {Adams}, {Barbarino},
  {Barlow}, {Bauer}, {Beck}, {Belicki}, {Biswas}, {Blagorodnova}, {Bodewits},
  {Bolin}, {Brinnel}, {Brooke}, {Bue}, {Bulla}, {Burruss}, {Cenko}, {Chang},
  {Connolly}, {Coughlin}, {Cromer}, {Cunningham}, {De}, {Delacroix}, {Desai},
  {Duev}, {Eadie}, {Farnham}, {Feeney}, {Feindt}, {Flynn}, {Franckowiak},
  {Frederick}, {Fremling}, {Gal-Yam}, {Gezari}, {Giomi}, {Goldstein},
  {Golkhou}, {Goobar}, {Groom}, {Hacopians}, {Hale}, {Henning}, {Ho}, {Hover},
  {Howell}, {Hung}, {Huppenkothen}, {Imel}, {Ip}, {Ivezi{\'c}}, {Jackson},
  {Jones}, {Juric}, {Kasliwal}, {Kaspi}, {Kaye}, {Kelley}, {Kowalski},
  {Kramer}, {Kupfer}, {Landry}, {Laher}, {Lee}, {Lin}, {Lin}, {Lunnan},
  {Giomi}, {Mahabal}, {Mao}, {Miller}, {Monkewitz}, {Murphy}, {Ngeow},
  {Nordin}, {Nugent}, {Ofek}, {Patterson}, {Penprase}, {Porter}, {Rauch},
  {Rebbapragada}, {Reiley}, {Rigault}, {Rodriguez}, {van Roestel}, {Rusholme},
  {van Santen}, {Schulze}, {Shupe}, {Singer}, {Soumagnac}, {Stein}, {Surace},
  {Sollerman}, {Szkody}, {Taddia}, {Terek}, {Van Sistine}, {van Velzen},
  {Vestrand}, {Walters}, {Ward}, {Ye}, {Yu}, {Yan}, \& {Zolkower}}]{Bellm19}
{Bellm}, E.~C., {Kulkarni}, S.~R., {Graham}, M.~J., {et~al.} 2019, \pasp, 131,
  018002, \dodoi{10.1088/1538-3873/aaecbe}

\bibitem[{{Bilicki} {et~al.}(2014){Bilicki}, {Jarrett}, {Peacock}, {Cluver}, \&
  {Steward}}]{Bilicki14}
{Bilicki}, M., {Jarrett}, T.~H., {Peacock}, J.~A., {Cluver}, M.~E., \&
  {Steward}, L. 2014, \apjs, 210, 9, \dodoi{10.1088/0067-0049/210/1/9}

\bibitem[{{Blondin} \& {Tonry}(2007)}]{Blondin07}
{Blondin}, S., \& {Tonry}, J.~L. 2007, \apj, 666, 1024, \dodoi{10.1086/520494}

\bibitem[{{Blondin} {et~al.}(2012){Blondin}, {Matheson}, {Kirshner}, {Mandel},
  {Berlind}, {Calkins}, {Challis}, {Garnavich}, {Jha}, {Modjaz}, {Riess}, \&
  {Schmidt}}]{Blondin12}
{Blondin}, S., {Matheson}, T., {Kirshner}, R.~P., {et~al.} 2012, \aj, 143, 126,
  \dodoi{10.1088/0004-6256/143/5/126}

\bibitem[{{Breeveld} {et~al.}(2011){Breeveld}, {Landsman}, {Holland}, {Roming},
  {Kuin}, \& {Page}}]{Breeveld11}
{Breeveld}, A.~A., {Landsman}, W., {Holland}, S.~T., {et~al.} 2011, in American
  Institute of Physics Conference Series, Vol. 1358, Gamma Ray Bursts 2010, ed.
  J.~E. {McEnery}, J.~L. {Racusin}, \& N.~{Gehrels}, 373--376,
  \dodoi{10.1063/1.3621807}

\bibitem[{{Brown} {et~al.}(2014{\natexlab{a}}){Brown}, {Breeveld}, {Holland},
  {Kuin}, \& {Pritchard}}]{Brown14b}
{Brown}, P.~J., {Breeveld}, A.~A., {Holland}, S., {Kuin}, P., \& {Pritchard},
  T. 2014{\natexlab{a}}, \apss, 354, 89, \dodoi{10.1007/s10509-014-2059-8}

\bibitem[{{Brown} \& {Crumpler}(2020)}]{Brown20}
{Brown}, P.~J., \& {Crumpler}, N.~R. 2020, \apj, 890, 45,
  \dodoi{10.3847/1538-4357/ab66b3}

\bibitem[{{Brown} {et~al.}(2012){Brown}, {Dawson}, {Harris}, {Olmstead},
  {Milne}, \& {Roming}}]{Brown12a}
{Brown}, P.~J., {Dawson}, K.~S., {Harris}, D.~W., {et~al.} 2012, \apj, 749, 18,
  \dodoi{10.1088/0004-637X/749/1/18}

\bibitem[{{Brown} {et~al.}(2017){Brown}, {Landez}, {Milne}, \&
  {Stritzinger}}]{Brown17}
{Brown}, P.~J., {Landez}, N.~J., {Milne}, P.~A., \& {Stritzinger}, M.~D. 2017,
  \apj, 836, 232, \dodoi{10.3847/1538-4357/aa5f5a}

\bibitem[{{Brown} {et~al.}(2018){Brown}, {Perry}, {Beeny}, {Milne}, \&
  {Wang}}]{Brown18}
{Brown}, P.~J., {Perry}, J.~M., {Beeny}, B.~A., {Milne}, P.~A., \& {Wang}, X.
  2018, \apj, 867, 56, \dodoi{10.3847/1538-4357/aae1ad}

\bibitem[{{Brown} {et~al.}(2023){Brown}, {Robertson}, {Devarakonda}, {Sarria},
  {Pooley}, \& {Stritzinger}}]{Brown23}
{Brown}, P.~J., {Robertson}, M., {Devarakonda}, Y., {et~al.} 2023, Universe, 9,
  218, \dodoi{10.3390/universe9050218}

\bibitem[{{Brown} {et~al.}(2010){Brown}, {Roming}, {Milne}, {Bufano},
  {Ciardullo}, {Elias-Rosa}, {Filippenko}, {Foley}, {Gehrels}, {Gronwall},
  {Hicken}, {Holland}, {Hoversten}, {Immler}, {Kirshner}, {Li}, {Mazzali},
  {Phillips}, {Pritchard}, {Still}, {Turatto}, \& {Vanden Berk}}]{Brown10}
{Brown}, P.~J., {Roming}, P. W.~A., {Milne}, P., {et~al.} 2010, \apj, 721,
  1608, \dodoi{10.1088/0004-637X/721/2/1608}

\bibitem[{{Brown} {et~al.}(2014{\natexlab{b}}){Brown}, {Kuin}, {Scalzo},
  {Smitka}, {de Pasquale}, {Holland}, {Krisciunas}, {Milne}, \&
  {Wang}}]{Brown14a}
{Brown}, P.~J., {Kuin}, P., {Scalzo}, R., {et~al.} 2014{\natexlab{b}}, \apj,
  787, 29, \dodoi{10.1088/0004-637X/787/1/29}

\bibitem[{{Brown} {et~al.}(2019){Brown}, {Hosseinzadeh}, {Jha}, {Sand},
  {Vieira}, {Wang}, {Dai}, {Dettman}, {Mould}, {Uddin}, {Wang}, {Arcavi},
  {Bento}, {Burns}, {Diamond}, {Hiramatsu}, {Howell}, {Hsiao}, {Marion},
  {McCully}, {Milne}, {Mirzaqulov}, {Ruiter}, {Valenti}, \& {Xiang}}]{Brown19}
{Brown}, P.~J., {Hosseinzadeh}, G., {Jha}, S.~W., {et~al.} 2019, \apj, 877,
  152, \dodoi{10.3847/1538-4357/ab1a3f}

\bibitem[{{Brown} {et~al.}(2013){Brown}, {Baliber}, {Bianco}, {Bowman},
  {Burleson}, {Conway}, {Crellin}, {Depagne}, {De Vera}, {Dilday}, {Dragomir},
  {Dubberley}, {Eastman}, {Elphick}, {Falarski}, {Foale}, {Ford}, {Fulton},
  {Garza}, {Gomez}, {Graham}, {Greene}, {Haldeman}, {Hawkins}, {Haworth},
  {Haynes}, {Hidas}, {Hjelstrom}, {Howell}, {Hygelund}, {Lister}, {Lobdill},
  {Martinez}, {Mullins}, {Norbury}, {Parrent}, {Paulson}, {Petry}, {Pickles},
  {Posner}, {Rosing}, {Ross}, {Sand}, {Saunders}, {Shobbrook}, {Shporer},
  {Street}, {Thomas}, {Tsapras}, {Tufts}, {Valenti}, {Vander Horst}, {Walker},
  {White}, \& {Willis}}]{Brown13}
{Brown}, T.~M., {Baliber}, N., {Bianco}, F.~B., {et~al.} 2013, \pasp, 125,
  1031, \dodoi{10.1086/673168}

\bibitem[{{Bulla} {et~al.}(2016){Bulla}, {Sim}, {Pakmor}, {Kromer},
  {Taubenberger}, {R{\"o}pke}, {Hillebrandt}, \& {Seitenzahl}}]{Bulla16}
{Bulla}, M., {Sim}, S.~A., {Pakmor}, R., {et~al.} 2016, \mnras, 455, 1060,
  \dodoi{10.1093/mnras/stv2402}

\bibitem[{{Bureau} {et~al.}(1996){Bureau}, {Mould}, \&
  {Staveley-Smith}}]{Bureau96}
{Bureau}, M., {Mould}, J.~R., \& {Staveley-Smith}, L. 1996, \apj, 463, 60,
  \dodoi{10.1086/177222}

\bibitem[{{Burgaz} {et~al.}(2021){Burgaz}, {Maeda}, {Kalomeni}, {Kawabata},
  {Yamanaka}, {Kawabata}, {Kawahara}, \& {Nakaoka}}]{Burgaz21}
{Burgaz}, U., {Maeda}, K., {Kalomeni}, B., {et~al.} 2021, \mnras, 502, 4112,
  \dodoi{10.1093/mnras/stab254}

\bibitem[{{Burke} {et~al.}(2022{\natexlab{a}}){Burke}, {Howell}, {McCully},
  {Newsome}, {Gonzalez}, {Pellegrino}, \& {Terreran}}]{22ilv_z}
{Burke}, J., {Howell}, D.~A., {McCully}, C., {et~al.} 2022{\natexlab{a}},
  Transient Name Server Classification Report, 2022-1137, 1

\bibitem[{{Burke} {et~al.}(2021){Burke}, {Howell}, {Sarbadhicary}, {Sand},
  {Amaro}, {Hiramatsu}, {McCully}, {Pellegrino}, {Andrews}, {Brown}, {Itagaki},
  {Shahbandeh}, {Bostroem}, {Chomiuk}, {Hsiao}, {Smith}, \&
  {Valenti}}]{Burke21}
{Burke}, J., {Howell}, D.~A., {Sarbadhicary}, S.~K., {et~al.} 2021, \apj, 919,
  142, \dodoi{10.3847/1538-4357/ac126b}

\bibitem[{{Burke} {et~al.}(2022{\natexlab{b}}){Burke}, {Howell}, {Sand},
  {Amaro}, {Brown}, {Andrews}, {Bostroem}, {Dong}, {Haislip}, {Hiramatsu},
  {Hosseinzadeh}, {Kouprianov}, {Lundquist}, {McCully}, {Pellegrino},
  {Reichart}, {Tartaglia}, {Valenti}, \& {Yang}}]{Burke22}
{Burke}, J., {Howell}, D.~A., {Sand}, D.~J., {et~al.} 2022{\natexlab{b}}, arXiv
  e-prints, arXiv:2207.07681, \dodoi{10.48550/arXiv.2207.07681}

\bibitem[{{Burns} {et~al.}(2021){Burns}, {Hsiao}, {Suntzeff}, {Baron},
  {Shappee}, {Aldoroty}, {Anderson}, {Ashall}, {Bersten}, {Brown}, {Burrow},
  {Clochiatti}, {Davis}, {DerKacy}, {Do}, {Folatelli}, {Forster Buron},
  {Galbany}, {Hoeflich}, {Holmbo}, {Karamehmetoglu}, {Krisciunas}, {Kumar},
  {Lu}, {Mazzali}, {Morrell}, {Pessi}, {Phillips}, {Pignata}, {Piro}, {Polin},
  {Shahbandeh}, {Stangl}, {Stritzinger}, {Teffs}, {Tonry}, {Tucker}, {Uddin},
  \& {Yang}}]{POISE_ATel}
{Burns}, C., {Hsiao}, E., {Suntzeff}, N., {et~al.} 2021, The Astronomer's
  Telegram, 14441, 1

\bibitem[{{Burns} {et~al.}(2011){Burns}, {Stritzinger}, {Phillips}, {Kattner},
  {Persson}, {Madore}, {Freedman}, {Boldt}, {Campillay}, {Contreras},
  {Folatelli}, {Gonzalez}, {Krzeminski}, {Morrell}, {Salgado}, \&
  {Suntzeff}}]{Burns11}
{Burns}, C.~R., {Stritzinger}, M., {Phillips}, M.~M., {et~al.} 2011, \aj, 141,
  19, \dodoi{10.1088/0004-6256/141/1/19}

\bibitem[{{Burns} {et~al.}(2014){Burns}, {Stritzinger}, {Phillips}, {Hsiao},
  {Contreras}, {Persson}, {Folatelli}, {Boldt}, {Campillay}, {Castell{\'o}n},
  {Freedman}, {Madore}, {Morrell}, {Salgado}, \& {Suntzeff}}]{Burns14}
---. 2014, \apj, 789, 32, \dodoi{10.1088/0004-637X/789/1/32}

\bibitem[{{Burns} {et~al.}(2018){Burns}, {Parent}, {Phillips}, {Stritzinger},
  {Krisciunas}, {Suntzeff}, {Hsiao}, {Contreras}, {Anais}, {Boldt}, {Busta},
  {Campillay}, {Castell{\'o}n}, {Folatelli}, {Freedman}, {Gonz{\'a}lez},
  {Hamuy}, {Heoflich}, {Krzeminski}, {Madore}, {Morrell}, {Persson}, {Roth},
  {Salgado}, {Ser{\'o}n}, \& {Torres}}]{Burns18}
{Burns}, C.~R., {Parent}, E., {Phillips}, M.~M., {et~al.} 2018, \apj, 869, 56,
  \dodoi{10.3847/1538-4357/aae51c}

\bibitem[{{Cao} {et~al.}(2016){Cao}, {Kulkarni}, {Gal-Yam}, {Papadogiannakis},
  {Nugent}, {Masci}, \& {Bue}}]{Cao16}
{Cao}, Y., {Kulkarni}, S.~R., {Gal-Yam}, A., {et~al.} 2016, \apj, 832, 86,
  \dodoi{10.3847/0004-637X/832/1/86}

\bibitem[{{Cao} {et~al.}(2015){Cao}, {Kulkarni}, {Howell}, {Gal-Yam},
  {Kasliwal}, {Valenti}, {Johansson}, {Amanullah}, {Goobar}, {Sollerman},
  {Taddia}, {Horesh}, {Sagiv}, {Cenko}, {Nugent}, {Arcavi}, {Surace},
  {Wo{\'z}niak}, {Moody}, {Rebbapragada}, {Bue}, \& {Gehrels}}]{Cao15}
{Cao}, Y., {Kulkarni}, S.~R., {Howell}, D.~A., {et~al.} 2015, \nat, 521, 328,
  \dodoi{10.1038/nature14440}

\bibitem[{{Cappellari} {et~al.}(2011){Cappellari}, {Emsellem}, {Krajnovi{\'c}},
  {McDermid}, {Scott}, {Verdoes Kleijn}, {Young}, {Alatalo}, {Bacon}, {Blitz},
  {Bois}, {Bournaud}, {Bureau}, {Davies}, {Davis}, {de Zeeuw}, {Duc},
  {Khochfar}, {Kuntschner}, {Lablanche}, {Morganti}, {Naab}, {Oosterloo},
  {Sarzi}, {Serra}, \& {Weijmans}}]{Cappellari11}
{Cappellari}, M., {Emsellem}, E., {Krajnovi{\'c}}, D., {et~al.} 2011, \mnras,
  413, 813, \dodoi{10.1111/j.1365-2966.2010.18174.x}

\bibitem[{{Cardelli} {et~al.}(1989){Cardelli}, {Clayton}, \& {Mathis}}]{CCM89}
{Cardelli}, J.~A., {Clayton}, G.~C., \& {Mathis}, J.~S. 1989, \apj, 345, 245,
  \dodoi{10.1086/167900}

\bibitem[{{Carrick} {et~al.}(2015){Carrick}, {Turnbull}, {Lavaux}, \&
  {Hudson}}]{Carrick15}
{Carrick}, J., {Turnbull}, S.~J., {Lavaux}, G., \& {Hudson}, M.~J. 2015,
  \mnras, 450, 317, \dodoi{10.1093/mnras/stv547}

\bibitem[{{Cartier} {et~al.}(2017){Cartier}, {Sullivan}, {Firth}, {Pignata},
  {Mazzali}, {Maguire}, {Childress}, {Arcavi}, {Ashall}, {Bassett}, {Crawford},
  {Frohmaier}, {Galbany}, {Gal-Yam}, {Hosseinzadeh}, {Howell}, {Inserra},
  {Johansson}, {Kasai}, {McCully}, {Prajs}, {Prentice}, {Schulze}, {Smartt},
  {Smith}, {Smith}, {Valenti}, \& {Young}}]{Cartier17}
{Cartier}, R., {Sullivan}, M., {Firth}, R.~E., {et~al.} 2017, \mnras, 464,
  4476, \dodoi{10.1093/mnras/stw2678}

\bibitem[{{Catinella} {et~al.}(2005){Catinella}, {Haynes}, \&
  {Giovanelli}}]{Catinella05}
{Catinella}, B., {Haynes}, M.~P., \& {Giovanelli}, R. 2005, \aj, 130, 1037,
  \dodoi{10.1086/432543}

\bibitem[{{Chakradhari} {et~al.}(2014){Chakradhari}, {Sahu}, {Srivastav}, \&
  {Anupama}}]{Chakradhari14}
{Chakradhari}, N.~K., {Sahu}, D.~K., {Srivastav}, S., \& {Anupama}, G.~C. 2014,
  \mnras, 443, 1663, \dodoi{10.1093/mnras/stu1258}

\bibitem[{{Chambers} {et~al.}(2016){Chambers}, {Magnier}, {Metcalfe},
  {Flewelling}, {Huber}, {Waters}, {Denneau}, {Draper}, {Farrow}, {Finkbeiner},
  {Holmberg}, {Koppenhoefer}, {Price}, {Rest}, {Saglia}, {Schlafly}, {Smartt},
  {Sweeney}, {Wainscoat}, {Burgett}, {Chastel}, {Grav}, {Heasley}, {Hodapp},
  {Jedicke}, {Kaiser}, {Kudritzki}, {Luppino}, {Lupton}, {Monet}, {Morgan},
  {Onaka}, {Shiao}, {Stubbs}, {Tonry}, {White}, {Ba{\~n}ados}, {Bell},
  {Bender}, {Bernard}, {Boegner}, {Boffi}, {Botticella}, {Calamida},
  {Casertano}, {Chen}, {Chen}, {Cole}, {Deacon}, {Frenk}, {Fitzsimmons},
  {Gezari}, {Gibbs}, {Goessl}, {Goggia}, {Gourgue}, {Goldman}, {Grant},
  {Grebel}, {Hambly}, {Hasinger}, {Heavens}, {Heckman}, {Henderson}, {Henning},
  {Holman}, {Hopp}, {Ip}, {Isani}, {Jackson}, {Keyes}, {Koekemoer}, {Kotak},
  {Le}, {Liska}, {Long}, {Lucey}, {Liu}, {Martin}, {Masci}, {McLean}, {Mindel},
  {Misra}, {Morganson}, {Murphy}, {Obaika}, {Narayan}, {Nieto-Santisteban},
  {Norberg}, {Peacock}, {Pier}, {Postman}, {Primak}, {Rae}, {Rai}, {Riess},
  {Riffeser}, {Rix}, {R{\"o}ser}, {Russel}, {Rutz}, {Schilbach}, {Schultz},
  {Scolnic}, {Strolger}, {Szalay}, {Seitz}, {Small}, {Smith}, {Soderblom},
  {Taylor}, {Thomson}, {Taylor}, {Thakar}, {Thiel}, {Thilker}, {Unger},
  {Urata}, {Valenti}, {Wagner}, {Walder}, {Walter}, {Watters}, {Werner},
  {Wood-Vasey}, \& {Wyse}}]{Chambers16}
{Chambers}, K.~C., {Magnier}, E.~A., {Metcalfe}, N., {et~al.} 2016, arXiv
  e-prints, arXiv:1612.05560, \dodoi{10.48550/arXiv.1612.05560}

\bibitem[{{Chen} {et~al.}(2019){Chen}, {Dong}, {Katz}, {Kochanek}, {Kollmeier},
  {Maguire}, {Phillips}, {Prieto}, {Shappee}, {Stritzinger}, {Bose}, {Brown},
  {Holoien}, {Galbany}, {Milne}, {Morrell}, {Piro}, {Stanek}, {Thompson}, \&
  {Young}}]{Chen19}
{Chen}, P., {Dong}, S., {Katz}, B., {et~al.} 2019, \apj, 880, 35,
  \dodoi{10.3847/1538-4357/ab2630}

\bibitem[{{Chen} \& {Li}(2009)}]{Chen09}
{Chen}, W.-C., \& {Li}, X.-D. 2009, \apj, 702, 686,
  \dodoi{10.1088/0004-637X/702/1/686}

\bibitem[{{Childress} {et~al.}(2013){Childress}, {Scalzo}, {Sim}, {Tucker},
  {Yuan}, {Schmidt}, {Cenko}, {Silverman}, {Contreras}, {Hsiao}, {Phillips},
  {Morrell}, {Jha}, {McCully}, {Filippenko}, {Anderson}, {Benetti}, {Bufano},
  {de Jaeger}, {Forster}, {Gal-Yam}, {Le Guillou}, {Maguire}, {Maund},
  {Mazzali}, {Pignata}, {Smartt}, {Spyromilio}, {Sullivan}, {Taddia},
  {Valenti}, {Bayliss}, {Bessell}, {Blanc}, {Carson}, {Clubb}, {de Burgh-Day},
  {Desjardins}, {Fang}, {Fox}, {Gates}, {Ho}, {Keller}, {Kelly}, {Lidman},
  {Loaring}, {Mould}, {Owers}, {Ozbilgen}, {Pei}, {Pickering}, {Pracy}, {Rich},
  {Schaefer}, {Scott}, {Stritzinger}, {Vogt}, \& {Zhou}}]{Childress13}
{Childress}, M.~J., {Scalzo}, R.~A., {Sim}, S.~A., {et~al.} 2013, \apj, 770,
  29, \dodoi{10.1088/0004-637X/770/1/29}

\bibitem[{{Contreras} {et~al.}(2018){Contreras}, {Phillips}, {Burns}, {Piro},
  {Shappee}, {Stritzinger}, {Baltay}, {Brown}, {Conseil}, {Klotz}, {Nugent},
  {Turpin}, {Parker}, {Rabinowitz}, {Hsiao}, {Morrell}, {Campillay},
  {Castell{\'o}n}, {Corco}, {Gonz{\'a}lez}, {Krisciunas}, {Ser{\'o}n},
  {Tucker}, {Walker}, {Baron}, {Cain}, {Childress}, {Folatelli}, {Freedman},
  {Hamuy}, {Hoeflich}, {Persson}, {Scalzo}, {Schmidt}, \&
  {Suntzeff}}]{Contreras18}
{Contreras}, C., {Phillips}, M.~M., {Burns}, C.~R., {et~al.} 2018, \apj, 859,
  24, \dodoi{10.3847/1538-4357/aabaf8}

\bibitem[{{Courtois} \& {Tully}(2012)}]{Courtois12}
{Courtois}, H.~M., \& {Tully}, R.~B. 2012, \apj, 749, 174,
  \dodoi{10.1088/0004-637X/749/2/174}

\bibitem[{{Das} \& {Mukhopadhyay}(2013)}]{Das13}
{Das}, U., \& {Mukhopadhyay}, B. 2013, \prl, 110, 071102,
  \dodoi{10.1103/PhysRevLett.110.071102}

\bibitem[{{de Vaucouleurs} {et~al.}(1991){de Vaucouleurs}, {de Vaucouleurs},
  {Corwin}, {Buta}, {Paturel}, \& {Fouque}}]{deVaucouleurs91}
{de Vaucouleurs}, G., {de Vaucouleurs}, A., {Corwin}, Herold~G., J., {et~al.}
  1991, {Third Reference Catalogue of Bright Galaxies}

\bibitem[{{DerKacy} {et~al.}(2020){DerKacy}, {Baron}, {Branch}, {Hoeflich},
  {Hauschildt}, {Brown}, \& {Wang}}]{DerKacy20}
{DerKacy}, J.~M., {Baron}, E., {Branch}, D., {et~al.} 2020, \apj, 901, 86,
  \dodoi{10.3847/1538-4357/abae67}

\bibitem[{{DerKacy} {et~al.}(2023){DerKacy}, {Paugh}, {Baron}, {Brown},
  {Ashall}, {Burns}, {Hsiao}, {Kumar}, {Lu}, {Morrell}, {Phillips},
  {Shahbandeh}, {Shappee}, {Stritzinger}, {Tucker}, {Yarbrough}, {Boutsia},
  {Hoeflich}, {Wang}, {Galbany}, {Karamehmetoglu}, {Krisciunas}, {Mazzali},
  {Piro}, {Suntzeff}, {Fiore}, {Guti{\'e}rrez}, {Lundqvist}, \&
  {Reguitti}}]{DerKacy23}
{DerKacy}, J.~M., {Paugh}, S., {Baron}, E., {et~al.} 2023, \mnras, 522, 3481,
  \dodoi{10.1093/mnras/stad1171}

\bibitem[{{Desai} {et~al.}(2023){Desai}, {Kochanek}, {Shappee}, {Jayasinghe},
  {Stanek}, {Holoien}, {Thompson}, {Ashall}, {Beacom}, {Do}, {Dong}, \&
  {Prieto}}]{Desai23}
{Desai}, D.~D., {Kochanek}, C.~S., {Shappee}, B.~J., {et~al.} 2023, arXiv
  e-prints, arXiv:2306.11100, \dodoi{10.48550/arXiv.2306.11100}

\bibitem[{{Dhawan} {et~al.}(2018){Dhawan}, {Bulla}, {Goobar}, {Lunnan},
  {Johansson}, {Fransson}, {Kulkarni}, {Papadogiannakis}, \&
  {Miller}}]{Dhawan18}
{Dhawan}, S., {Bulla}, M., {Goobar}, A., {et~al.} 2018, \mnras, 480, 1445,
  \dodoi{10.1093/mnras/sty1908}

\bibitem[{{Dimitriadis} {et~al.}(2019){Dimitriadis}, {Foley}, {Rest}, {Kasen},
  {Piro}, {Polin}, {Jones}, {Villar}, {Narayan}, {Coulter}, {Kilpatrick},
  {Pan}, {Rojas-Bravo}, {Fox}, {Jha}, {Nugent}, {Riess}, {Scolnic}, {Drout},
  {K2 Mission Team}, {Barentsen}, {Dotson}, {Gully-Santiago}, {Hedges}, {Cody},
  {Barclay}, {Howell}, {KEGS}, {Garnavich}, {Tucker}, {Shaya}, {Mushotzky},
  {Olling}, {Margheim}, {Zenteno}, {Kepler spacecraft Team}, {Coughlin}, {Van
  Cleve}, {Cardoso}, {Larson}, {McCalmont-Everton}, {Peterson}, {Ross},
  {Reedy}, {Osborne}, {McGinn}, {Kohnert}, {Migliorini}, {Wheaton}, {Spencer},
  {Labonde}, {Castillo}, {Beerman}, {Steward}, {Hanley}, {Larsen},
  {Gangopadhyay}, {Kloetzel}, {Weschler}, {Nystrom}, {Moffatt}, {Redick},
  {Griest}, {Packard}, {Muszynski}, {Kampmeier}, {Bjella}, {Flynn},
  {Elsaesser}, {Pan-STARRS}, {Chambers}, {Flewelling}, {Huber}, {Magnier},
  {Waters}, {Schultz}, {Bulger}, {Lowe}, {Willman}, {Smartt}, {Smith}, {DECam},
  {Points}, {Strampelli}, {ASAS-SN}, {Brimacombe}, {Chen}, {Mu{\~n}oz},
  {Mutel}, {Shields}, {Vallely}, {Villanueva}, {PTSS/TNTS}, {Li}, {Wang},
  {Zhang}, {Lin}, {Mo}, {Zhao}, {Sai}, {Zhang}, {Zhang}, {Zhang}, {Wang},
  {Zhang}, {Baron}, {DerKacy}, {Li}, {Chen}, {Xiang}, {Rui}, {Wang}, {Huang},
  {Li}, {Cumbres Observatory}, {Hosseinzadeh}, {Howell}, {Arcavi}, {Hiramatsu},
  {Burke}, {Valenti}, {ATLAS}, {Tonry}, {Denneau}, {Heinze}, {Weiland},
  {Stalder}, {Konkoly}, {Vink{\'o}}, {S{\'a}rneczky}, {P{\'a}l}, {B{\'o}di},
  {Bogn{\'a}r}, {Cs{\'a}k}, {Cseh}, {Cs{\"o}rnyei}, {Hanyecz}, {Ign{\'a}cz},
  {Kalup}, {K{\"o}nyves-T{\'o}th}, {Kriskovics}, {Ordasi}, {Rajmon},
  {S{\'o}dor}, {Szab{\'o}}, {Szak{\'a}ts}, {Zsidi}, {ePESSTO}, {Williams},
  {Nordin}, {Cartier}, {Frohmaier}, {Galbany}, {Guti{\'e}rrez}, {Hook},
  {Inserra}, {Smith}, {Arizona}, {Sand}, {Andrews}, {Smith}, \&
  {Bilinski}}]{Dimitriadis19}
{Dimitriadis}, G., {Foley}, R.~J., {Rest}, A., {et~al.} 2019, \apjl, 870, L1,
  \dodoi{10.3847/2041-8213/aaedb0}

\bibitem[{{Dimitriadis} {et~al.}(2023){Dimitriadis}, {Maguire}, {Karambelkar},
  {Lebron}, {Liu}, {Kozyreva}, {Miller}, {Ridden-Harper}, {Anderson}, {Chen},
  {Coughlin}, {Della Valle}, {Drake}, {Galbany}, {Gromadzki}, {Groom},
  {Guti{\'e}rrez}, {Ihanec}, {Inserra}, {Johansson}, {M{\"u}ller-Bravo},
  {Nicholl}, {Polin}, {Rusholme}, {Schulze}, {Sollerman}, {Srivastav},
  {Taggart}, {Wang}, {Yang}, \& {Young}}]{Dimitriadis23}
{Dimitriadis}, G., {Maguire}, K., {Karambelkar}, V.~R., {et~al.} 2023, \mnras,
  521, 1162, \dodoi{10.1093/mnras/stad536}

\bibitem[{{Ellis} {et~al.}(2008){Ellis}, {Sullivan}, {Nugent}, {Howell},
  {Gal-Yam}, {Astier}, {Balam}, {Balland}, {Basa}, {Carlberg}, {Conley},
  {Fouchez}, {Guy}, {Hardin}, {Hook}, {Pain}, {Perrett}, {Pritchet}, \&
  {Regnault}}]{Ellis08}
{Ellis}, R.~S., {Sullivan}, M., {Nugent}, P.~E., {et~al.} 2008, \apj, 674, 51,
  \dodoi{10.1086/524981}

\bibitem[{{Epinat} {et~al.}(2008){Epinat}, {Amram}, {Marcelin}, {Balkowski},
  {Daigle}, {Hernandez}, {Chemin}, {Carignan}, {Gach}, \& {Balard}}]{Epinat08}
{Epinat}, B., {Amram}, P., {Marcelin}, M., {et~al.} 2008, \mnras, 388, 500,
  \dodoi{10.1111/j.1365-2966.2008.13422.x}

\bibitem[{{Falco} {et~al.}(1999){Falco}, {Kurtz}, {Geller}, {Huchra}, {Peters},
  {Berlind}, {Mink}, {Tokarz}, \& {Elwell}}]{Falco99}
{Falco}, E.~E., {Kurtz}, M.~J., {Geller}, M.~J., {et~al.} 1999, \pasp, 111,
  438, \dodoi{10.1086/316343}

\bibitem[{{Fausnaugh} {et~al.}(2023){Fausnaugh}, {Valleley}, {Tucker},
  {Kochanek}, {Shappee}, {Ricker}, {Vanderspek}, {Agarwal}, {Daylan},
  {Jayaraman}, {Hounsell}, \& {Muthukrishna}}]{Fausnaugh23}
{Fausnaugh}, M.~M., {Valleley}, P.~J., {Tucker}, M.~A., {et~al.} 2023, arXiv
  e-prints, arXiv:2307.11815.
\newblock \doarXiv{2307.11815}

\bibitem[{{Ferretti} {et~al.}(2016){Ferretti}, {Amanullah}, {Goobar},
  {Johansson}, {Vreeswijk}, {Butler}, {Cao}, {Cenko}, {Doran}, {Filippenko},
  {Freeland}, {Hosseinzadeh}, {Howell}, {Lundqvist}, {Mattila}, {Nordin},
  {Nugent}, {Petrushevska}, {Valenti}, {Vogt}, \& {Wozniak}}]{Ferretti16}
{Ferretti}, R., {Amanullah}, R., {Goobar}, A., {et~al.} 2016, \aap, 592, A40,
  \dodoi{10.1051/0004-6361/201628351}

\bibitem[{{Ferretti} {et~al.}(2017){Ferretti}, {Amanullah}, {Goobar},
  {Petrushevska}, {Borthakur}, {Bulla}, {Fox}, {Freeland}, {Fremling},
  {Hangard}, \& {Hayes}}]{Ferretti17}
---. 2017, \aap, 606, A111, \dodoi{10.1051/0004-6361/201731409}

\bibitem[{{Filippenko} {et~al.}(1992{\natexlab{a}}){Filippenko}, {Richmond},
  {Matheson}, {Shields}, {Burbidge}, {Cohen}, {Dickinson}, {Malkan}, {Nelson},
  {Pietz}, {Schlegel}, {Schmeer}, {Spinrad}, {Steidel}, {Tran}, \&
  {Wren}}]{fil92b}
{Filippenko}, A.~V., {Richmond}, M.~W., {Matheson}, T., {et~al.}
  1992{\natexlab{a}}, \apjl, 384, L15, \dodoi{10.1086/186252}

\bibitem[{{Filippenko} {et~al.}(1992{\natexlab{b}}){Filippenko}, {Richmond},
  {Branch}, {Gaskell}, {Herbst}, {Ford}, {Treffers}, {Matheson}, {Ho}, {Dey},
  {Sargent}, {Small}, \& {van Breugel}}]{fil92a}
{Filippenko}, A.~V., {Richmond}, M.~W., {Branch}, D., {et~al.}
  1992{\natexlab{b}}, \aj, 104, 1543, \dodoi{10.1086/116339}

\bibitem[{{Fisher}(1970)}]{Fisher70}
{Fisher}, R.~A. 1970, {Statistical methods for research workers}

\bibitem[{{Fitzpatrick}(1999)}]{Fitzpatrick99}
{Fitzpatrick}, E.~L. 1999, \pasp, 111, 63, \dodoi{10.1086/316293}

\bibitem[{{Foley} {et~al.}(2008){Foley}, {Filippenko}, \& {Jha}}]{Foley08}
{Foley}, R.~J., {Filippenko}, A.~V., \& {Jha}, S.~W. 2008, \apj, 686, 117,
  \dodoi{10.1086/590467}

\bibitem[{{Foley} {et~al.}(2010){Foley}, {Narayan}, {Challis}, {Filippenko},
  {Kirshner}, {Silverman}, \& {Steele}}]{Foley10}
{Foley}, R.~J., {Narayan}, G., {Challis}, P.~J., {et~al.} 2010, \apj, 708,
  1748, \dodoi{10.1088/0004-637X/708/2/1748}

\bibitem[{{Foley} {et~al.}(2011){Foley}, {Sanders}, \& {Kirshner}}]{Foley11}
{Foley}, R.~J., {Sanders}, N.~E., \& {Kirshner}, R.~P. 2011, \apj, 742, 89,
  \dodoi{10.1088/0004-637X/742/2/89}

\bibitem[{{Foley} {et~al.}(2012){Foley}, {Challis}, {Filippenko},
  {Ganeshalingam}, {Landsman}, {Li}, {Marion}, {Silverman}, {Beaton},
  {Bennert}, {Cenko}, {Childress}, {Guhathakurta}, {Jiang}, {Kalirai},
  {Kirshner}, {Stockton}, {Tollerud}, {Vink{\'o}}, {Wheeler}, \&
  {Woo}}]{Foley12}
{Foley}, R.~J., {Challis}, P.~J., {Filippenko}, A.~V., {et~al.} 2012, \apj,
  744, 38, \dodoi{10.1088/0004-637X/744/1/38}

\bibitem[{{Foley} {et~al.}(2014){Foley}, {Fox}, {McCully}, {Phillips}, {Sand},
  {Zheng}, {Challis}, {Filippenko}, {Folatelli}, {Hillebrandt}, {Hsiao}, {Jha},
  {Kirshner}, {Kromer}, {Marion}, {Nelson}, {Pakmor}, {Pignata}, {R{\"o}pke},
  {Seitenzahl}, {Silverman}, {Skrutskie}, \& {Stritzinger}}]{Foley14}
{Foley}, R.~J., {Fox}, O.~D., {McCully}, C., {et~al.} 2014, \mnras, 443, 2887,
  \dodoi{10.1093/mnras/stu1378}

\bibitem[{{Foreman-Mackey} {et~al.}(2013){Foreman-Mackey}, {Hogg}, {Lang}, \&
  {Goodman}}]{Foreman-Mackey13}
{Foreman-Mackey}, D., {Hogg}, D.~W., {Lang}, D., \& {Goodman}, J. 2013, \pasp,
  125, 306, \dodoi{10.1086/670067}

\bibitem[{{Fremling} {et~al.}(2020){Fremling}, {Miller}, {Sharma}, {Dugas},
  {Perley}, {Taggart}, {Sollerman}, {Goobar}, {Graham}, {Neill}, {Nordin},
  {Rigault}, {Walters}, {Andreoni}, {Bagdasaryan}, {Belicki}, {Cannella},
  {Bellm}, {Cenko}, {De}, {Dekany}, {Frederick}, {Golkhou}, {Graham}, {Helou},
  {Ho}, {Kasliwal}, {Kupfer}, {Laher}, {Mahabal}, {Masci}, {Riddle},
  {Rusholme}, {Schulze}, {Shupe}, {Smith}, {van Velzen}, {Yan}, {Yao},
  {Zhuang}, \& {Kulkarni}}]{Fremling20}
{Fremling}, C., {Miller}, A.~A., {Sharma}, Y., {et~al.} 2020, \apj, 895, 32,
  \dodoi{10.3847/1538-4357/ab8943}

\bibitem[{{Ganeshalingam} {et~al.}(2012){Ganeshalingam}, {Li}, {Filippenko},
  {Silverman}, {Chornock}, {Foley}, {Matheson}, {Kirshner}, {Milne}, {Calkins},
  \& {Shen}}]{Ganeshalingam12}
{Ganeshalingam}, M., {Li}, W., {Filippenko}, A.~V., {et~al.} 2012, \apj, 751,
  142, \dodoi{10.1088/0004-637X/751/2/142}

\bibitem[{{Goobar} {et~al.}(2014){Goobar}, {Johansson}, {Amanullah}, {Cao},
  {Perley}, {Kasliwal}, {Ferretti}, {Nugent}, {Harris}, {Gal-Yam}, {Ofek},
  {Tendulkar}, {Dennefeld}, {Valenti}, {Arcavi}, {Banerjee}, {Venkataraman},
  {Joshi}, {Ashok}, {Cenko}, {Diaz}, {Fremling}, {Horesh}, {Howell},
  {Kulkarni}, {Papadogiannakis}, {Petrushevska}, {Sand}, {Sollerman},
  {Stanishev}, {Bloom}, {Surace}, {Dupuy}, \& {Liu}}]{Goobar14}
{Goobar}, A., {Johansson}, J., {Amanullah}, R., {et~al.} 2014, \apjl, 784, L12,
  \dodoi{10.1088/2041-8205/784/1/L12}

\bibitem[{{Goobar} {et~al.}(2015){Goobar}, {Kromer}, {Siverd}, {Stassun},
  {Pepper}, {Amanullah}, {Kasliwal}, {Sollerman}, \& {Taddia}}]{Goobar15}
{Goobar}, A., {Kromer}, M., {Siverd}, R., {et~al.} 2015, \apj, 799, 106,
  \dodoi{10.1088/0004-637X/799/1/106}

\bibitem[{{Graham} {et~al.}(2022){Graham}, {Kennedy}, {Kumar}, {Amaro}, {Sand},
  {Jha}, {Galbany}, {Vinko}, {Wheeler}, {Hsiao}, {Bostroem}, {Burke},
  {Hiramatsu}, {Hosseinzadeh}, {McCully}, {Howell}, {Diamond}, {Hoeflich},
  {Wang}, \& {Li}}]{Graham22}
{Graham}, M.~L., {Kennedy}, T.~D., {Kumar}, S., {et~al.} 2022, \mnras, 511,
  3682, \dodoi{10.1093/mnras/stac192}

\bibitem[{{Guy} {et~al.}(2007){Guy}, {Astier}, {Baumont}, {Hardin}, {Pain},
  {Regnault}, {Basa}, {Carlberg}, {Conley}, {Fabbro}, {Fouchez}, {Hook},
  {Howell}, {Perrett}, {Pritchet}, {Rich}, {Sullivan}, {Antilogus}, {Aubourg},
  {Bazin}, {Bronder}, {Filiol}, {Palanque-Delabrouille}, {Ripoche}, \&
  {Ruhlmann-Kleider}}]{Guy07}
{Guy}, J., {Astier}, P., {Baumont}, S., {et~al.} 2007, \aap, 466, 11,
  \dodoi{10.1051/0004-6361:20066930}

\bibitem[{{Hachisu} {et~al.}(2012){Hachisu}, {Kato}, \& {Nomoto}}]{Hachisu12}
{Hachisu}, I., {Kato}, M., \& {Nomoto}, K. 2012, \apjl, 756, L4,
  \dodoi{10.1088/2041-8205/756/1/L4}

\bibitem[{{Han} {et~al.}(2020){Han}, {Zheng}, {Stahl}, {Burke}, {Vinko}, {de
  Jaeger}, {Arcavi}, {Brink}, {Cseh}, {Hiramatsu}, {Hosseinzadeh}, {Howell},
  {Ignacz}, {Konyves-Toth}, {Krezinger}, {McCully}, {Ordasi}, {Pinter},
  {Sarneczky}, {Szakats}, {Tang}, {Vida}, {Wang}, {Wei}, {Wheeler}, {Xin}, \&
  {Filippenko}}]{Han20}
{Han}, X., {Zheng}, W., {Stahl}, B.~E., {et~al.} 2020, \apj, 892, 142,
  \dodoi{10.3847/1538-4357/ab7a27}

\bibitem[{{Hart} {et~al.}(2023){Hart}, {Shappee}, {Hey}, {Kochanek}, {Stanek},
  {Lim}, {Dobbs}, {Tucker}, {Jayasinghe}, {Beacom}, {Boright}, {Holoien},
  {Ong}, {Prieto}, {Thompson}, \& {Will}}]{Hart23}
{Hart}, K., {Shappee}, B.~J., {Hey}, D., {et~al.} 2023, arXiv e-prints,
  arXiv:2304.03791, \dodoi{10.48550/arXiv.2304.03791}

\bibitem[{{Hicken} {et~al.}(2007){Hicken}, {Garnavich}, {Prieto}, {Blondin},
  {DePoy}, {Kirshner}, \& {Parrent}}]{Hicken07}
{Hicken}, M., {Garnavich}, P.~M., {Prieto}, J.~L., {et~al.} 2007, \apjl, 669,
  L17, \dodoi{10.1086/523301}

\bibitem[{{Hoeflich} \& {Khokhlov}(1996)}]{Hoeflich:Khokhlov:96}
{Hoeflich}, P., \& {Khokhlov}, A. 1996, \apj, 457, 500, \dodoi{10.1086/176748}

\bibitem[{{Hoeflich} {et~al.}(2017){Hoeflich}, {Hsiao}, {Ashall}, {Burns},
  {Diamond}, {Phillips}, {Sand}, {Stritzinger}, {Suntzeff}, {Contreras},
  {Krisciunas}, {Morrell}, \& {Wang}}]{Hoeflich17}
{Hoeflich}, P., {Hsiao}, E.~Y., {Ashall}, C., {et~al.} 2017, \apj, 846, 58,
  \dodoi{10.3847/1538-4357/aa84b2}

\bibitem[{{Holmbo} {et~al.}(2019){Holmbo}, {Stritzinger}, {Shappee}, {Tucker},
  {Zheng}, {Ashall}, {Phillips}, {Contreras}, {Filippenko}, {Hoeflich},
  {Huber}, {Piro}, {Wang}, {Zhang}, {Anais}, {Baron}, {Burns}, {Campillay},
  {Castell{\'o}n}, {Corco}, {Hsiao}, {Krisciunas}, {Morrell}, {Nielsen},
  {Persson}, {Taddia}, {Tomasella}, {Zhang}, \& {Zhao}}]{Holmbo19}
{Holmbo}, S., {Stritzinger}, M.~D., {Shappee}, B.~J., {et~al.} 2019, \aap, 627,
  A174, \dodoi{10.1051/0004-6361/201834389}

\bibitem[{{Holoien} {et~al.}(2017{\natexlab{a}}){Holoien}, {Stanek},
  {Kochanek}, {Shappee}, {Prieto}, {Brimacombe}, {Bersier}, {Bishop}, {Dong},
  {Brown}, {Danilet}, {Simonian}, {Basu}, {Beacom}, {Falco}, {Pojmanski},
  {Skowron}, {Wo{\'z}niak}, {{\'A}vila}, {Conseil}, {Contreras}, {Cruz},
  {Fern{\'a}ndez}, {Koff}, {Guo}, {Herczeg}, {Hissong}, {Hsiao}, {Jose},
  {Kiyota}, {Long}, {Monard}, {Nicholls}, {Nicolas}, \& {Wiethoff}}]{Holoien17}
{Holoien}, T.~W.~S., {Stanek}, K.~Z., {Kochanek}, C.~S., {et~al.}
  2017{\natexlab{a}}, \mnras, 464, 2672, \dodoi{10.1093/mnras/stw2273}

\bibitem[{{Holoien} {et~al.}(2017{\natexlab{b}}){Holoien}, {Brown}, {Stanek},
  {Kochanek}, {Shappee}, {Prieto}, {Dong}, {Brimacombe}, {Bishop}, {Bose},
  {Beacom}, {Bersier}, {Chen}, {Chomiuk}, {Falco}, {Godoy-Rivera}, {Morrell},
  {Pojmanski}, {Shields}, {Strader}, {Stritzinger}, {Thompson}, {Wo{\'z}niak},
  {Bock}, {Cacella}, {Conseil}, {Cruz}, {Fernandez}, {Kiyota}, {Koff},
  {Krannich}, {Marples}, {Masi}, {Monard}, {Nicholls}, {Nicolas}, {Post},
  {Stone}, \& {Wiethoff}}]{Holoien17B}
{Holoien}, T.~W.~S., {Brown}, J.~S., {Stanek}, K.~Z., {et~al.}
  2017{\natexlab{b}}, \mnras, 471, 4966, \dodoi{10.1093/mnras/stx1544}

\bibitem[{{Holoien} {et~al.}(2017{\natexlab{c}}){Holoien}, {Brown}, {Stanek},
  {Kochanek}, {Shappee}, {Prieto}, {Dong}, {Brimacombe}, {Bishop}, {Basu},
  {Beacom}, {Bersier}, {Chen}, {Danilet}, {Falco}, {Godoy-Rivera}, {Goss},
  {Pojmanski}, {Simonian}, {Skowron}, {Thompson}, {Wo{\'z}niak}, {{\'A}vila},
  {Bock}, {Carballo}, {Conseil}, {Contreras}, {Cruz}, {And{\'u}jar}, {Guo},
  {Hsiao}, {Kiyota}, {Koff}, {Krannich}, {Madore}, {Marples}, {Masi},
  {Morrell}, {Monard}, {Munoz-Mateos}, {Nicholls}, {Nicolas}, {Wagner}, \&
  {Wiethoff}}]{Holoien17C}
---. 2017{\natexlab{c}}, \mnras, 467, 1098, \dodoi{10.1093/mnras/stx057}

\bibitem[{{Holoien} {et~al.}(2019){Holoien}, {Brown}, {Vallely}, {Stanek},
  {Kochanek}, {Shappee}, {Prieto}, {Dong}, {Brimacombe}, {Bishop}, {Bose},
  {Beacom}, {Bersier}, {Chen}, {Chomiuk}, {Falco}, {Holmbo}, {Jayasinghe},
  {Morrell}, {Pojmanski}, {Shields}, {Strader}, {Stritzinger}, {Thompson},
  {Wo{\'z}niak}, {Bock}, {Cacella}, {Carballo}, {Cruz}, {Conseil}, {Farfan},
  {Fernandez}, {Kiyota}, {Koff}, {Krannich}, {Marples}, {Masi}, {Monard},
  {Mu{\~n}oz}, {Nicholls}, {Post}, {Stone}, {Trappett}, \&
  {Wiethoff}}]{Holoien19}
{Holoien}, T.~W.~S., {Brown}, J.~S., {Vallely}, P.~J., {et~al.} 2019, \mnras,
  484, 1899, \dodoi{10.1093/mnras/stz073}

\bibitem[{{Hoogendam} {et~al.}(2022){Hoogendam}, {Ashall}, {Galbany},
  {Shappee}, {Burns}, {Lu}, {Phillips}, {Baron}, {Holmbo}, {Hsiao}, {Morrell},
  {Stritzinger}, {Suntzeff}, {Taddia}, {Young}, {Lyman}, {Benetti}, {Mazzali},
  {Delgado Manche{\~n}o}, {D{\'\i}az}, \& {Torres}}]{Hoogendam22}
{Hoogendam}, W.~B., {Ashall}, C., {Galbany}, L., {et~al.} 2022, \apj, 928, 103,
  \dodoi{10.3847/1538-4357/ac54aa}

\bibitem[{{Hosseinzadeh} {et~al.}(2017){Hosseinzadeh}, {Sand}, {Valenti},
  {Brown}, {Howell}, {McCully}, {Kasen}, {Arcavi}, {Bostroem}, {Tartaglia},
  {Hsiao}, {Davis}, {Shahbandeh}, \& {Stritzinger}}]{Hosseinzadeh17}
{Hosseinzadeh}, G., {Sand}, D.~J., {Valenti}, S., {et~al.} 2017, \apjl, 845,
  L11, \dodoi{10.3847/2041-8213/aa8402}

\bibitem[{{Hosseinzadeh} {et~al.}(2022){Hosseinzadeh}, {Sand}, {Lundqvist},
  {Andrews}, {Bostroem}, {Dong}, {Janzen}, {Jencson}, {Lundquist}, {Meza
  Retamal}, {Pearson}, {Valenti}, {Wyatt}, {Burke}, {Howell}, {McCully},
  {Newsome}, {Gonzalez}, {Pellegrino}, {Terreran}, {Kwok}, {Jha}, {Strader},
  {Kundu}, {Ryder}, {Haislip}, {Kouprianov}, \& {Reichart}}]{Hosseinzadeh22}
{Hosseinzadeh}, G., {Sand}, D.~J., {Lundqvist}, P., {et~al.} 2022, \apjl, 933,
  L45, \dodoi{10.3847/2041-8213/ac7cef}

\bibitem[{{Hosseinzadeh} {et~al.}(2023){Hosseinzadeh}, {Sand}, {Sarbadhicary},
  {Ryder}, {Jha}, {Dong}, {Bostroem}, {Andrews}, {Hoang}, {Janzen}, {Jencson},
  {Lundquist}, {Meza Retamal}, {Pearson}, {Shrestha}, {Valenti}, {Wyatt},
  {Farah}, {Howell}, {McCully}, {Newsome}, {Padilla Gonzalez}, {Pellegrino},
  {Terreran}, {Alzaabi}, {Green}, {Gurney}, {Milne}, {Ridenhour}, {Smith},
  {Robles}, {Kwok}, {Schwab}, {Gromadzki}, {Buckley}, {Itagaki}, {Hiramatsu},
  {Chomiuk}, {Lundqvist}, {Haislip}, {Kouprianov}, \&
  {Reichart}}]{Hosseinzadeh23}
{Hosseinzadeh}, G., {Sand}, D.~J., {Sarbadhicary}, S.~K., {et~al.} 2023, arXiv
  e-prints, arXiv:2305.03071, \dodoi{10.48550/arXiv.2305.03071}

\bibitem[{{Howell} {et~al.}(2006){Howell}, {Sullivan}, {Nugent}, {Ellis},
  {Conley}, {Le Borgne}, {Carlberg}, {Guy}, {Balam}, {Basa}, {Fouchez}, {Hook},
  {Hsiao}, {Neill}, {Pain}, {Perrett}, \& {Pritchet}}]{Howell06}
{Howell}, D.~A., {Sullivan}, M., {Nugent}, P.~E., {et~al.} 2006, \nat, 443,
  308, \dodoi{10.1038/nature05103}

\bibitem[{{Hoyle} \& {Fowler}(1960)}]{hoy60}
{Hoyle}, F., \& {Fowler}, W.~A. 1960, \apj, 132, 565, \dodoi{10.1086/146963}

\bibitem[{{Hoyt} {et~al.}(2021){Hoyt}, {Beaton}, {Freedman}, {Jang}, {Lee},
  {Madore}, {Monson}, {Neeley}, {Rich}, \& {Seibert}}]{Hoyt21}
{Hoyt}, T.~J., {Beaton}, R.~L., {Freedman}, W.~L., {et~al.} 2021, \apj, 915,
  34, \dodoi{10.3847/1538-4357/abfe5a}

\bibitem[{{Hsiao} {et~al.}(2015){Hsiao}, {Burns}, {Contreras}, {H{\"o}flich},
  {Sand}, {Marion}, {Phillips}, {Stritzinger}, {Gonz{\'a}lez-Gait{\'a}n},
  {Mason}, {Folatelli}, {Parent}, {Gall}, {Amanullah}, {Anupama}, {Arcavi},
  {Banerjee}, {Beletsky}, {Blanc}, {Bloom}, {Brown}, {Campillay}, {Cao}, {De
  Cia}, {Diamond}, {Freedman}, {Gonzalez}, {Goobar}, {Holmbo}, {Howell},
  {Johansson}, {Kasliwal}, {Kirshner}, {Krisciunas}, {Kulkarni}, {Maguire},
  {Milne}, {Morrell}, {Nugent}, {Ofek}, {Osip}, {Palunas}, {Perley}, {Persson},
  {Piro}, {Rabus}, {Roth}, {Schiefelbein}, {Srivastav}, {Sullivan}, {Suntzeff},
  {Surace}, {Wo{\'z}niak}, \& {Yaron}}]{Hsiao15}
{Hsiao}, E.~Y., {Burns}, C.~R., {Contreras}, C., {et~al.} 2015, \aap, 578, A9,
  \dodoi{10.1051/0004-6361/201425297}

\bibitem[{{Hsiao} {et~al.}(2020){Hsiao}, {Hoeflich}, {Ashall}, {Lu},
  {Contreras}, {Burns}, {Phillips}, {Galbany}, {Anderson}, {Baltay}, {Baron},
  {Castell{\'o}n}, {Davis}, {Freedman}, {Gall}, {Gonzalez}, {Graham}, {Hamuy},
  {Holoien}, {Karamehmetoglu}, {Krisciunas}, {Kumar}, {Kuncarayakti},
  {Morrell}, {Moriya}, {Nugent}, {Perlmutter}, {Persson}, {Piro}, {Rabinowitz},
  {Roth}, {Shahbandeh}, {Shappee}, {Stritzinger}, {Suntzeff}, {Taddia}, \&
  {Uddin}}]{Hsiao20}
{Hsiao}, E.~Y., {Hoeflich}, P., {Ashall}, C., {et~al.} 2020, \apj, 900, 140,
  \dodoi{10.3847/1538-4357/abaf4c}

\bibitem[{{Iben} \& {Tutukov}(1984)}]{iben84}
{Iben}, I., J., \& {Tutukov}, A.~V. 1984, \apjs, 54, 335,
  \dodoi{10.1086/190932}

\bibitem[{{Im} {et~al.}(2015){Im}, {Choi}, {Yoon}, {Kim}, {Ehgamberdiev},
  {Monard}, \& {Sung}}]{Im15}
{Im}, M., {Choi}, C., {Yoon}, S.-C., {et~al.} 2015, \apjs, 221, 22,
  \dodoi{10.1088/0067-0049/221/1/22}

\bibitem[{{Jensen} {et~al.}(2021){Jensen}, {Blakeslee}, {Ma}, {Milne}, {Brown},
  {Cantiello}, {Garnavich}, {Greene}, {Lucey}, {Phan}, {Tully}, \&
  {Wood}}]{Jensen21}
{Jensen}, J.~B., {Blakeslee}, J.~P., {Ma}, C.-P., {et~al.} 2021, arXiv
  e-prints, arXiv:2105.08299.
\newblock \doarXiv{2105.08299}

\bibitem[{{Jha} {et~al.}(2007){Jha}, {Riess}, \& {Kirshner}}]{Jha07}
{Jha}, S., {Riess}, A.~G., \& {Kirshner}, R.~P. 2007, \apj, 659, 122,
  \dodoi{10.1086/512054}

\bibitem[{{Jha} {et~al.}(2019){Jha}, {Maguire}, \& {Sullivan}}]{Jha19}
{Jha}, S.~W., {Maguire}, K., \& {Sullivan}, M. 2019, Nature Astronomy, 3, 706,
  \dodoi{10.1038/s41550-019-0858-0}

\bibitem[{{Jiang} {et~al.}(2018){Jiang}, {Doi}, {Maeda}, \&
  {Shigeyama}}]{Jiang18}
{Jiang}, J.-a., {Doi}, M., {Maeda}, K., \& {Shigeyama}, T. 2018, \apj, 865,
  149, \dodoi{10.3847/1538-4357/aadb9a}

\bibitem[{{Jiang} {et~al.}(2021){Jiang}, {Maeda}, {Kawabata}, {Doi},
  {Shigeyama}, {Tanaka}, {Tominaga}, {Nomoto}, {Niino}, {Sako}, {Ohsawa},
  {Schramm}, {Yamanaka}, {Kobayashi}, {Takahashi}, {Nakaoka}, {Kawabata},
  {Isogai}, {Aoki}, {Kondo}, {Mori}, {Arimatsu}, {Kasuga}, {Okumura},
  {Urakawa}, {Reichart}, {Taguchi}, {Arima}, {Beniyama}, {Uno}, \&
  {Hamada}}]{Jiang21}
{Jiang}, J.-a., {Maeda}, K., {Kawabata}, M., {et~al.} 2021, \apjl, 923, L8,
  \dodoi{10.3847/2041-8213/ac375f}

\bibitem[{{Jones} {et~al.}(2009){Jones}, {Read}, {Saunders}, {Colless},
  {Jarrett}, {Parker}, {Fairall}, {Mauch}, {Sadler}, {Watson}, {Burton},
  {Campbell}, {Cass}, {Croom}, {Dawe}, {Fiegert}, {Frankcombe}, {Hartley},
  {Huchra}, {James}, {Kirby}, {Lahav}, {Lucey}, {Mamon}, {Moore}, {Peterson},
  {Prior}, {Proust}, {Russell}, {Safouris}, {Wakamatsu}, {Westra}, \&
  {Williams}}]{Jones09}
{Jones}, D.~H., {Read}, M.~A., {Saunders}, W., {et~al.} 2009, \mnras, 399, 683,
  \dodoi{10.1111/j.1365-2966.2009.15338.x}

\bibitem[{{Jones} {et~al.}(2019){Jones}, {Scolnic}, {Foley}, {Rest}, {Kessler},
  {Challis}, {Chambers}, {Coulter}, {Dettman}, {Foley}, {Huber}, {Jha},
  {Johnson}, {Kilpatrick}, {Kirshner}, {Manuel}, {Narayan}, {Pan}, {Riess},
  {Schultz}, {Siebert}, {Berger}, {Chornock}, {Flewelling}, {Magnier},
  {Smartt}, {Smith}, {Wainscoat}, {Waters}, \& {Willman}}]{Jones19}
{Jones}, D.~O., {Scolnic}, D.~M., {Foley}, R.~J., {et~al.} 2019, \apj, 881, 19,
  \dodoi{10.3847/1538-4357/ab2bec}

\bibitem[{{Jones} {et~al.}(2021){Jones}, {Foley}, {Narayan}, {Hjorth}, {Huber},
  {Aleo}, {Alexander}, {Angus}, {Auchettl}, {Baldassare}, {Bruun}, {Chambers},
  {Chatterjee}, {Coppejans}, {Coulter}, {DeMarchi}, {Dimitriadis}, {Drout},
  {Engel}, {French}, {Gagliano}, {Gall}, {Hung}, {Izzo}, {Jacobson-Gal{\'a}n},
  {Kilpatrick}, {Korhonen}, {Margutti}, {Raimundo}, {Ramirez-Ruiz}, {Rest},
  {Rojas-Bravo}, {Siebert}, {Smartt}, {Smith}, {Terreran}, {Wang}, {Wojtak},
  {Agnello}, {Ansari}, {Arendse}, {Baldeschi}, {Blanchard}, {Brethauer},
  {Bright}, {Brown}, {de Boer}, {Dodd}, {Fairlamb}, {Grillo}, {Hajela}, {Hede},
  {Kolborg}, {Law-Smith}, {Lin}, {Magnier}, {Malanchev}, {Matthews}, {Mockler},
  {Muthukrishna}, {Pan}, {Pfister}, {Ramanah}, {Rest}, {Sarangi},
  {Schr{\o}der}, {Stauffer}, {Stroh}, {Taggart}, {Tinyanont}, {Wainscoat}, \&
  {Young Supernova Experiment}}]{Jones21}
{Jones}, D.~O., {Foley}, R.~J., {Narayan}, G., {et~al.} 2021, \apj, 908, 143,
  \dodoi{10.3847/1538-4357/abd7f5}

\bibitem[{{Kasen}(2010)}]{Kasen10}
{Kasen}, D. 2010, \apj, 708, 1025, \dodoi{10.1088/0004-637X/708/2/1025}

\bibitem[{{Kasen} \& {Plewa}(2007)}]{Kasen07}
{Kasen}, D., \& {Plewa}, T. 2007, \apj, 662, 459, \dodoi{10.1086/516834}

\bibitem[{{Kashi} \& {Soker}(2011)}]{Kashi11}
{Kashi}, A., \& {Soker}, N. 2011, \mnras, 417, 1466,
  \dodoi{10.1111/j.1365-2966.2011.19361.x}

\bibitem[{{Kawabata} {et~al.}(2020){Kawabata}, {Maeda}, {Yamanaka}, {Nakaoka},
  {Kawabata}, {Adachi}, {Akitaya}, {Burgaz}, {Hanayama}, {Horiuchi},
  {Hosokawa}, {Iida}, {Imazato}, {Isogai}, {Jiang}, {Katoh}, {Kimura}, {Kino},
  {Kuroda}, {Maehara}, {Matsubayashi}, {Morihana}, {Murata}, {Nagao}, {Niwano},
  {Nogami}, {Oeda}, {Ono}, {Onozato}, {Otsuka}, {Saito}, {Sasada}, {Shiraishi},
  {Sugiyama}, {Taguchi}, {Takahashi}, {Takagi}, {Takagi}, {Takayama}, {Tozuka},
  \& {Sekiguchi}}]{Kawabata20}
{Kawabata}, M., {Maeda}, K., {Yamanaka}, M., {et~al.} 2020, \apj, 893, 143,
  \dodoi{10.3847/1538-4357/ab8236}

\bibitem[{{Kent} {et~al.}(2008){Kent}, {Giovanelli}, {Haynes}, {Martin},
  {Saintonge}, {Stierwalt}, {Balonek}, {Brosch}, \& {Koopmann}}]{Kent08}
{Kent}, B.~R., {Giovanelli}, R., {Haynes}, M.~P., {et~al.} 2008, \aj, 136, 713,
  \dodoi{10.1088/0004-6256/136/2/713}

\bibitem[{{Kerr} \& {Lynden-Bell}(1986)}]{Kerr86}
{Kerr}, F.~J., \& {Lynden-Bell}, D. 1986, \mnras, 221, 1023,
  \dodoi{10.1093/mnras/221.4.1023}

\bibitem[{{Khokhlov}(1991)}]{Khokhlov91}
{Khokhlov}, A.~M. 1991, \aap, 245, 114

\bibitem[{{Kochanek} {et~al.}(2017){Kochanek}, {Shappee}, {Stanek}, {Holoien},
  {Thompson}, {Prieto}, {Dong}, {Shields}, {Will}, {Britt}, {Perzanowski}, \&
  {Pojma{\'n}ski}}]{Kochanek17}
{Kochanek}, C.~S., {Shappee}, B.~J., {Stanek}, K.~Z., {et~al.} 2017, \pasp,
  129, 104502, \dodoi{10.1088/1538-3873/aa80d9}

\bibitem[{{Koribalski} {et~al.}(2004){Koribalski}, {Staveley-Smith}, {Kilborn},
  {Ryder}, {Kraan-Korteweg}, {Ryan-Weber}, {Ekers}, {Jerjen}, {Henning},
  {Putman}, {Zwaan}, {de Blok}, {Calabretta}, {Disney}, {Minchin}, {Bhathal},
  {Boyce}, {Drinkwater}, {Freeman}, {Gibson}, {Green}, {Haynes}, {Juraszek},
  {Kesteven}, {Knezek}, {Mader}, {Marquarding}, {Meyer}, {Mould}, {Oosterloo},
  {O'Brien}, {Price}, {Sadler}, {Schr{\"o}der}, {Stewart}, {Stootman}, {Waugh},
  {Warren}, {Webster}, \& {Wright}}]{Koribalski04}
{Koribalski}, B.~S., {Staveley-Smith}, L., {Kilborn}, V.~A., {et~al.} 2004,
  \aj, 128, 16, \dodoi{10.1086/421744}

\bibitem[{{Krisciunas} {et~al.}(2004{\natexlab{a}}){Krisciunas}, {Phillips},
  {Suntzeff}, {Persson}, {Hamuy}, {Antezana}, {Candia}, {Clocchiatti}, {DePoy},
  {Germany}, {Gonzalez}, {Gonzalez}, {Krzeminski}, {Maza}, {Nugent}, {Qiu},
  {Rest}, {Roth}, {Stritzinger}, {Strolger}, {Thompson}, {Williams}, \&
  {Wischnjewsky}}]{Krisciunas04a}
{Krisciunas}, K., {Phillips}, M.~M., {Suntzeff}, N.~B., {et~al.}
  2004{\natexlab{a}}, \aj, 127, 1664, \dodoi{10.1086/381911}

\bibitem[{{Krisciunas} {et~al.}(2004{\natexlab{b}}){Krisciunas}, {Suntzeff},
  {Phillips}, {Candia}, {Prieto}, {Antezana}, {Chassagne}, {Chen}, {Dickinson},
  {Eisenhardt}, {Espinoza}, {Garnavich}, {Gonz{\'a}lez}, {Harrison}, {Hamuy},
  {Ivanov}, {Krzemi{\'n}ski}, {Kulesa}, {McCarthy}, {Moro-Mart{\'\i}n},
  {Muena}, {Noriega-Crespo}, {Persson}, {Pinto}, {Roth}, {Rubenstein},
  {Stanford}, {Stringfellow}, {Zapata}, {Porter}, \&
  {Wischnjewsky}}]{Krisciunas04b}
{Krisciunas}, K., {Suntzeff}, N.~B., {Phillips}, M.~M., {et~al.}
  2004{\natexlab{b}}, \aj, 128, 3034, \dodoi{10.1086/425629}

\bibitem[{{Kromer} {et~al.}(2013){Kromer}, {Pakmor}, {Taubenberger}, {Pignata},
  {Fink}, {R{\"o}pke}, {Seitenzahl}, {Sim}, \& {Hillebrandt}}]{Kromer13}
{Kromer}, M., {Pakmor}, R., {Taubenberger}, S., {et~al.} 2013, \apjl, 778, L18,
  \dodoi{10.1088/2041-8205/778/1/L18}

\bibitem[{{Kromer} {et~al.}(2015){Kromer}, {Ohlmann}, {Pakmor}, {Ruiter},
  {Hillebrandt}, {Marquardt}, {R{\"o}pke}, {Seitenzahl}, {Sim}, \&
  {Taubenberger}}]{Kromer15}
{Kromer}, M., {Ohlmann}, S.~T., {Pakmor}, R., {et~al.} 2015, \mnras, 450, 3045,
  \dodoi{10.1093/mnras/stv886}

\bibitem[{{Kromer} {et~al.}(2016){Kromer}, {Fremling}, {Pakmor},
  {Taubenberger}, {Amanullah}, {Cenko}, {Fransson}, {Goobar}, {Leloudas},
  {Taddia}, {R{\"o}pke}, {Seitenzahl}, {Sim}, \& {Sollerman}}]{Kromer16}
{Kromer}, M., {Fremling}, C., {Pakmor}, R., {et~al.} 2016, \mnras, 459, 4428,
  \dodoi{10.1093/mnras/stw962}

\bibitem[{{Kwok} {et~al.}(2023){Kwok}, {Siebert}, {Johansson}, {Jha},
  {Blondin}, {Dessart}, {Foley}, {Hillier}, {Larison}, {Pakmor}, {Temim},
  {Andrews}, {Auchettl}, {Badenes}, {Barna}, {Bostroem}, {Brenner Newman},
  {Brink}, {Bustamante-Rosell}, {Camacho-Neves}, {Clocchiatti}, {Coulter},
  {Davis}, {Deckers}, {Dimitriadis}, {Dong}, {Farah}, {Filippenko}, {Flors},
  {Fox}, {Garnavich}, {Padilla Gonzalez}, {Graur}, {Hambsch}, {Hosseinzadeh},
  {Howell}, {Hughes}, {Kerzendorf}, {Le Saux}, {Maeda}, {Maguire}, {McCully},
  {Mihalenko}, {Newsome}, {O'Brien}, {Pearson}, {Pellegrino}, {Pierel},
  {Polin}, {Rest}, {Rojas-Bravo}, {Sand}, {Schwab}, {Shahbandeh}, {Shrestha},
  {Smith}, {Strolger}, {Szalai}, {Taggart}, {Terreran}, {Terwel}, {Tinyanont},
  {Valenti}, {Vinko}, {Wheeler}, {Yang}, {Zheng}, {Ashall}, {DerKacy},
  {Galbany}, {Hoeflich}, {de Jaeger}, {Lu}, {Maund}, {Medler}, {Morrell},
  {Shappee}, {Stritzinger}, {Suntzeff}, {Tucker}, \& {Wang}}]{Kwok23}
{Kwok}, L.~A., {Siebert}, M.~R., {Johansson}, J., {et~al.} 2023, arXiv
  e-prints, arXiv:2308.12450.
\newblock \doarXiv{2308.12450}

\bibitem[{{Langer} {et~al.}(2000){Langer}, {Deutschmann}, {Wellstein}, \&
  {H{\"o}flich}}]{Langer00}
{Langer}, N., {Deutschmann}, A., {Wellstein}, S., \& {H{\"o}flich}, P. 2000,
  \aap, 362, 1046, \dodoi{10.48550/arXiv.astro-ph/0008444}

\bibitem[{{Lentz} {et~al.}(2000){Lentz}, {Baron}, {Branch}, {Hauschildt}, \&
  {Nugent}}]{Lentz00}
{Lentz}, E.~J., {Baron}, E., {Branch}, D., {Hauschildt}, P.~H., \& {Nugent},
  P.~E. 2000, \apj, 530, 966, \dodoi{10.1086/308400}

\bibitem[{{Li} {et~al.}(2022){Li}, {Zhang}, {Dai}, {Li}, {Wang}, {Zhai}, \&
  {Bai}}]{Li22}
{Li}, L., {Zhang}, J., {Dai}, B., {et~al.} 2022, \apj, 924, 35,
  \dodoi{10.3847/1538-4357/ac323f}

\bibitem[{{Li} {et~al.}(2011){Li}, {Leaman}, {Chornock}, {Filippenko},
  {Poznanski}, {Ganeshalingam}, {Wang}, {Modjaz}, {Jha}, {Foley}, \&
  {Smith}}]{Li2011}
{Li}, W., {Leaman}, J., {Chornock}, R., {et~al.} 2011, \mnras, 412, 1441,
  \dodoi{10.1111/j.1365-2966.2011.18160.x}

\bibitem[{{Li} {et~al.}(2019){Li}, {Wang}, {Vink{\'o}}, {Mo}, {Hosseinzadeh},
  {Sand}, {Zhang}, {Lin}, {PTSS/TNTS}, {Zhang}, {Wang}, {Zhang}, {Chen},
  {Xiang}, {Rui}, {Huang}, {Li}, {Zhang}, {Li}, {Baron}, {Derkacy}, {Zhao},
  {Sai}, {Zhang}, {Wang}, {LCO}, {Howell}, {McCully}, {Arcavi}, {Valenti},
  {Hiramatsu}, {Burke}, {KEGS}, {Rest}, {Garnavich}, {Tucker}, {Narayan},
  {Shaya}, {Margheim}, {Zenteno}, {Villar}, {UCSC}, {Dimitriadis}, {Foley},
  {Pan}, {Coulter}, {Fox}, {Jha}, {Jones}, {Kasen}, {Kilpatrick}, {Piro},
  {Riess}, {Rojas-Bravo}, {ASAS-SN}, {Shappee}, {Holoien}, {Stanek}, {Drout},
  {Auchettl}, {Kochanek}, {Brown}, {Bose}, {Bersier}, {Brimacombe}, {Chen},
  {Dong}, {Holmbo}, {Mu{\~n}oz}, {Mutel}, {Post}, {Prieto}, {Shields},
  {Tallon}, {Thompson}, {Vallely}, {Villanueva}, {Pan-STARRS}, {Smartt},
  {Smith}, {Chambers}, {Flewelling}, {Huber}, {Magnier}, {Waters}, {Schultz},
  {Bulger}, {Lowe}, {Willman}, {Konkoly/Texas}, {S{\'a}rneczky}, {P{\'a}l},
  {Wheeler}, {B{\'o}di}, {Bogn{\'a}r}, {Cs{\'a}k}, {Cseh}, {Cs{\"o}rnyei},
  {Hanyecz}, {Ign{\'a}cz}, {Kalup}, {K{\"o}nyves-T{\'o}th}, {Kriskovics},
  {Ordasi}, {Rajmon}, {S{\'o}dor}, {Szab{\'o}}, {Szak{\'a}ts}, {Zsidi},
  {Arizona}, {Milne}, {Andrews}, {Smith}, {Bilinski}, {Swift}, {Brown},
  {ePESSTO}, {Nordin}, {Williams}, {Galbany}, {Palmerio}, {Hook}, {Inserra},
  {Maguire}, {Cartier}, {Razza}, {Guti{\'e}rrez}, {North Carolina}, {Hermes},
  {Reding}, {Kaiser}, {ATLAS}, {Tonry}, {Heinze}, {Denneau}, {Weiland},
  {Stalder}, {K2 Mission Team}, {Barentsen}, {Dotson}, {Barclay},
  {Gully-Santiago}, {Hedges}, {Cody}, {Howell}, {Kepler Spacecraft Team},
  {Coughlin}, {Van Cleve}, {Cardoso}, {Larson}, {McCalmont-Everton},
  {Peterson}, {Ross}, {Reedy}, {Osborne}, {McGinn}, {Kohnert}, {Migliorini},
  {Wheaton}, {Spencer}, {Labonde}, {Castillo}, {Beerman}, {Steward}, {Hanley},
  {Larsen}, {Gangopadhyay}, {Kloetzel}, {Weschler}, {Nystrom}, {Moffatt},
  {Redick}, {Griest}, {Packard}, {Muszynski}, {Kampmeier}, {Bjella}, {Flynn},
  \& {Elsaesser}}]{Li19}
{Li}, W., {Wang}, X., {Vink{\'o}}, J., {et~al.} 2019, \apj, 870, 12,
  \dodoi{10.3847/1538-4357/aaec74}

\bibitem[{{Li} {et~al.}(2023){Li}, {Zhang}, {Wang}, {Zhang}, {Galbany},
  {Filippenko}, {Brink}, {Ashall}, {Zheng}, {de Jaeger}, {Ragosta}, {Deckers},
  {Gromadzki}, {Young}, {Xi}, {Chen}, {Zhao}, {Sai}, {Yan}, {Xiang}, {Chen},
  {Li}, {Wang}, {Zou}, {Sui}, {Wang}, {Ma}, {Nie}, {Xue}, {Zhou}, \&
  {Zhou}}]{Li23}
{Li}, Z., {Zhang}, T., {Wang}, X., {et~al.} 2023, \apj, 950, 17,
  \dodoi{10.3847/1538-4357/accde3}

\bibitem[{{Lim} {et~al.}(2023){Lim}, {Im}, {Paek}, {Yoon}, {Choi}, {Kim},
  {Wheeler}, {Thomas}, {Vink{\'o}}, {Kim}, {Seo}, {Kang}, {Kim}, {Sung}, {Kim},
  {Yoon}, {Kim}, {Kim}, {Bae}, {Ehgamberdiev}, {Burhonov}, \&
  {Mirzaqulov}}]{Lim23}
{Lim}, G., {Im}, M., {Paek}, G. S.~H., {et~al.} 2023, \apj, 949, 33,
  \dodoi{10.3847/1538-4357/acc10c}

\bibitem[{{Lira} {et~al.}(1998){Lira}, {Suntzeff}, {Phillips}, {Hamuy}, {Maza},
  {Schommer}, {Smith}, {Wells}, {Avil{\'e}s}, {Baldwin}, {Elias},
  {Gonz{\'a}lez}, {Layden}, {Navarrete}, {Ugarte}, {Walker}, {Williger},
  {Baganoff}, {Crotts}, {Rich}, {Tyson}, {Dey}, {Guhathakurta}, {Hibbard},
  {Kim}, {Rehner}, {Siciliano}, {Roth}, {Seitzer}, \& {Williams}}]{Lira98}
{Lira}, P., {Suntzeff}, N.~B., {Phillips}, M.~M., {et~al.} 1998, \aj, 115, 234,
  \dodoi{10.1086/300175}

\bibitem[{{Liu} {et~al.}(2023){Liu}, {R{\"o}pke}, \& {Han}}]{Liu23}
{Liu}, Z.-W., {R{\"o}pke}, F.~K., \& {Han}, Z. 2023, Research in Astronomy and
  Astrophysics, 23, 082001, \dodoi{10.1088/1674-4527/acd89e}

\bibitem[{{Livio} \& {Mazzali}(2018)}]{Livio18}
{Livio}, M., \& {Mazzali}, P. 2018, \physrep, 736, 1,
  \dodoi{10.1016/j.physrep.2018.02.002}

\bibitem[{{Livne}(1990)}]{Livne90}
{Livne}, E. 1990, \apjl, 354, L53, \dodoi{10.1086/185721}

\bibitem[{{Loveday} {et~al.}(1996){Loveday}, {Peterson}, {Maddox}, \&
  {Efstathiou}}]{Loveday96}
{Loveday}, J., {Peterson}, B.~A., {Maddox}, S.~J., \& {Efstathiou}, G. 1996,
  \apjs, 107, 201, \dodoi{10.1086/192360}

\bibitem[{{Lu} {et~al.}(2021){Lu}, {Ashall}, {Hsiao}, {Hoeflich}, {Galbany},
  {Baron}, {Phillips}, {Contreras}, {Burns}, {Suntzeff}, {Stritzinger},
  {Anais}, {Anderson}, {Brown}, {Busta}, {Castell{\'o}n}, {Davis}, {Diamond},
  {Falco}, {Gonzalez}, {Hamuy}, {Holmbo}, {Holoien}, {Krisciunas}, {Kirshner},
  {Kumar}, {Kuncarayakti}, {Marion}, {Morrell}, {Persson}, {Piro}, {Prieto},
  {Sand}, {Shahbandeh}, {Shappee}, \& {Taddia}}]{Lu21}
{Lu}, J., {Ashall}, C., {Hsiao}, E.~Y., {et~al.} 2021, \apj, 920, 107,
  \dodoi{10.3847/1538-4357/ac1606}

\bibitem[{{Lundqvist} {et~al.}(2015){Lundqvist}, {Nyholm}, {Taddia},
  {Sollerman}, {Johansson}, {Kozma}, {Lundqvist}, {Fransson}, {Garnavich},
  {Kromer}, {Shappee}, \& {Goobar}}]{Lundqvist15}
{Lundqvist}, P., {Nyholm}, A., {Taddia}, F., {et~al.} 2015, \aap, 577, A39,
  \dodoi{10.1051/0004-6361/201525719}

\bibitem[{{Maeda} {et~al.}(2023){Maeda}, {Jiang}, {Doi}, {Kawabata}, \&
  {Shigeyama}}]{Maeda23}
{Maeda}, K., {Jiang}, J.-a., {Doi}, M., {Kawabata}, M., \& {Shigeyama}, T.
  2023, \mnras, 521, 1897, \dodoi{10.1093/mnras/stad618}

\bibitem[{{Maeda} {et~al.}(2018){Maeda}, {Jiang}, {Shigeyama}, \&
  {Doi}}]{Maeda18}
{Maeda}, K., {Jiang}, J.-a., {Shigeyama}, T., \& {Doi}, M. 2018, \apj, 861, 78,
  \dodoi{10.3847/1538-4357/aac8d8}

\bibitem[{{Magee} \& {Maguire}(2020)}]{Magee20b}
{Magee}, M.~R., \& {Maguire}, K. 2020, \aap, 642, A189,
  \dodoi{10.1051/0004-6361/202037870}

\bibitem[{{Maguire} {et~al.}(2011){Maguire}, {Sullivan}, {Thomas}, {Nugent},
  {Howell}, {Gal-Yam}, {Arcavi}, {Ben-Ami}, {Blake}, {Botyanszki}, {Buton},
  {Cooke}, {Ellis}, {Hook}, {Kasliwal}, {Pan}, {Pereira}, {Podsiadlowski},
  {Sternberg}, {Suzuki}, {Xu}, {Yaron}, {Bloom}, {Cenko}, {Kulkarni}, {Law},
  {Ofek}, {Poznanski}, \& {Quimby}}]{Maguire11}
{Maguire}, K., {Sullivan}, M., {Thomas}, R.~C., {et~al.} 2011, \mnras, 418,
  747, \dodoi{10.1111/j.1365-2966.2011.19526.x}

\bibitem[{{Maguire} {et~al.}(2023){Maguire}, {Magee}, {Leloudas}, {Miller},
  {Dimitriadis}, {Pursiainen}, {Bulla}, {De}, {Gal-Yam}, {Perley}, {Fremling},
  {Karambelkar}, {Nordin}, {Reusch}, {Schulze}, {Sollerman}, {Terreran},
  {Yang}, {Bellm}, {Groom}, {Kasliwal}, {Kulkarni}, {Lacroix}, {Masci},
  {Purdum}, {Sharma}, \& {Smith}}]{Maguire23}
{Maguire}, K., {Magee}, M.~R., {Leloudas}, G., {et~al.} 2023, arXiv e-prints,
  arXiv:2304.12361, \dodoi{10.48550/arXiv.2304.12361}

\bibitem[{{Mandel} {et~al.}(2022){Mandel}, {Thorp}, {Narayan}, {Friedman}, \&
  {Avelino}}]{Mandel22}
{Mandel}, K.~S., {Thorp}, S., {Narayan}, G., {Friedman}, A.~S., \& {Avelino},
  A. 2022, \mnras, 510, 3939, \dodoi{10.1093/mnras/stab3496}

\bibitem[{{Maoz} {et~al.}(2014){Maoz}, {Mannucci}, \& {Nelemans}}]{Maoz14}
{Maoz}, D., {Mannucci}, F., \& {Nelemans}, G. 2014, \araa, 52, 107,
  \dodoi{10.1146/annurev-astro-082812-141031}

\bibitem[{{Marion} {et~al.}(2016){Marion}, {Brown}, {Vink{\'o}}, {Silverman},
  {Sand}, {Challis}, {Kirshner}, {Wheeler}, {Berlind}, {Brown}, {Calkins},
  {Camacho}, {Dhungana}, {Foley}, {Friedman}, {Graham}, {Howell}, {Hsiao},
  {Irwin}, {Jha}, {Kehoe}, {Macri}, {Maeda}, {Mandel}, {McCully}, {Pandya},
  {Rines}, {Wilhelmy}, \& {Zheng}}]{Marion16}
{Marion}, G.~H., {Brown}, P.~J., {Vink{\'o}}, J., {et~al.} 2016, \apj, 820, 92,
  \dodoi{10.3847/0004-637X/820/2/92}

\bibitem[{{Matteucci} \& {Recchi}(2001)}]{Matteucci01}
{Matteucci}, F., \& {Recchi}, S. 2001, \apj, 558, 351, \dodoi{10.1086/322472}

\bibitem[{{Mazzali} {et~al.}(2022){Mazzali}, {Benetti}, {Stritzinger}, \&
  {Ashall}}]{Mazzali22}
{Mazzali}, P.~A., {Benetti}, S., {Stritzinger}, M., \& {Ashall}, C. 2022,
  \mnras, 511, 5560, \dodoi{10.1093/mnras/stac409}

\bibitem[{{Meyer} {et~al.}(2004){Meyer}, {Zwaan}, {Webster}, {Staveley-Smith},
  {Ryan-Weber}, {Drinkwater}, {Barnes}, {Howlett}, {Kilborn}, {Stevens},
  {Waugh}, {Pierce}, {Bhathal}, {de Blok}, {Disney}, {Ekers}, {Freeman},
  {Garcia}, {Gibson}, {Harnett}, {Henning}, {Jerjen}, {Kesteven}, {Knezek},
  {Koribalski}, {Mader}, {Marquarding}, {Minchin}, {O'Brien}, {Oosterloo},
  {Price}, {Putman}, {Ryder}, {Sadler}, {Stewart}, {Stootman}, \&
  {Wright}}]{Meyer04}
{Meyer}, M.~J., {Zwaan}, M.~A., {Webster}, R.~L., {et~al.} 2004, \mnras, 350,
  1195, \dodoi{10.1111/j.1365-2966.2004.07710.x}

\bibitem[{{Miller} {et~al.}(2018){Miller}, {Cao}, {Piro}, {Blagorodnova},
  {Bue}, {Cenko}, {Dhawan}, {Ferretti}, {Fox}, {Fremling}, {Goobar}, {Howell},
  {Hosseinzadeh}, {Kasliwal}, {Laher}, {Lunnan}, {Masci}, {McCully}, {Nugent},
  {Sollerman}, {Taddia}, \& {Kulkarni}}]{Miller18}
{Miller}, A.~A., {Cao}, Y., {Piro}, A.~L., {et~al.} 2018, \apj, 852, 100,
  \dodoi{10.3847/1538-4357/aaa01f}

\bibitem[{{Miller} {et~al.}(2020){Miller}, {Magee}, {Polin}, {Maguire},
  {Zimmerman}, {Yao}, {Sollerman}, {Schulze}, {Perley}, {Kromer}, {Dhawan},
  {Bulla}, {Andreoni}, {Bellm}, {De}, {Dekany}, {Delacroix}, {Fremling},
  {Gal-Yam}, {Goldstein}, {Golkhou}, {Goobar}, {Graham}, {Irani}, {Kasliwal},
  {Kaye}, {Kim}, {Laher}, {Mahabal}, {Masci}, {Nugent}, {Ofek}, {Phinney},
  {Prentice}, {Riddle}, {Rigault}, {Rusholme}, {Schweyer}, {Shupe},
  {Soumagnac}, {Terreran}, {Walters}, {Yan}, {Zolkower}, \&
  {Kulkarni}}]{Miller20}
{Miller}, A.~A., {Magee}, M.~R., {Polin}, A., {et~al.} 2020, \apj, 898, 56,
  \dodoi{10.3847/1538-4357/ab9e05}

\bibitem[{{Milne} {et~al.}(2013){Milne}, {Brown}, {Roming}, {Bufano}, \&
  {Gehrels}}]{Milne13}
{Milne}, P.~A., {Brown}, P.~J., {Roming}, P. W.~A., {Bufano}, F., \& {Gehrels},
  N. 2013, \apj, 779, 23, \dodoi{10.1088/0004-637X/779/1/23}

\bibitem[{{Moll} {et~al.}(2014){Moll}, {Raskin}, {Kasen}, \&
  {Woosley}}]{Moll14}
{Moll}, R., {Raskin}, C., {Kasen}, D., \& {Woosley}, S.~E. 2014, \apj, 785,
  105, \dodoi{10.1088/0004-637X/785/2/105}

\bibitem[{{Moriya} {et~al.}(2023){Moriya}, {Mazzali}, {Ashall}, \&
  {Pian}}]{Moriya23}
{Moriya}, T.~J., {Mazzali}, P.~A., {Ashall}, C., \& {Pian}, E. 2023, \mnras,
  522, 6035, \dodoi{10.1093/mnras/stad1386}

\bibitem[{{Munari} {et~al.}(2013){Munari}, {Henden}, {Belligoli}, {Castellani},
  {Cherini}, {Righetti}, \& {Vagnozzi}}]{Munari13}
{Munari}, U., {Henden}, A., {Belligoli}, R., {et~al.} 2013, \na, 20, 30,
  \dodoi{10.1016/j.newast.2012.09.003}

\bibitem[{{Neumann} {et~al.}(2023){Neumann}, {Holoien}, {Kochanek}, {Stanek},
  {Vallely}, {Shappee}, {Prieto}, {Pessi}, {Jayasinghe}, {Brimacombe},
  {Bersier}, {Aydi}, {Basinger}, {Beacom}, {Bose}, {Brown}, {Chen},
  {Clocchiatti}, {Desai}, {Dong}, {Falco}, {Holmbo}, {Morrell}, {Shields},
  {Sokolovsky}, {Strader}, {Stritzinger}, {Swihart}, {Thompson}, {Way},
  {Aslan}, {Bishop}, {Bock}, {Bradshaw}, {Cacella}, {Castro-Morales},
  {Conseil}, {Cornect}, {Cruz}, {Farfan}, {Fernandez}, {Gabuya},
  {Gonzalez-Carballo}, {Kendurkar}, {Kiyota}, {Koff}, {Krannich}, {Marples},
  {Masi}, {Monard}, {Mu{\~n}oz}, {Nicholls}, {Post}, {Pujic}, {Stone},
  {Tomasella}, {Trappett}, \& {Wiethoff}}]{Neumann23}
{Neumann}, K.~D., {Holoien}, T.~W.~S., {Kochanek}, C.~S., {et~al.} 2023,
  \mnras, 520, 4356, \dodoi{10.1093/mnras/stad355}

\bibitem[{{Ni} {et~al.}(2022){Ni}, {Moon}, {Drout}, {Polin}, {Sand},
  {Gonz{\'a}lez-Gait{\'a}n}, {Kim}, {Lee}, {Park}, {Howell}, {Nugent}, {Piro},
  {Brown}, {Galbany}, {Burke}, {Hiramatsu}, {Hosseinzadeh}, {Valenti},
  {Afsariardchi}, {Andrews}, {Antoniadis}, {Arcavi}, {Beaton}, {Bostroem},
  {Carlberg}, {Cenko}, {Cha}, {Dong}, {Gal-Yam}, {Haislip}, {Holoien},
  {Johnson}, {Kouprianov}, {Lee}, {Matzner}, {Morrell}, {McCully}, {Pignata},
  {Reichart}, {Rich}, {Ryder}, {Smith}, {Wyatt}, \& {Yang}}]{Ni22}
{Ni}, Y.~Q., {Moon}, D.-S., {Drout}, M.~R., {et~al.} 2022, Nature Astronomy, 6,
  568, \dodoi{10.1038/s41550-022-01603-4}

\bibitem[{{Ni} {et~al.}(2023){Ni}, {Moon}, {Drout}, {Polin}, {Sand},
  {Gonz{\'a}lez-Gait{\'a}n}, {Kim}, {Lee}, {Park}, {Howell}, {Nugent}, {Piro},
  {Brown}, {Galbany}, {Burke}, {Hiramatsu}, {Hosseinzadeh}, {Valenti},
  {Afsariardchi}, {Andrews}, {Antoniadis}, {Beaton}, {Bostroem}, {Carlberg},
  {Cenko}, {Cha}, {Dong}, {Gal-Yam}, {Haislip}, {Holoien}, {Johnson},
  {Kouprianov}, {Lee}, {Matzner}, {Morrell}, {McCully}, {Pignata}, {Reichart},
  {Rich}, {Ryder}, {Smith}, {Wyatt}, \& {Yang}}]{Ni23}
---. 2023, \apj, 946, 7, \dodoi{10.3847/1538-4357/aca9be}

\bibitem[{{Noebauer} {et~al.}(2016){Noebauer}, {Taubenberger}, {Blinnikov},
  {Sorokina}, \& {Hillebrandt}}]{Noebauer16}
{Noebauer}, U.~M., {Taubenberger}, S., {Blinnikov}, S., {Sorokina}, E., \&
  {Hillebrandt}, W. 2016, \mnras, 463, 2972, \dodoi{10.1093/mnras/stw2197}

\bibitem[{{Nomoto}(1980)}]{Nomoto80}
{Nomoto}, K. 1980, in Texas Workshop on Type I Supernovae, ed. J.~C. {Wheeler},
  164--181

\bibitem[{{Nomoto}(1982)}]{Nomoto82}
{Nomoto}, K. 1982, \apj, 253, 798, \dodoi{10.1086/159682}

\bibitem[{{Norris} \& {Kannappan}(2011)}]{Norris11}
{Norris}, M.~A., \& {Kannappan}, S.~J. 2011, \mnras, 414, 739,
  \dodoi{10.1111/j.1365-2966.2011.18440.x}

\bibitem[{{Nugent} {et~al.}(2011){Nugent}, {Sullivan}, {Cenko}, {Thomas},
  {Kasen}, {Howell}, {Bersier}, {Bloom}, {Kulkarni}, {Kandrashoff},
  {Filippenko}, {Silverman}, {Marcy}, {Howard}, {Isaacson}, {Maguire},
  {Suzuki}, {Tarlton}, {Pan}, {Bildsten}, {Fulton}, {Parrent}, {Sand},
  {Podsiadlowski}, {Bianco}, {Dilday}, {Graham}, {Lyman}, {James}, {Kasliwal},
  {Law}, {Quimby}, {Hook}, {Walker}, {Mazzali}, {Pian}, {Ofek}, {Gal-Yam}, \&
  {Poznanski}}]{Nugent11}
{Nugent}, P.~E., {Sullivan}, M., {Cenko}, S.~B., {et~al.} 2011, \nat, 480, 344,
  \dodoi{10.1038/nature10644}

\bibitem[{{Pakmor} {et~al.}(2010){Pakmor}, {Kromer}, {R{\"o}pke}, {Sim},
  {Ruiter}, \& {Hillebrandt}}]{Pakmor10}
{Pakmor}, R., {Kromer}, M., {R{\"o}pke}, F.~K., {et~al.} 2010, \nat, 463, 61,
  \dodoi{10.1038/nature08642}

\bibitem[{{Pakmor} {et~al.}(2012){Pakmor}, {Kromer}, {Taubenberger}, {Sim},
  {R{\"o}pke}, \& {Hillebrandt}}]{Pakmor12}
{Pakmor}, R., {Kromer}, M., {Taubenberger}, S., {et~al.} 2012, \apjl, 747, L10,
  \dodoi{10.1088/2041-8205/747/1/L10}

\bibitem[{{Pakmor} {et~al.}(2013){Pakmor}, {Kromer}, {Taubenberger}, \&
  {Springel}}]{Pakmor13}
{Pakmor}, R., {Kromer}, M., {Taubenberger}, S., \& {Springel}, V. 2013, \apjl,
  770, L8, \dodoi{10.1088/2041-8205/770/1/L8}

\bibitem[{{Pan} {et~al.}(2020){Pan}, {Foley}, {Jones}, {Filippenko}, \&
  {Kuin}}]{Pan20}
{Pan}, Y.~C., {Foley}, R.~J., {Jones}, D.~O., {Filippenko}, A.~V., \& {Kuin},
  N.~P.~M. 2020, \mnras, 491, 5897, \dodoi{10.1093/mnras/stz3391}

\bibitem[{{Pejcha} {et~al.}(2013){Pejcha}, {Antognini}, {Shappee}, \&
  {Thompson}}]{Pejcha13}
{Pejcha}, O., {Antognini}, J.~M., {Shappee}, B.~J., \& {Thompson}, T.~A. 2013,
  \mnras, 435, 943, \dodoi{10.1093/mnras/stt1281}

\bibitem[{{Pellegrino} {et~al.}(2020){Pellegrino}, {Howell}, {Sarbadhicary},
  {Burke}, {Hiramatsu}, {McCully}, {Milne}, {Andrews}, {Brown}, {Chomiuk},
  {Hsiao}, {Sand}, {Shahbandeh}, {Smith}, {Valenti}, {Vink{\'o}}, {Wheeler},
  {Wyatt}, \& {Yang}}]{Pellegrino20}
{Pellegrino}, C., {Howell}, D.~A., {Sarbadhicary}, S.~K., {et~al.} 2020, \apj,
  897, 159, \dodoi{10.3847/1538-4357/ab8e3f}

\bibitem[{{Pepper} {et~al.}(2007){Pepper}, {Pogge}, {DePoy}, {Marshall},
  {Stanek}, {Stutz}, {Poindexter}, {Siverd}, {O'Brien}, {Trueblood}, \&
  {Trueblood}}]{Pepper07}
{Pepper}, J., {Pogge}, R.~W., {DePoy}, D.~L., {et~al.} 2007, \pasp, 119, 923,
  \dodoi{10.1086/521836}

\bibitem[{{Pereira} {et~al.}(2013){Pereira}, {Thomas}, {Aldering}, {Antilogus},
  {Baltay}, {Benitez-Herrera}, {Bongard}, {Buton}, {Canto}, {Cellier-Holzem},
  {Chen}, {Childress}, {Chotard}, {Copin}, {Fakhouri}, {Fink}, {Fouchez},
  {Gangler}, {Guy}, {Hillebrandt}, {Hsiao}, {Kerschhaggl}, {Kowalski},
  {Kromer}, {Nordin}, {Nugent}, {Paech}, {Pain}, {P{\'e}contal}, {Perlmutter},
  {Rabinowitz}, {Rigault}, {Runge}, {Saunders}, {Smadja}, {Tao},
  {Taubenberger}, {Tilquin}, \& {Wu}}]{Pereira13}
{Pereira}, R., {Thomas}, R.~C., {Aldering}, G., {et~al.} 2013, \aap, 554, A27,
  \dodoi{10.1051/0004-6361/201221008}

\bibitem[{{Perlmutter} {et~al.}(1999){Perlmutter}, {Aldering}, {Goldhaber},
  {Knop}, {Nugent}, {Castro}, {Deustua}, {Fabbro}, {Goobar}, {Groom}, {Hook},
  {Kim}, {Kim}, {Lee}, {Nunes}, {Pain}, {Pennypacker}, {Quimby}, {Lidman},
  {Ellis}, {Irwin}, {McMahon}, {Ruiz-Lapuente}, {Walton}, {Schaefer}, {Boyle},
  {Filippenko}, {Matheson}, {Fruchter}, {Panagia}, {Newberg}, {Couch}, \&
  {Project}}]{Perlmutter99}
{Perlmutter}, S., {Aldering}, G., {Goldhaber}, G., {et~al.} 1999, \apj, 517,
  565, \dodoi{10.1086/307221}

\bibitem[{{Peterson} {et~al.}(2023){Peterson}, {Jones}, {Scolnic},
  {S{\'a}nchez}, {Do}, {Riess}, {Ward}, {Dwomoh}, {de Jaeger}, {Jha}, {Mandel},
  {Pierel}, {Popovic}, {Rose}, {Rubin}, {Shappee}, {Thorp}, {Tonry}, {Tully},
  \& {Vincenzi}}]{Peterson23}
{Peterson}, E.~R., {Jones}, D.~O., {Scolnic}, D., {et~al.} 2023, \mnras, 522,
  2478, \dodoi{10.1093/mnras/stad1077}

\bibitem[{{Phillips}(1993)}]{Phillips93}
{Phillips}, M.~M. 1993, \apjl, 413, L105, \dodoi{10.1086/186970}

\bibitem[{{Phillips} {et~al.}(1999){Phillips}, {Lira}, {Suntzeff}, {Schommer},
  {Hamuy}, \& {Maza}}]{Phillips99}
{Phillips}, M.~M., {Lira}, P., {Suntzeff}, N.~B., {et~al.} 1999, \aj, 118,
  1766, \dodoi{10.1086/301032}

\bibitem[{{Phillips} {et~al.}(1992){Phillips}, {Wells}, {Suntzeff}, {Hamuy},
  {Leibundgut}, {Kirshner}, \& {Foltz}}]{Phillips92}
{Phillips}, M.~M., {Wells}, L.~A., {Suntzeff}, N.~B., {et~al.} 1992, \aj, 103,
  1632, \dodoi{10.1086/116177}

\bibitem[{{Phillips} {et~al.}(2013){Phillips}, {Simon}, {Morrell}, {Burns},
  {Cox}, {Foley}, {Karakas}, {Patat}, {Sternberg}, {Williams}, {Gal-Yam},
  {Hsiao}, {Leonard}, {Persson}, {Stritzinger}, {Thompson}, {Campillay},
  {Contreras}, {Folatelli}, {Freedman}, {Hamuy}, {Roth}, {Shields}, {Suntzeff},
  {Chomiuk}, {Ivans}, {Madore}, {Penprase}, {Perley}, {Pignata}, {Preston}, \&
  {Soderberg}}]{Phillips13}
{Phillips}, M.~M., {Simon}, J.~D., {Morrell}, N., {et~al.} 2013, \apj, 779, 38,
  \dodoi{10.1088/0004-637X/779/1/38}

\bibitem[{{Phillips} {et~al.}(2019){Phillips}, {Contreras}, {Hsiao}, {Morrell},
  {Burns}, {Stritzinger}, {Ashall}, {Freedman}, {Hoeflich}, {Persson}, {Piro},
  {Suntzeff}, {Uddin}, {Anais}, {Baron}, {Busta}, {Campillay}, {Castell{\'o}n},
  {Corco}, {Diamond}, {Gall}, {Gonzalez}, {Holmbo}, {Krisciunas}, {Roth},
  {Ser{\'o}n}, {Taddia}, {Torres}, {Anderson}, {Baltay}, {Folatelli},
  {Galbany}, {Goobar}, {Hadjiyska}, {Hamuy}, {Kasliwal}, {Lidman}, {Nugent},
  {Perlmutter}, {Rabinowitz}, {Ryder}, {Schmidt}, {Shappee}, \&
  {Walker}}]{Phillips19}
{Phillips}, M.~M., {Contreras}, C., {Hsiao}, E.~Y., {et~al.} 2019, \pasp, 131,
  014001, \dodoi{10.1088/1538-3873/aae8bd}

\bibitem[{{Piersanti} {et~al.}(2003){Piersanti}, {Gagliardi}, {Iben}, \&
  {Tornamb{\'e}}}]{Piersanti03}
{Piersanti}, L., {Gagliardi}, S., {Iben}, Icko, J., \& {Tornamb{\'e}}, A. 2003,
  \apj, 598, 1229, \dodoi{10.1086/378952}

\bibitem[{{Piro} \& {Morozova}(2016)}]{Piro16}
{Piro}, A.~L., \& {Morozova}, V.~S. 2016, \apj, 826, 96,
  \dodoi{10.3847/0004-637X/826/1/96}

\bibitem[{{Piro} \& {Nakar}(2013)}]{Piro13}
{Piro}, A.~L., \& {Nakar}, E. 2013, \apj, 769, 67,
  \dodoi{10.1088/0004-637X/769/1/67}

\bibitem[{{Piro} \& {Nakar}(2014)}]{Piro14}
---. 2014, \apj, 784, 85, \dodoi{10.1088/0004-637X/784/1/85}

\bibitem[{{Polin} {et~al.}(2019){Polin}, {Nugent}, \& {Kasen}}]{Polin19}
{Polin}, A., {Nugent}, P., \& {Kasen}, D. 2019, \apj, 873, 84,
  \dodoi{10.3847/1538-4357/aafb6a}

\bibitem[{{Poole} {et~al.}(2008){Poole}, {Breeveld}, {Page}, {Landsman},
  {Holland}, {Roming}, {Kuin}, {Brown}, {Gronwall}, {Hunsberger}, {Koch},
  {Mason}, {Schady}, {vanden Berk}, {Blustin}, {Boyd}, {Broos}, {Carter},
  {Chester}, {Cucchiara}, {Hancock}, {Huckle}, {Immler}, {Ivanushkina},
  {Kennedy}, {Marshall}, {Morgan}, {Pandey}, {de Pasquale}, {Smith}, \&
  {Still}}]{Poole08}
{Poole}, T.~S., {Breeveld}, A.~A., {Page}, M.~J., {et~al.} 2008, \mnras, 383,
  627, \dodoi{10.1111/j.1365-2966.2007.12563.x}

\bibitem[{{Poznanski} {et~al.}(2012){Poznanski}, {Prochaska}, \&
  {Bloom}}]{Poznanski12}
{Poznanski}, D., {Prochaska}, J.~X., \& {Bloom}, J.~S. 2012, \mnras, 426, 1465,
  \dodoi{10.1111/j.1365-2966.2012.21796.x}

\bibitem[{{Raiteri} {et~al.}(1996){Raiteri}, {Villata}, \&
  {Navarro}}]{Raiteri96}
{Raiteri}, C.~M., {Villata}, M., \& {Navarro}, J.~F. 1996, \aap, 315, 105

\bibitem[{{Raskin} \& {Kasen}(2013)}]{Raskin13}
{Raskin}, C., \& {Kasen}, D. 2013, \apj, 772, 1,
  \dodoi{10.1088/0004-637X/772/1/1}

\bibitem[{{Raskin} {et~al.}(2014){Raskin}, {Kasen}, {Moll}, {Schwab}, \&
  {Woosley}}]{Raskin14}
{Raskin}, C., {Kasen}, D., {Moll}, R., {Schwab}, J., \& {Woosley}, S. 2014,
  \apj, 788, 75, \dodoi{10.1088/0004-637X/788/1/75}

\bibitem[{{Reindl} {et~al.}(2005){Reindl}, {Tammann}, {Sandage}, \&
  {Saha}}]{Reindl05}
{Reindl}, B., {Tammann}, G.~A., {Sandage}, A., \& {Saha}, A. 2005, \apj, 624,
  532, \dodoi{10.1086/429218}

\bibitem[{{Rhee} \& {van Albada}(1996)}]{Rhee96}
{Rhee}, M.~H., \& {van Albada}, T.~S. 1996, \aaps, 115, 407

\bibitem[{{Riess} {et~al.}(1998){Riess}, {Filippenko}, {Challis},
  {Clocchiatti}, {Diercks}, {Garnavich}, {Gilliland}, {Hogan}, {Jha},
  {Kirshner}, {Leibundgut}, {Phillips}, {Reiss}, {Schmidt}, {Schommer},
  {Smith}, {Spyromilio}, {Stubbs}, {Suntzeff}, \& {Tonry}}]{riess98}
{Riess}, A.~G., {Filippenko}, A.~V., {Challis}, P., {et~al.} 1998, \aj, 116,
  1009, \dodoi{10.1086/300499}

\bibitem[{{Riess} {et~al.}(2022){Riess}, {Yuan}, {Macri}, {Scolnic}, {Brout},
  {Casertano}, {Jones}, {Murakami}, {Anand}, {Breuval}, {Brink}, {Filippenko},
  {Hoffmann}, {Jha}, {D'arcy Kenworthy}, {Mackenty}, {Stahl}, \&
  {Zheng}}]{Riess22}
{Riess}, A.~G., {Yuan}, W., {Macri}, L.~M., {et~al.} 2022, \apjl, 934, L7,
  \dodoi{10.3847/2041-8213/ac5c5b}

\bibitem[{{Rines} {et~al.}(2016){Rines}, {Geller}, {Diaferio}, \&
  {Hwang}}]{Rines16}
{Rines}, K.~J., {Geller}, M.~J., {Diaferio}, A., \& {Hwang}, H.~S. 2016, \apj,
  819, 63, \dodoi{10.3847/0004-637X/819/1/63}

\bibitem[{{R{\"o}pke} \& {Niemeyer}(2007)}]{Ropke07}
{R{\"o}pke}, F.~K., \& {Niemeyer}, J.~C. 2007, \aap, 464, 683,
  \dodoi{10.1051/0004-6361:20066585}

\bibitem[{{R{\"o}pke} {et~al.}(2012){R{\"o}pke}, {Kromer}, {Seitenzahl},
  {Pakmor}, {Sim}, {Taubenberger}, {Ciaraldi-Schoolmann}, {Hillebrandt},
  {Aldering}, {Antilogus}, {Baltay}, {Benitez-Herrera}, {Bongard}, {Buton},
  {Canto}, {Cellier-Holzem}, {Childress}, {Chotard}, {Copin}, {Fakhouri},
  {Fink}, {Fouchez}, {Gangler}, {Guy}, {Hachinger}, {Hsiao}, {Chen},
  {Kerschhaggl}, {Kowalski}, {Nugent}, {Paech}, {Pain}, {Pecontal}, {Pereira},
  {Perlmutter}, {Rabinowitz}, {Rigault}, {Runge}, {Saunders}, {Smadja},
  {Suzuki}, {Tao}, {Thomas}, {Tilquin}, \& {Wu}}]{Ropke12}
{R{\"o}pke}, F.~K., {Kromer}, M., {Seitenzahl}, I.~R., {et~al.} 2012, \apjl,
  750, L19, \dodoi{10.1088/2041-8205/750/1/L19}

\bibitem[{{Rothberg} \& {Joseph}(2006)}]{Rothberg06}
{Rothberg}, B., \& {Joseph}, R.~D. 2006, \aj, 131, 185, \dodoi{10.1086/498452}

\bibitem[{{Sabbi} {et~al.}(2018){Sabbi}, {Calzetti}, {Ubeda}, {Adamo},
  {Cignoni}, {Thilker}, {Aloisi}, {Elmegreen}, {Elmegreen}, {Gouliermis},
  {Grebel}, {Messa}, {Smith}, {Tosi}, {Dolphin}, {Andrews}, {Ashworth},
  {Bright}, {Brown}, {Chandar}, {Christian}, {Clayton}, {Cook}, {Dale}, {de
  Mink}, {Dobbs}, {Evans}, {Fumagalli}, {Gallagher}, {Grasha}, {Herrero},
  {Hunter}, {Johnson}, {Kahre}, {Kennicutt}, {Kim}, {Krumholz}, {Lee},
  {Lennon}, {Martin}, {Nair}, {Nota}, {{\"O}stlin}, {Pellerin}, {Prieto},
  {Regan}, {Ryon}, {Sacchi}, {Schaerer}, {Schiminovich}, {Shabani}, {Van Dyk},
  {Walterbos}, {Whitmore}, \& {Wofford}}]{Sabbi18}
{Sabbi}, E., {Calzetti}, D., {Ubeda}, L., {et~al.} 2018, \apjs, 235, 23,
  \dodoi{10.3847/1538-4365/aaa8e5}

\bibitem[{{Sagiv} {et~al.}(2014){Sagiv}, {Gal-Yam}, {Ofek}, {Waxman},
  {Aharonson}, {Kulkarni}, {Nakar}, {Maoz}, {Trakhtenbrot}, {Phinney}, {Topaz},
  {Beichman}, {Murthy}, \& {Worden}}]{Sagiv14}
{Sagiv}, I., {Gal-Yam}, A., {Ofek}, E.~O., {et~al.} 2014, \aj, 147, 79,
  \dodoi{10.1088/0004-6256/147/4/79}

\bibitem[{{Sai} {et~al.}(2022){Sai}, {Wang}, {Elias-Rosa}, {Yang}, {Zhang},
  {Lin}, {Mo}, {Piro}, {Zeng}, {Reguitti}, {Brown}, {Burns}, {Cai}, {Fiore},
  {Hsiao}, {Isern}, {Itagaki}, {Li}, {Li}, {Pessi}, {Phillips}, {Schuldt},
  {Shahbandeh}, {Stritzinger}, {Tomasella}, {Vogl}, {Wang}, {Wang}, {Wu},
  {Yang}, {Zhang}, {Zhang}, \& {Zhang}}]{Sai22}
{Sai}, H., {Wang}, X., {Elias-Rosa}, N., {et~al.} 2022, \mnras, 514, 3541,
  \dodoi{10.1093/mnras/stac1525}

\bibitem[{{Sand} {et~al.}(2021){Sand}, {Sarbadhicary}, {Pellegrino}, {Misra},
  {Dastidar}, {Brown}, {Itagaki}, {Valenti}, {Swift}, {Andrews}, {Bostroem},
  {Burke}, {Chomiuk}, {Dong}, {Galbany}, {Graham}, {Hiramatsu}, {Howell},
  {Hsiao}, {Janzen}, {Jencson}, {Lundquist}, {McCully}, {Reichart}, {Smith},
  {Wang}, \& {Wyatt}}]{Sand21}
{Sand}, D.~J., {Sarbadhicary}, S.~K., {Pellegrino}, C., {et~al.} 2021, \apj,
  922, 21, \dodoi{10.3847/1538-4357/ac20da}

\bibitem[{{Saulder} {et~al.}(2016){Saulder}, {van Kampen}, {Chilingarian},
  {Mieske}, \& {Zeilinger}}]{Saulder16}
{Saulder}, C., {van Kampen}, E., {Chilingarian}, I.~V., {Mieske}, S., \&
  {Zeilinger}, W.~W. 2016, \aap, 596, A14, \dodoi{10.1051/0004-6361/201526711}

\bibitem[{{Scalzo} {et~al.}(2012){Scalzo}, {Aldering}, {Antilogus}, {Aragon},
  {Bailey}, {Baltay}, {Bongard}, {Buton}, {Canto}, {Cellier-Holzem},
  {Childress}, {Chotard}, {Copin}, {Fakhouri}, {Gangler}, {Guy}, {Hsiao},
  {Kerschhaggl}, {Kowalski}, {Nugent}, {Paech}, {Pain}, {Pecontal}, {Pereira},
  {Perlmutter}, {Rabinowitz}, {Rigault}, {Runge}, {Smadja}, {Tao}, {Thomas},
  {Weaver}, {Wu}, \& {Nearby Supernova Factory}}]{Scalzo12}
{Scalzo}, R., {Aldering}, G., {Antilogus}, P., {et~al.} 2012, \apj, 757, 12,
  \dodoi{10.1088/0004-637X/757/1/12}

\bibitem[{{Scalzo} {et~al.}(2010){Scalzo}, {Aldering}, {Antilogus}, {Aragon},
  {Bailey}, {Baltay}, {Bongard}, {Buton}, {Childress}, {Chotard}, {Copin},
  {Fakhouri}, {Gal-Yam}, {Gangler}, {Hoyer}, {Kasliwal}, {Loken}, {Nugent},
  {Pain}, {P{\'e}contal}, {Pereira}, {Perlmutter}, {Rabinowitz}, {Rau},
  {Rigaudier}, {Runge}, {Smadja}, {Tao}, {Thomas}, {Weaver}, \&
  {Wu}}]{Scalzo10}
{Scalzo}, R.~A., {Aldering}, G., {Antilogus}, P., {et~al.} 2010, \apj, 713,
  1073, \dodoi{10.1088/0004-637X/713/2/1073}

\bibitem[{{Scalzo} {et~al.}(2014){Scalzo}, {Childress}, {Tucker}, {Yuan},
  {Schmidt}, {Brown}, {Contreras}, {Morrell}, {Hsiao}, {Burns}, {Phillips},
  {Campillay}, {Gonzalez}, {Krisciunas}, {Stritzinger}, {Graham}, {Parrent},
  {Valenti}, {Lidman}, {Schaefer}, {Scott}, {Fraser}, {Gal-Yam}, {Inserra},
  {Maguire}, {Smartt}, {Sollerman}, {Sullivan}, {Taddia}, {Yaron}, {Young},
  {Taubenberger}, {Baltay}, {Ellman}, {Feindt}, {Hadjiyska}, {McKinnon},
  {Nugent}, {Rabinowitz}, \& {Walker}}]{Scalzo14}
{Scalzo}, R.~A., {Childress}, M., {Tucker}, B., {et~al.} 2014, \mnras, 445, 30,
  \dodoi{10.1093/mnras/stu1723}

\bibitem[{{Schlafly} \& {Finkbeiner}(2011)}]{Schlafly11}
{Schlafly}, E.~F., \& {Finkbeiner}, D.~P. 2011, \apj, 737, 103,
  \dodoi{10.1088/0004-637X/737/2/103}

\bibitem[{{Schneider} {et~al.}(1990){Schneider}, {Thuan}, {Magri}, \&
  {Wadiak}}]{Schneider90}
{Schneider}, S.~E., {Thuan}, T.~X., {Magri}, C., \& {Wadiak}, J.~E. 1990,
  \apjs, 72, 245, \dodoi{10.1086/191416}

\bibitem[{{Schneider} {et~al.}(1992){Schneider}, {Thuan}, {Mangum}, \&
  {Miller}}]{Schneider92}
{Schneider}, S.~E., {Thuan}, T.~X., {Mangum}, J.~G., \& {Miller}, J. 1992,
  \apjs, 81, 5, \dodoi{10.1086/191684}

\bibitem[{{Shappee} {et~al.}(2013{\natexlab{a}}){Shappee}, {Kochanek}, \&
  {Stanek}}]{Shappee13b}
{Shappee}, B.~J., {Kochanek}, C.~S., \& {Stanek}, K.~Z. 2013{\natexlab{a}},
  \apj, 765, 150, \dodoi{10.1088/0004-637X/765/2/150}

\bibitem[{{Shappee} {et~al.}(2018){Shappee}, {Piro}, {Stanek}, {Patel},
  {Margutti}, {Lipunov}, \& {Pogge}}]{Shappee18}
{Shappee}, B.~J., {Piro}, A.~L., {Stanek}, K.~Z., {et~al.} 2018, \apj, 855, 6,
  \dodoi{10.3847/1538-4357/aaa1e9}

\bibitem[{{Shappee} \& {Stanek}(2011)}]{Shappee11}
{Shappee}, B.~J., \& {Stanek}, K.~Z. 2011, \apj, 733, 124,
  \dodoi{10.1088/0004-637X/733/2/124}

\bibitem[{{Shappee} {et~al.}(2017){Shappee}, {Stanek}, {Kochanek}, \&
  {Garnavich}}]{Shappee17}
{Shappee}, B.~J., {Stanek}, K.~Z., {Kochanek}, C.~S., \& {Garnavich}, P.~M.
  2017, \apj, 841, 48, \dodoi{10.3847/1538-4357/aa6eab}

\bibitem[{{Shappee} {et~al.}(2013{\natexlab{b}}){Shappee}, {Stanek}, {Pogge},
  \& {Garnavich}}]{Shappee13a}
{Shappee}, B.~J., {Stanek}, K.~Z., {Pogge}, R.~W., \& {Garnavich}, P.~M.
  2013{\natexlab{b}}, \apjl, 762, L5, \dodoi{10.1088/2041-8205/762/1/L5}

\bibitem[{{Shappee} \& {Thompson}(2013)}]{Shappee13c}
{Shappee}, B.~J., \& {Thompson}, T.~A. 2013, \apj, 766, 64,
  \dodoi{10.1088/0004-637X/766/1/64}

\bibitem[{{Shappee} {et~al.}(2014){Shappee}, {Prieto}, {Grupe}, {Kochanek},
  {Stanek}, {De Rosa}, {Mathur}, {Zu}, {Peterson}, {Pogge}, {Komossa}, {Im},
  {Jencson}, {Holoien}, {Basu}, {Beacom}, {Szczygie{\l}}, {Brimacombe},
  {Adams}, {Campillay}, {Choi}, {Contreras}, {Dietrich}, {Dubberley},
  {Elphick}, {Foale}, {Giustini}, {Gonzalez}, {Hawkins}, {Howell}, {Hsiao},
  {Koss}, {Leighly}, {Morrell}, {Mudd}, {Mullins}, {Nugent}, {Parrent},
  {Phillips}, {Pojmanski}, {Rosing}, {Ross}, {Sand}, {Terndrup}, {Valenti},
  {Walker}, \& {Yoon}}]{Shappee14}
{Shappee}, B.~J., {Prieto}, J.~L., {Grupe}, D., {et~al.} 2014, \apj, 788, 48,
  \dodoi{10.1088/0004-637X/788/1/48}

\bibitem[{{Shappee} {et~al.}(2016){Shappee}, {Piro}, {Holoien}, {Prieto},
  {Contreras}, {Itagaki}, {Burns}, {Kochanek}, {Stanek}, {Alper}, {Basu},
  {Beacom}, {Bersier}, {Brimacombe}, {Conseil}, {Danilet}, {Dong}, {Falco},
  {Grupe}, {Hsiao}, {Kiyota}, {Morrell}, {Nicolas}, {Phillips}, {Pojmanski},
  {Simonian}, {Stritzinger}, {Szczygie{\l}}, {Taddia}, {Thompson},
  {Thorstensen}, {Wagner}, \& {Wo{\'z}niak}}]{Shappee16}
{Shappee}, B.~J., {Piro}, A.~L., {Holoien}, T.~W.~S., {et~al.} 2016, \apj, 826,
  144, \dodoi{10.3847/0004-637X/826/2/144}

\bibitem[{{Shappee} {et~al.}(2019){Shappee}, {Holoien}, {Drout}, {Auchettl},
  {Stritzinger}, {Kochanek}, {Stanek}, {Shaya}, {Narayan}, {ASAS-SN}, {Brown},
  {Bose}, {Bersier}, {Brimacombe}, {Chen}, {Dong}, {Holmbo}, {Katz},
  {Mu{\~n}oz}, {Mutel}, {Post}, {Prieto}, {Shields}, {Tallon}, {Thompson},
  {Vallely}, {Villanueva}, {ATLAS}, {Denneau}, {Flewelling}, {Heinze}, {Smith},
  {Stalder}, {Tonry}, {Weiland}, {Kepler/K2}, {Barclay}, {Barentsen}, {Cody},
  {Dotson}, {Foerster}, {Garnavich}, {Gully-Santiago}, {Hedges}, {Howell},
  {Kasen}, {Margheim}, {Mushotzky}, {Rest}, {Tucker}, {Villar}, {Zenteno},
  {Kepler Spacecraft Team}, {Beerman}, {Bjella}, {Castillo}, {Coughlin},
  {Elsaesser}, {Flynn}, {Gangopadhyay}, {Griest}, {Hanley}, {Kampmeier},
  {Kloetzel}, {Kohnert}, {Labonde}, {Larsen}, {Larson}, {McCalmont-Everton},
  {McGinn}, {Migliorini}, {Moffatt}, {Muszynski}, {Nystrom}, {Osborne},
  {Packard}, {Peterson}, {Redick}, {Reedy}, {Ross}, {Spencer}, {Steward}, {Van
  Cleve}, {Cardoso}, {Weschler}, {Wheaton}, {Pan-STARRS}, {Bulger}, {Chambers},
  {Flewelling}, {Huber}, {Lowe}, {Magnier}, {Schultz}, {Waters}, {Willman},
  {PTSS/TNTS}, {Baron}, {Chen}, {Derkacy}, {Huang}, {Li}, {Li}, {Li}, {Mo},
  {Rui}, {Sai}, {Wang}, {Wang}, {Wang}, {Xiang}, {Zhang}, {Zhang}, {Zhang},
  {Zhang}, {Zhang}, {Zhao}, {Brown}, {Hermes}, {Nordin}, {Points}, {S{\'o}dor},
  {Strampelli}, \& {Zenteno}}]{Shappee19}
{Shappee}, B.~J., {Holoien}, T.~W.~S., {Drout}, M.~R., {et~al.} 2019, \apj,
  870, 13, \dodoi{10.3847/1538-4357/aaec79}

\bibitem[{{Siebert} {et~al.}(2020){Siebert}, {Dimitriadis}, {Polin}, \&
  {Foley}}]{Siebert20}
{Siebert}, M.~R., {Dimitriadis}, G., {Polin}, A., \& {Foley}, R.~J. 2020,
  \apjl, 900, L27, \dodoi{10.3847/2041-8213/abae6e}

\bibitem[{{Siebert} {et~al.}(2023){Siebert}, {Kwok}, {Johansson}, {Jha},
  {Blondin}, {Dessart}, {Foley}, {Hillier}, {Larison}, {Pakmor}, {Temim},
  {Andrews}, {Auchettl}, {Badenes}, {Barna}, {Bostroem}, {Brenner Newman},
  {Brink}, {Jos{\'e} Bustamante-Rosell}, {Camacho-Neves}, {Clocchiatti},
  {Coulter}, {Davis}, {Deckers}, {Dimitriadis}, {Dong}, {Farah}, {Filippenko},
  {Fl{\"o}rs}, {Fox}, {Garnavich}, {Padilla Gonzalez}, {Graur}, {Hambsch},
  {Hosseinzadeh}, {Howell}, {Hughes}, {Kerzendorf}, {Le Saux}, {Maeda},
  {Maguire}, {McCully}, {Mihalenko}, {Newsome}, {O'Brien}, {Pearson},
  {Pellegrino}, {Pierel}, {Polin}, {Rest}, {Rojas-Bravo}, {Sand}, {Schwab},
  {Shahbandeh}, {Shrestha}, {Smith}, {Strolger}, {Szalai}, {Taggart},
  {Terreran}, {Terwel}, {Tinyanont}, {Valenti}, {Vink{\'o}}, {Wheeler}, {Yang},
  {Zheng}, {Ashall}, {Derkacy}, {Galbany}, {Hoeflich}, {Hsiao}, {De Jaeger},
  {Lu}, {Maund}, {Medler}, {Morrell}, {Shappee}, {Stritzinger}, {Suntzeff},
  {Tucker}, \& {Wang}}]{Siebert23b}
{Siebert}, M.~R., {Kwok}, L.~A., {Johansson}, J., {et~al.} 2023, arXiv
  e-prints, arXiv:2308.12449.
\newblock \doarXiv{2308.12449}

\bibitem[{{Silverman} {et~al.}(2011){Silverman}, {Ganeshalingam}, {Li},
  {Filippenko}, {Miller}, \& {Poznanski}}]{Silverman11}
{Silverman}, J.~M., {Ganeshalingam}, M., {Li}, W., {et~al.} 2011, \mnras, 410,
  585, \dodoi{10.1111/j.1365-2966.2010.17474.x}

\bibitem[{{Silverman} {et~al.}(2012){Silverman}, {Foley}, {Filippenko},
  {Ganeshalingam}, {Barth}, {Chornock}, {Griffith}, {Kong}, {Lee}, {Leonard},
  {Matheson}, {Miller}, {Steele}, {Barris}, {Bloom}, {Cobb}, {Coil},
  {Desroches}, {Gates}, {Ho}, {Jha}, {Kandrashoff}, {Li}, {Mandel}, {Modjaz},
  {Moore}, {Mostardi}, {Papenkova}, {Park}, {Perley}, {Poznanski}, {Reuter},
  {Scala}, {Serduke}, {Shields}, {Swift}, {Tonry}, {Van Dyk}, {Wang}, \&
  {Wong}}]{Silverman12}
{Silverman}, J.~M., {Foley}, R.~J., {Filippenko}, A.~V., {et~al.} 2012, \mnras,
  425, 1789, \dodoi{10.1111/j.1365-2966.2012.21270.x}

\bibitem[{{Siverd} {et~al.}(2015){Siverd}, {Goobar}, {Stassun}, \&
  {Pepper}}]{Siverd15}
{Siverd}, R.~J., {Goobar}, A., {Stassun}, K.~G., \& {Pepper}, J. 2015, \apj,
  799, 105, \dodoi{10.1088/0004-637X/799/1/105}

\bibitem[{{Siverd} {et~al.}(2012){Siverd}, {Beatty}, {Pepper}, {Eastman},
  {Collins}, {Bieryla}, {Latham}, {Buchhave}, {Jensen}, {Crepp}, {Street},
  {Stassun}, {Gaudi}, {Berlind}, {Calkins}, {DePoy}, {Esquerdo}, {Fulton},
  {F{\H{u}}r{\'e}sz}, {Geary}, {Gould}, {Hebb}, {Kielkopf}, {Marshall},
  {Pogge}, {Stanek}, {Stefanik}, {Szentgyorgyi}, {Trueblood}, {Trueblood},
  {Stutz}, \& {van Saders}}]{Siverd12}
{Siverd}, R.~J., {Beatty}, T.~G., {Pepper}, J., {et~al.} 2012, \apj, 761, 123,
  \dodoi{10.1088/0004-637X/761/2/123}

\bibitem[{{Smartt} {et~al.}(2015){Smartt}, {Valenti}, {Fraser}, {Inserra},
  {Young}, {Sullivan}, {Pastorello}, {Benetti}, {Gal-Yam}, {Knapic},
  {Molinaro}, {Smareglia}, {Smith}, {Taubenberger}, {Yaron}, {Anderson},
  {Ashall}, {Balland}, {Baltay}, {Barbarino}, {Bauer}, {Baumont}, {Bersier},
  {Blagorodnova}, {Bongard}, {Botticella}, {Bufano}, {Bulla}, {Cappellaro},
  {Campbell}, {Cellier-Holzem}, {Chen}, {Childress}, {Clocchiatti},
  {Contreras}, {Dall'Ora}, {Danziger}, {de Jaeger}, {De Cia}, {Della Valle},
  {Dennefeld}, {Elias-Rosa}, {Elman}, {Feindt}, {Fleury}, {Gall},
  {Gonzalez-Gaitan}, {Galbany}, {Morales Garoffolo}, {Greggio}, {Guillou},
  {Hachinger}, {Hadjiyska}, {Hage}, {Hillebrandt}, {Hodgkin}, {Hsiao}, {James},
  {Jerkstrand}, {Kangas}, {Kankare}, {Kotak}, {Kromer}, {Kuncarayakti},
  {Leloudas}, {Lundqvist}, {Lyman}, {Hook}, {Maguire}, {Manulis}, {Margheim},
  {Mattila}, {Maund}, {Mazzali}, {McCrum}, {McKinnon}, {Moreno-Raya},
  {Nicholl}, {Nugent}, {Pain}, {Pignata}, {Phillips}, {Polshaw}, {Pumo},
  {Rabinowitz}, {Reilly}, {Romero-Ca{\~n}izales}, {Scalzo}, {Schmidt},
  {Schulze}, {Sim}, {Sollerman}, {Taddia}, {Tartaglia}, {Terreran},
  {Tomasella}, {Turatto}, {Walker}, {Walton}, {Wyrzykowski}, {Yuan}, \&
  {Zampieri}}]{Smartt15}
{Smartt}, S.~J., {Valenti}, S., {Fraser}, M., {et~al.} 2015, \aap, 579, A40,
  \dodoi{10.1051/0004-6361/201425237}

\bibitem[{{Smith} {et~al.}(2000){Smith}, {Lucey}, {Hudson}, {Schlegel}, \&
  {Davies}}]{Smith2000}
{Smith}, R.~J., {Lucey}, J.~R., {Hudson}, M.~J., {Schlegel}, D.~J., \&
  {Davies}, R.~L. 2000, \mnras, 313, 469,
  \dodoi{10.1046/j.1365-8711.2000.03251.x}

\bibitem[{{Smitka} {et~al.}(2015){Smitka}, {Brown}, {Suntzeff}, {Zhang},
  {Zhai}, {Wang}, {Mo}, \& {Zhang}}]{Smitka15}
{Smitka}, M.~T., {Brown}, P.~J., {Suntzeff}, N.~B., {et~al.} 2015, \apj, 813,
  30, \dodoi{10.1088/0004-637X/813/1/30}

\bibitem[{{Springob} {et~al.}(2005){Springob}, {Haynes}, {Giovanelli}, \&
  {Kent}}]{Springob05}
{Springob}, C.~M., {Haynes}, M.~P., {Giovanelli}, R., \& {Kent}, B.~R. 2005,
  \apjs, 160, 149, \dodoi{10.1086/431550}

\bibitem[{{Springob} {et~al.}(2014){Springob}, {Magoulas}, {Colless}, {Mould},
  {Erdo{\u{g}}du}, {Jones}, {Lucey}, {Campbell}, \& {Fluke}}]{Springob14}
{Springob}, C.~M., {Magoulas}, C., {Colless}, M., {et~al.} 2014, \mnras, 445,
  2677, \dodoi{10.1093/mnras/stu1743}

\bibitem[{{Srivastav} {et~al.}(2023{\natexlab{a}}){Srivastav}, {Smartt},
  {Huber}, {Dimitriadis}, {Chambers}, {Fulton}, {Moore}, {Callan},
  {Gillanders}, {Maguire}, {Nicholl}, {Shingles}, {Sim}, {Smith}, {Anderson},
  {de Boer}, {Chen}, {Gao}, \& {Young}}]{Srivastav23a}
{Srivastav}, S., {Smartt}, S.~J., {Huber}, M.~E., {et~al.} 2023{\natexlab{a}},
  \apjl, 943, L20, \dodoi{10.3847/2041-8213/acb2ce}

\bibitem[{{Srivastav} {et~al.}(2023{\natexlab{b}}){Srivastav}, {Moore},
  {Nicholl}, {Magee}, {Smartt}, {Fulton}, {Sim}, {Pollin}, {Galbany},
  {Inserra}, {Kozyreva}, {Moriya}, {Callan}, {Sheng}, {Smith}, {Sommer},
  {Anderson}, {Deckers}, {Gromadzki}, {M{\"u}ller-Bravo}, {Pignata}, {Rest}, \&
  {Young}}]{Srivastav23b}
{Srivastav}, S., {Moore}, T., {Nicholl}, M., {et~al.} 2023{\natexlab{b}}, arXiv
  e-prints, arXiv:2308.06019, \dodoi{10.48550/arXiv.2308.06019}

\bibitem[{{Tanaka} {et~al.}(2010){Tanaka}, {Kawabata}, {Yamanaka}, {Maeda},
  {Hattori}, {Aoki}, {Nomoto}, {Iye}, {Sasaki}, {Mazzali}, \&
  {Pian}}]{Tanaka10}
{Tanaka}, M., {Kawabata}, K.~S., {Yamanaka}, M., {et~al.} 2010, \apj, 714,
  1209, \dodoi{10.1088/0004-637X/714/2/1209}

\bibitem[{{Taubenberger}(2017)}]{Taubenberger17}
{Taubenberger}, S. 2017, {The Extremes of Thermonuclear Supernovae}, ed. A.~W.
  {Alsabti} \& P.~{Murdin}, 317, \dodoi{10.1007/978-3-319-21846-5\_37}

\bibitem[{{Taubenberger} {et~al.}(2013{\natexlab{a}}){Taubenberger}, {Kromer},
  {Pakmor}, {Pignata}, {Maeda}, {Hachinger}, {Leibundgut}, \&
  {Hillebrandt}}]{Taubenberger13b}
{Taubenberger}, S., {Kromer}, M., {Pakmor}, R., {et~al.} 2013{\natexlab{a}},
  \apjl, 775, L43, \dodoi{10.1088/2041-8205/775/2/L43}

\bibitem[{{Taubenberger} {et~al.}(2011){Taubenberger}, {Benetti}, {Childress},
  {Pakmor}, {Hachinger}, {Mazzali}, {Stanishev}, {Elias-Rosa}, {Agnoletto},
  {Bufano}, {Ergon}, {Harutyunyan}, {Inserra}, {Kankare}, {Kromer},
  {Navasardyan}, {Nicolas}, {Pastorello}, {Prosperi}, {Salgado}, {Sollerman},
  {Stritzinger}, {Turatto}, {Valenti}, \& {Hillebrandt}}]{Taubenberger11}
{Taubenberger}, S., {Benetti}, S., {Childress}, M., {et~al.} 2011, \mnras, 412,
  2735, \dodoi{10.1111/j.1365-2966.2010.18107.x}

\bibitem[{{Taubenberger} {et~al.}(2013{\natexlab{b}}){Taubenberger}, {Kromer},
  {Hachinger}, {Mazzali}, {Benetti}, {Nugent}, {Scalzo}, {Pakmor}, {Stanishev},
  {Spyromilio}, {Bufano}, {Sim}, {Leibundgut}, \&
  {Hillebrandt}}]{Taubenberger13a}
{Taubenberger}, S., {Kromer}, M., {Hachinger}, S., {et~al.} 2013{\natexlab{b}},
  \mnras, 432, 3117, \dodoi{10.1093/mnras/stt668}

\bibitem[{{Taubenberger} {et~al.}(2019){Taubenberger}, {Floers}, {Vogl},
  {Kromer}, {Spyromilio}, {Aldering}, {Antilogus}, {Bailey}, {Baltay},
  {Bongard}, {Boone}, {Buton}, {Chotard}, {Copin}, {Dixon}, {Fouchez},
  {Fransson}, {Gangler}, {Gupta}, {Hachinger}, {Hayden}, {Hillebrandt}, {Kim},
  {Kowalski}, {Leget}, {Leibundgut}, {Mazzali}, {Noebauer}, {Nordin}, {Pain},
  {Pakmor}, {Pecontal}, {Pereira}, {Perlmutter}, {Ponder}, {Rabinowitz},
  {Rigault}, {Rubin}, {Runge}, {Saunders}, {Smadja}, {Tao}, \&
  {Thomas}}]{Taubenberger19}
{Taubenberger}, S., {Floers}, A., {Vogl}, C., {et~al.} 2019, \mnras, 488, 5473,
  \dodoi{10.1093/mnras/stz1977}

\bibitem[{{Theureau} {et~al.}(2007){Theureau}, {Hanski}, {Coudreau}, {Hallet},
  \& {Martin}}]{Theureau07}
{Theureau}, G., {Hanski}, M.~O., {Coudreau}, N., {Hallet}, N., \& {Martin},
  J.~M. 2007, \aap, 465, 71, \dodoi{10.1051/0004-6361:20066187}

\bibitem[{{Theureau} {et~al.}(2005){Theureau}, {Coudreau}, {Hallet}, {Hanski},
  {Alsac}, {Bottinelli}, {Gouguenheim}, {Martin}, \& {Paturel}}]{Theureau05}
{Theureau}, G., {Coudreau}, N., {Hallet}, N., {et~al.} 2005, \aap, 430, 373,
  \dodoi{10.1051/0004-6361:20047152}

\bibitem[{{Thompson}(2011)}]{thom11}
{Thompson}, T.~A. 2011, \apj, 741, 82, \dodoi{10.1088/0004-637X/741/2/82}

\bibitem[{{Thorp} {et~al.}(2021){Thorp}, {Mandel}, {Jones}, {Ward}, \&
  {Narayan}}]{Thorp21}
{Thorp}, S., {Mandel}, K.~S., {Jones}, D.~O., {Ward}, S.~M., \& {Narayan}, G.
  2021, \mnras, 508, 4310, \dodoi{10.1093/mnras/stab2849}

\bibitem[{{Tonry} {et~al.}(2018){Tonry}, {Denneau}, {Heinze}, {Stalder},
  {Smith}, {Smartt}, {Stubbs}, {Weiland}, \& {Rest}}]{Tonry18}
{Tonry}, J.~L., {Denneau}, L., {Heinze}, A.~N., {et~al.} 2018, \pasp, 130,
  064505, \dodoi{10.1088/1538-3873/aabadf}

\bibitem[{{Tucker} {et~al.}(2022{\natexlab{a}}){Tucker}, {Ashall}, {Shappee},
  {Kochanek}, {Stanek}, \& {Garnavich}}]{Tucker22a}
{Tucker}, M.~A., {Ashall}, C., {Shappee}, B.~J., {et~al.} 2022{\natexlab{a}},
  \apjl, 926, L25, \dodoi{10.3847/2041-8213/ac4fbd}

\bibitem[{{Tucker} {et~al.}(2022{\natexlab{b}}){Tucker}, {Shappee}, {Kochanek},
  {Stanek}, {Ashall}, {Anand}, \& {Garnavich}}]{Tucker22b}
{Tucker}, M.~A., {Shappee}, B.~J., {Kochanek}, C.~S., {et~al.}
  2022{\natexlab{b}}, \mnras, 517, 4119, \dodoi{10.1093/mnras/stac2873}

\bibitem[{{Tucker} {et~al.}(2020){Tucker}, {Shappee}, {Vallely}, {Stanek},
  {Prieto}, {Botyanszki}, {Kochanek}, {Anderson}, {Brown}, {Galbany},
  {Holoien}, {Hsiao}, {Kumar}, {Kuncarayakti}, {Morrell}, {Phillips},
  {Stritzinger}, \& {Thompson}}]{Tucker20}
{Tucker}, M.~A., {Shappee}, B.~J., {Vallely}, P.~J., {et~al.} 2020, \mnras,
  493, 1044, \dodoi{10.1093/mnras/stz3390}

\bibitem[{{Tucker} {et~al.}(2021){Tucker}, {Ashall}, {Shappee}, {Vallely},
  {Kochanek}, {Huber}, {Anand}, {Keane}, {Hsiao}, \& {Holoien}}]{Tucker21}
{Tucker}, M.~A., {Ashall}, C., {Shappee}, B.~J., {et~al.} 2021, \apj, 914, 50,
  \dodoi{10.3847/1538-4357/abf93b}

\bibitem[{{Tucker} {et~al.}(2022{\natexlab{c}}){Tucker}, {Shappee}, {Huber},
  {Payne}, {Do}, {Hinkle}, {de Jaeger}, {Ashall}, {Desai}, {Hoogendam},
  {Aldering}, {Auchettl}, {Baranec}, {Bulger}, {Chambers}, {Chun}, {Hodapp},
  {Lowe}, {McKay}, {Rampy}, {Rubin}, \& {Tonry}}]{Tucker22c}
{Tucker}, M.~A., {Shappee}, B.~J., {Huber}, M.~E., {et~al.} 2022{\natexlab{c}},
  \pasp, 134, 124502, \dodoi{10.1088/1538-3873/aca719}

\bibitem[{{Tully} {et~al.}(2016){Tully}, {Courtois}, \& {Sorce}}]{Tully16}
{Tully}, R.~B., {Courtois}, H.~M., \& {Sorce}, J.~G. 2016, \aj, 152, 50,
  \dodoi{10.3847/0004-6256/152/2/50}

\bibitem[{{Tully} {et~al.}(2013){Tully}, {Courtois}, {Dolphin}, {Fisher},
  {H{\'e}raudeau}, {Jacobs}, {Karachentsev}, {Makarov}, {Makarova},
  {Mitronova}, {Rizzi}, {Shaya}, {Sorce}, \& {Wu}}]{Tully13}
{Tully}, R.~B., {Courtois}, H.~M., {Dolphin}, A.~E., {et~al.} 2013, \aj, 146,
  86, \dodoi{10.1088/0004-6256/146/4/86}

\bibitem[{{van den Bosch} {et~al.}(2015){van den Bosch}, {Gebhardt},
  {G{\"u}ltekin}, {Y{\i}ld{\i}r{\i}m}, \& {Walsh}}]{Bosch_v/d15}
{van den Bosch}, R. C.~E., {Gebhardt}, K., {G{\"u}ltekin}, K.,
  {Y{\i}ld{\i}r{\i}m}, A., \& {Walsh}, J.~L. 2015, \apjs, 218, 10,
  \dodoi{10.1088/0067-0049/218/1/10}

\bibitem[{{van der Tak} {et~al.}(2008){van der Tak}, {Aalto}, \&
  {Meijerink}}]{Tak_v/d08}
{van der Tak}, F.~F.~S., {Aalto}, S., \& {Meijerink}, R. 2008, \aap, 477, L5,
  \dodoi{10.1051/0004-6361:20078824}

\bibitem[{{van Driel} {et~al.}(2001){van Driel}, {Marcum}, {Gallagher},
  {Wilcots}, {Guidoux}, \& {Monnier Ragaigne}}]{Dried_van01}
{van Driel}, W., {Marcum}, P., {Gallagher}, J.~S., I., {et~al.} 2001, \aap,
  378, 370, \dodoi{10.1051/0004-6361:20011241}

\bibitem[{{van Driel} {et~al.}(2016){van Driel}, {Butcher}, {Schneider},
  {Lehnert}, {Minchin}, {Blyth}, {Chemin}, {Hallet}, {Joseph}, {Kotze},
  {Kraan-Korteweg}, {Olofsson}, \& {Ramatsoku}}]{van_Driel16}
{van Driel}, W., {Butcher}, Z., {Schneider}, S., {et~al.} 2016, \aap, 595,
  A118, \dodoi{10.1051/0004-6361/201528048}

\bibitem[{{van Kerkwijk} {et~al.}(2010){van Kerkwijk}, {Chang}, \&
  {Justham}}]{vanKerkwijk10}
{van Kerkwijk}, M.~H., {Chang}, P., \& {Justham}, S. 2010, \apjl, 722, L157,
  \dodoi{10.1088/2041-8205/722/2/L157}

\bibitem[{{Walker} {et~al.}(2012){Walker}, {Hachinger}, {Mazzali}, {Ellis},
  {Sullivan}, {Gal Yam}, \& {Howell}}]{Walker12}
{Walker}, E.~S., {Hachinger}, S., {Mazzali}, P.~A., {et~al.} 2012, \mnras, 427,
  103, \dodoi{10.1111/j.1365-2966.2012.21928.x}

\bibitem[{{Wang} {et~al.}(2020){Wang}, {Contreras}, {Hu}, {Hamuy}, {Hsiao},
  {Sand}, {Anderson}, {Ashall}, {Burns}, {Chen}, {Diamond}, {Davis},
  {F{\"o}rster}, {Galbany}, {Gonz{\'a}lez-Gait{\'a}n}, {Gromadzki}, {Hoeflich},
  {Li}, {Marion}, {Morrell}, {Pignata}, {Prieto}, {Phillips}, {Shahbandeh},
  {Suntzeff}, {Valenti}, {Wang}, {Wang}, {Young}, {Yu}, \& {Zhang}}]{Wang20}
{Wang}, L., {Contreras}, C., {Hu}, M., {et~al.} 2020, \apj, 904, 14,
  \dodoi{10.3847/1538-4357/abba82}

\bibitem[{{Wang} {et~al.}(2021){Wang}, {Rest}, {Zenati}, {Ridden-Harper},
  {Dimitriadis}, {Narayan}, {Villar}, {Magee}, {Foley}, {Shaya}, {Garnavich},
  {Wang}, {Hu}, {B{\'o}di}, {Armstrong}, {Auchettl}, {Barclay}, {Barentsen},
  {Bogn{\'a}r}, {Brimacombe}, {Bulger}, {Burke}, {Challis}, {Chambers},
  {Coulter}, {Cs{\"o}rnyei}, {Cseh}, {Deckers}, {Dotson}, {Galbany},
  {Gonz{\'a}lez-Gait{\'a}n}, {Gromadzki}, {Gully-Santiago}, {Hanyecz},
  {Hedges}, {Hiramatsu}, {Hosseinzadeh}, {Howell}, {Howell}, {Huber}, {Jha},
  {Jones}, {K{\"o}nyves-T{\'o}th}, {Kalup}, {Kilpatrick}, {Kriskovics}, {Li},
  {Lowe}, {Margheim}, {McCully}, {Mitra}, {Mu{\~n}oz}, {Nicholl}, {Nordin},
  {P{\'a}l}, {Pan}, {Piro}, {Rest}, {Rino-Silvestre}, {Rojas-Bravo},
  {S{\'a}rneczky}, {Siebert}, {Smartt}, {Smith}, {S{\'o}dor}, {Stritzinger},
  {Szab{\'o}}, {Szak{\'a}ts}, {Tucker}, {Vink{\'o}}, {Wang}, {Wheeler},
  {Young}, {Zenteno}, {Zhang}, \& {Zsidi}}]{Wang21}
{Wang}, Q., {Rest}, A., {Zenati}, Y., {et~al.} 2021, \apj, 923, 167,
  \dodoi{10.3847/1538-4357/ac2c84}

\bibitem[{{Wang} {et~al.}(2023){Wang}, {Rest}, {Dimitriadis}, {Ridden-harper},
  {Siebert}, {Magee}, {Angus}, {Auchettl}, {Davis}, {Foley}, {Fox}, {Gomez},
  {Jencson}, {Jones}, {Kilpatrick}, {Pierel}, {Piro}, {Polin}, {Politsch},
  {Rojas-bravo}, {Shahbandeh}, {Villar}, {Zenati}, {Ashall}, {Chambers},
  {Coulter}, {De Boer}, {Dilullo}, {Gall}, {Gao}, {Hsiao}, {Huber}, {Izzo},
  {Khetan}, {Lebaron}, {Magnier}, {Mandel}, {Mcgill}, {Miao}, {Pan}, {Stevens},
  {Swift}, {Taggart}, \& {Yang}}]{Wang23}
{Wang}, Q., {Rest}, A., {Dimitriadis}, G., {et~al.} 2023, arXiv e-prints,
  arXiv:2305.03779, \dodoi{10.48550/arXiv.2305.03779}

\bibitem[{{Ward} {et~al.}(2022){Ward}, {Thorp}, {Mandel}, {Dhawan}, {Jones},
  {Taggart}, {Foley}, {Narayan}, {Chambers}, {Coulter}, {Davis}, {de Boer}, {de
  Soto}, {Earl}, {Gagliano}, {Gao}, {Hjorth}, {Huber}, {Izzo}, {Langeroodi},
  {Magnier}, {McGill}, {Rest}, {Rojas-Bravo}, \& {Wojtak}}]{Ward22}
{Ward}, S.~M., {Thorp}, S., {Mandel}, K.~S., {et~al.} 2022, arXiv e-prints,
  arXiv:2209.10558, \dodoi{10.48550/arXiv.2209.10558}

\bibitem[{{Webbink}(1984)}]{Webbink84}
{Webbink}, R.~F. 1984, \apj, 277, 355, \dodoi{10.1086/161701}

\bibitem[{{Wee} {et~al.}(2018){Wee}, {Chakraborty}, {Wang}, \&
  {Penprase}}]{Wee18}
{Wee}, J., {Chakraborty}, N., {Wang}, J., \& {Penprase}, B.~E. 2018, \apj, 863,
  90, \dodoi{10.3847/1538-4357/aacd4e}

\bibitem[{{Whelan} \& {Iben}(1973)}]{Whelan73}
{Whelan}, J., \& {Iben}, Icko, J. 1973, \apj, 186, 1007, \dodoi{10.1086/152565}

\bibitem[{{White} {et~al.}(2015){White}, {Kasliwal}, {Nugent}, {Gal-Yam},
  {Howell}, {Sullivan}, {Goobar}, {Piro}, {Bloom}, {Kulkarni}, {Laher},
  {Masci}, {Ofek}, {Surace}, {Ben-Ami}, {Cao}, {Cenko}, {Hook}, {J{\"o}nsson},
  {Matheson}, {Sternberg}, {Quimby}, \& {Yaron}}]{White15}
{White}, C.~J., {Kasliwal}, M.~M., {Nugent}, P.~E., {et~al.} 2015, \apj, 799,
  52, \dodoi{10.1088/0004-637X/799/1/52}

\bibitem[{{Woosley} \& {Kasen}(2011)}]{Woosley11}
{Woosley}, S.~E., \& {Kasen}, D. 2011, \apj, 734, 38,
  \dodoi{10.1088/0004-637X/734/1/38}

\bibitem[{{Woosley} \& {Weaver}(1994)}]{Woosley94}
{Woosley}, S.~E., \& {Weaver}, T.~A. 1994, \apj, 423, 371,
  \dodoi{10.1086/173813}

\bibitem[{{Xi} {et~al.}(2023){Xi}, {Wang}, {Li}, {Liu}, {Yan}, {Lin}, {Zhao},
  {Filippenko}, {Zheng}, {Brink}, {Yang}, {Ehgamberdiev}, {Mirzaqulov},
  {Reguitti}, {Pastorello}, {Tomasella}, {Cai}, {Zhang}, {Li}, {Zhang}, {Chen},
  {Liu}, {Ma}, \& {Xiang}}]{Xi23}
{Xi}, G., {Wang}, X., {Li}, G., {et~al.} 2023, arXiv e-prints,
  arXiv:2309.09213.
\newblock \doarXiv{2309.09213}

\bibitem[{{Yamanaka} {et~al.}(2009){Yamanaka}, {Kawabata}, {Kinugasa},
  {Tanaka}, {Imada}, {Maeda}, {Nomoto}, {Arai}, {Chiyonobu}, {Fukazawa},
  {Hashimoto}, {Honda}, {Ikejiri}, {Itoh}, {Kamata}, {Kawai}, {Komatsu},
  {Konishi}, {Kuroda}, {Miyamoto}, {Miyazaki}, {Nagae}, {Nakaya}, {Ohsugi},
  {Omodaka}, {Sakai}, {Sasada}, {Suzuki}, {Taguchi}, {Takahashi}, {Tanaka},
  {Uemura}, {Yamashita}, {Yanagisawa}, \& {Yoshida}}]{Yamanaka09}
{Yamanaka}, M., {Kawabata}, K.~S., {Kinugasa}, K., {et~al.} 2009, \apjl, 707,
  L118, \dodoi{10.1088/0004-637X/707/2/L118}

\bibitem[{{Yamanaka} {et~al.}(2014){Yamanaka}, {Maeda}, {Kawabata}, {Tanaka},
  {Takaki}, {Ueno}, {Masumoto}, {Kawabata}, {Itoh}, {Moritani}, {Akitaya},
  {Arai}, {Honda}, {Nishiyama}, {Kabashima}, {Matsumoto}, {Nogami}, \&
  {Yoshida}}]{Yamanaka14}
{Yamanaka}, M., {Maeda}, K., {Kawabata}, M., {et~al.} 2014, \apjl, 782, L35,
  \dodoi{10.1088/2041-8205/782/2/L35}

\bibitem[{{Yang} {et~al.}(2020){Yang}, {Hoeflich}, {Baade}, {Maund}, {Wang},
  {Brown}, {Stevance}, {Arcavi}, {Burke}, {Cikota}, {Clocchiatti}, {Gal-Yam},
  {Graham}, {Hiramatsu}, {Hosseinzadeh}, {Howell}, {Jha}, {McCully}, {Patat},
  {Sand}, {Schulze}, {Spyromilio}, {Valenti}, {Vink{\'o}}, {Wang}, {Wheeler},
  {Yaron}, \& {Zhang}}]{Yang20}
{Yang}, Y., {Hoeflich}, P., {Baade}, D., {et~al.} 2020, \apj, 902, 46,
  \dodoi{10.3847/1538-4357/aba759}

\bibitem[{{Yoon} \& {Langer}(2005)}]{Yoon05}
{Yoon}, S.~C., \& {Langer}, N. 2005, \aap, 435, 967,
  \dodoi{10.1051/0004-6361:20042542}

\bibitem[{{Zeng} {et~al.}(2021){Zeng}, {Wang}, {Esamdin}, {Pellegrino},
  {Burke}, {Stahl}, {Zheng}, {Filippenko}, {Howell}, {Sand}, {Valenti}, {Mo},
  {Xi}, {Liu}, {Zhang}, {Li}, {Iskandar}, {Zhang}, {Lin}, {Sai}, {Xiang},
  {Wei}, {Zhang}, {Reichart}, {Brink}, {McCully}, {Hiramatsu}, {Hosseinzadeh},
  {Jeffers}, {Ross}, {Stegman}, {Wang}, {Zhang}, \& {Ma}}]{Zeng21}
{Zeng}, X., {Wang}, X., {Esamdin}, A., {et~al.} 2021, \apj, 919, 49,
  \dodoi{10.3847/1538-4357/ac0e9c}

\bibitem[{{Zhang} {et~al.}(2022){Zhang}, {Zhang}, {Danzengluobu}, {Li}, {Zhao},
  {Zhang}, {Du}, {Zhu}, \& {Wu}}]{Zhang22}
{Zhang}, Y., {Zhang}, T., {Danzengluobu}, {et~al.} 2022, \pasp, 134, 074201,
  \dodoi{10.1088/1538-3873/ac7583}

\bibitem[{{Zheng} {et~al.}(2013){Zheng}, {Silverman}, {Filippenko}, {Kasen},
  {Nugent}, {Graham}, {Wang}, {Valenti}, {Ciabattari}, {Kelly}, {Fox},
  {Shivvers}, {Clubb}, {Cenko}, {Balam}, {Howell}, {Hsiao}, {Li}, {Marion},
  {Sand}, {Vinko}, {Wheeler}, \& {Zhang}}]{Zheng13}
{Zheng}, W., {Silverman}, J.~M., {Filippenko}, A.~V., {et~al.} 2013, \apjl,
  778, L15, \dodoi{10.1088/2041-8205/778/1/L15}

\bibitem[{{Zheng} {et~al.}(2014){Zheng}, {Shivvers}, {Filippenko}, {Itagaki},
  {Clubb}, {Fox}, {Graham}, {Kelly}, \& {Mauerhan}}]{Zheng14}
{Zheng}, W., {Shivvers}, I., {Filippenko}, A.~V., {et~al.} 2014, \apjl, 783,
  L24, \dodoi{10.1088/2041-8205/783/1/L24}

\end{thebibliography}
\bibliographystyle{aasjournal}

\begin{acknowledgements}
We thank the anonymous referee for providing useful comments and feedback which improved the draft. We thank Jason Hinkle, Aaron Do, Dhvanil Desai, Michael Fausnaugh, JJ Hermes, Jing Lu, and Josh Shields for insightful discussions and Mark Phillips for helpful comments that improved the draft. We thank Takashi Moriya and Keiichi Maeda for providing their supernovae model data. This material is based upon work supported by the National Science Foundation Graduate Research Fellowship Program under Grant No. 2236415. Any opinions, findings, and conclusions or recommendations expressed in this material are those of the author(s) and do not necessarily reflect the views of the National Science Foundation.

This research has made use of data obtained through the High Energy Astrophysics Science Archive Research Center Online Service, provided by the NASA/Goddard Space Flight Center.
\end{acknowledgements}

\appendix
\restartappendixnumbering

\section{Specific Comments on SNe Ia in Our Sample}\label{sec:individual_comments}
The majority of SNe Ia in our sample are well documented in the literature due to their early discoveries and intense follow-up campaigns compared to the majority of \Ia. In this section, we comment on our early-time light curve categorization for each SN Ia, which categorizes the \Ia\ sample into three groups: single, double, and bump as discussed in Section \ref{sec:data}.  We also provide a discussion on adopted $A_V$ values as well as $t_{B}^{max}$. For some \Ia, the only literature extinction estimate is from the \citet{Poznanski12} \NaI\ D pEW relationship, which we adopt with the reported uncertainties (although see \citealt{Phillips13}). Lastly, spectroscopic classification data were taken from TNS or individual object papers, as available.

\subsection{SN~2009ig}
\citet{Foley12} report $A_V = 0.01\pm0.01$ mag for the host galaxy of SN~2009ig and $t_{B}^{max}$ on MJD 55080.04. The Milky Way extinction is $E(B-V)_{MW}=0.03$~mag \citep{Schlafly11}. 

\citet{Foley12} reported early-time observations of SN~2009ig, a normal SN Ia. After subtracting a \citet{Arnett82} $f\propto t^2$ fireball model, SN~2009ig has positive residuals which are indicative of excess flux above what is expected with the fireball model. However, \citet{Foley12} found the rise was well fit with a single component power law. Given this acceptable single-component fit, we categorize SN~2009ig as an SN Ia \emph{without} early-time excess flux (i.e., as a single SN Ia). This determination is similar to \citet{Jiang18}, who also categorize SN~2009ig as having no early excess in either the UV or the optical. 

\subsection{SN~2011fe}
\citet{Pereira13} report $t_{B}^{max}$ on MJD 55814.51 and a host galaxy extinction of $E(B-V)=0.03\pm0.04$ mag. The Milky Way extinction is $E(B-V)_{MW}=0.01$~mag \citep{Schlafly11}.

SN~2011fe was an incredibly nearby normal SN Ia \citet{Nugent11}. Located in M~101 (NGC~5457) at 6.4 Mpc \citep{Shappee11}, SN~2011fe is one of the most nearby SN Ia to date. As such, extensive searches for companion interaction under the \citet{Kasen10} models have been performed, all yielding no evidence of companion interaction \citep{Li2011,Ropke12,Brown12a,Shappee13a,Shappee17,Tucker22a,Tucker22b} or even a surviving companion \citep{Lundqvist15} which are predicted to be overluminous \citep{Shappee13b}. Given the large sample of early data indicating strong agreement with the fireball model, we categorize SN~2011fe as a single SN Ia. 

\subsection{SN~2012cg}
SN~2012cg was initially reported by \citet{Silverman12}, and they found $t_{B}^{max}$ occurred on MJD =56080.0 and $E(B-V)=0.18$ mag. \citet{Marion16} also studied SN~2012cg and found $t_{B}^{max}$ on MJD 56081.3. Finally, \citet{Munari13} determined $t_{B}^{max}$ happened on MJD 56082.0. We elect to use the \citet{Marion16} value for $t_{B}^{max}$, which is consistent with the \citet{Munari13} value. The Milky Way extinction is $E(B-V)_{MW}=0.02$~mag \citep{Schlafly11}. 

SN~2012cg was initially reported by \citet{Silverman12}. Subsequent studies found evidence for \citep{Marion16} and against \citep{Shappee18} companion interaction. Irrespective to the mechanism of the early-time emission, it is clear that emission beyond the predicted $f\propto t^n$ fireball model was detected. Thus, we categorize SN~2012 in the doube category.

\subsection{SN~2012fr}
\citet{Childress13} presented the first study of SN~2012fr and derived $t_{B}^{max}$ on MJD 56243.0 and an upper limit of $E(B-V) < 0.015$~mag via the \NaI\ D line. A later study by \citet{Contreras18} found a similar $t_{B}^{max}$ on MJD 56242.6. \citet{Contreras18} examined the host-galaxy extinction using both the \NaI\ D line from different spectra than \citet{Childress13} as well as high quality Carnegie Supernova Project II \citep{Phillips19} photometry. This analysis by \citet{Contreras18} resulted in a final $E(B-V) = 0.03\pm0.03$ mag. We use the values from \citet{Contreras18} in our analysis. The Milky Way extinction is $E(B-V)_{MW}=0.02$~mag \citep{Schlafly11}.

There is a similarity between SN~2012fr and SN~2014J shown in \citet{Contreras18}, which suggests that the broken power law fitted to SN~2014J by \citet{Zheng14} matches the data of SN~2012fr well. We categorize SN~2012fr as a double SN Ia, which is different than the categorization of \citet{Jiang18}, who categorize SN~2012fr as having no excess. Our determination is based on information provided in \citet{Contreras18} which \citet{Jiang18} may not have had available to the,.

\subsection{SN~2012ht}
\citet{Yamanaka14} found $t_{B}^{max}$ to be on MJD 56295.6, and they claim host-galaxy extinction is negligible. We accept their claim as valid given the presented $B-V$ color curve in their Figure 1, as well as the lack of visible \NaI\ D in their spectra. However, a negligible extinction in the optical will be larger in the UV, so we assume a host galaxy extinction of $A_V = 0.01\pm0.01$~mag to extrapolate to the \emph{Swift} UV filters. The Milky Way line of sight extinction from \citet{Schlafly11} is $E(B-V)_{MW}=0.02$ mag.

\citet{Yamanaka14} present a smooth, single component rise, so we categorize SN~2012ht as a single SN Ia, similarly to \citet{Jiang18}.

\subsection{LSQ12gdj}
\citet{Scalzo14} find \timeB\ on MJD 56252.5 and a host-galaxy extinction of $E(B-V)_{HG}=0.02\pm0.08$ from the \citep{Lira98} law. Using \verb|SNooPy|, they find $E(B-V)_{HG}=0.01\pm0.01$ with $R_V=1.66\pm1.66$. We opt to use the extinction from the \citep{Lira98} law assuming an $R_V=3.1$. The Milky Way extinction is $E(B-V)_{MW}=0.02$~mag \citep{Schlafly11}.

\citet{Scalzo14} fit a two-component \citet{Arnett82} with a shock component model to the bolometric light curve of LSQ12gdj. Thus, we categorize LSQ12gdj as a double SN Ia. 

\subsection{SN~2013dy}
\citet{Zheng13} derive $t_{B}^{max}$ to be on MJD 56500.7 and a host-galaxy extinction from fitting the \NaI\ D line \citep{Poznanski12} to be $E(B-V) = 0.15$ mag. The Milky Way extinction is $E(B-V)_{MW}=0.13$~mag \citep{Schlafly11}.

\citet{Zheng13} found the best fit to the early-time light curve of SN~2013dy was a broken power law. Thus, we categorize SN~2013dy as a double SN Ia. In their work, \citet{Jiang18} also classify SN~2013dy as an early-excess SN Ia. 

\subsection{SN~2013gy}
\citet{Holmbo19} present the discovery and an analysis of SN~2013gy where they find $E(B-V)_{HG} = 0.11\pm0.06$~mag and $t_{B}^{max}$ on MJD 56648.5 from \verb|SNooPy| \citep{Burns11,Burns14} fits. The \citet{Schlafly11} Milky Way extinction toward SN~2013gy is $E(B-V)_{MW}=0.05$~mag. 

A single power-law rise fits the early-time light curve of SN~2013gy \citep{Holmbo19}, so we categorize it as a single SN Ia.

\subsection{iPTF13dge}
iPTF13dge was studied by \citet{Ferretti16} who found $t_{B}^{max}$ occurred on MJD 56558.0. \citet{Ferretti16} determined there was minimal host-galaxy toward iPTF13dge. They derived a value of $E(B-V)_{HG}=0.03\pm0.04$~mag. The Milky Way extinction for SN~2013gh is $E(B-V)_{MW}=0.08$ mag \citep{Schlafly11}.

In their analysis, \citet{Ferretti16} did not include fits to the early-time light curve, however, we find no evidence for a two-component power-law rise hence we categorize iPTF13dge as a single SN Ia.

\subsection{iPTF13ebh}
\citet{Hsiao15} fit iPTF13ebh with \verb|SNooPY| and fit $t_{B}^{max}$ on MJD 56622.9 and $E(B-V)_{HG}=0.05\pm0.02$ mag. The Milky Way extinction is $E(B-V)_{MW}=0.07$~mag \citep{Schlafly11}.

While \citet{Hsiao15} did not directly fit the early-time rise of iPTF13ebh, a single-component power-law fit is a reasonable conclusion from their comparison to normal and 1991bg-like models. Thus, like \citet{Jiang18}, we categorize iPTF13ebh as a single SN Ia. 

\subsection{ASASSN-14lp}
\verb|SNooPY| fits performed by \citet{Shappee16} show that the $B$-band maximum $t_{B}^{max}$ of ASASSN-14lp was on MJD 57015.3 and had a host-galaxy extinction of $E(B-V)_{HG}=0.33\pm0.06$ mag. Fits performed by \citet{Shappee16} find that ASASSN-14lp is in good agreement with a single-component power-law early-time light curve rise. The Milky Way extinction is $E(B-V)_{MW}=0.02$~mag \citep{Schlafly11}.

\subsection{iPTF14atg}
Determinations for the $t_{B}^{max}$ and $E(B-V)_{HG}$ for iPTF14atg come from two different sources. First, \citet{Cao15} determine that $t_{B}^{max}$ of iPTF14atg occurred on MJD 56799.2, but they do not provide an estimate for the host-galaxy extinction in their manuscript. Second, \citet{Kromer16} determine the host-galaxy extinction for iPTF14atg is $A_B = 0.00\pm0.02$~mag based on the \NaI\ D line. We adopt a slightly different value of $A_V = 0.01\pm0.02$~mag, which is consistent with the \citet{Kromer15} value but is in the same band as all the other \Ia\ in our sample. The Milky Way extinction is $E(B-V)_{MW}=0.01$~mag \citep{Schlafly11}.

\citet{Cao15} and \citet{Kromer16} report early-time light curve bumps, thus we classify iPTF14atg as a bump \Ia. 

\subsection{iPTF14bdn}
\citet{Smitka15} find no evidence of extinction in the spectra of iPTF14bdn, thus we assume $A_V=0.01\pm0.01$ mag. $t_{B}^{max}$ is on MJD 56822.5 \citep{Smitka15}. The Milky Way extinction is $E(B-V)_{MW}=0.01$~mag \citep{Schlafly11}.

While the early-time \emph{Swift} photometry in \citet{Smitka15} may have an early-time bump, we found no bump after redoing the photometry. Thus, we categorize iPTF14bdn as double SN Ia. 

\subsection{SN~2014J}
We adopt $t_{B}^{max}$ from \citet{Foley14} which is MJD 56690.0. Unfortunately, determining the host-galaxy extinction is not so straightforward. 

SN~2014J is one of the most heavily extincted \Ia\ to date. As such, there are a plethora of estimates on the host-galaxy extinction for this object. \citet{Amanullah14} perform various extinction fits to their data. Their power-law fit ($\frac{A_\lambda}{A_V}=\left(\frac{\lambda}{\lambda_V}\right)^p$) yielded $A_V=1.85\pm0.11$ mag, whereas their MW-like fit based on the \citet{Fitzpatrick99} parametrization yielded $E(B-V)=1.37\pm0.03$ mag with $R_V=1.4\pm0.1$.
Alternatively, \citet{Ashall14} determine the host-galaxy extinction via selecting the $A_V$ and $R_V$ values which optimize their abundance tomography models. This yields $E(B-V)=1.2$ mag and $R_V=1.38$. 
\citet{Goobar14} present two extinction values, one from \verb|SNooPy| fits and the other based on the \citet{Phillips13} method. The \verb|SNooPy| fits yield $E(B-V)=1.22\pm0.05$ mag with $R_V=1.4\pm0.15$ and the \citet{Phillips13} method yields $A_V=2.5\pm1.3$ mag 
Finally, \citet{Foley14} presents several further calculations of $A_V$ for SN~2014J. First, by using distant independent optical-infrared colors, they derive $A_V=1.95\pm0.09$ mag. Second, using the \citet{Phillips13} relation, they derive $A_V=1.8\pm0.9$ mag from their high-resolution spectrum. Finally, using the color excess method, they derive $A_V=1.91$ mag using a \citet{Fitzpatrick99} parametrization and $A_V=1.82$ mag using a \citet{CCM89} parametrization.
We elect to adopt the \citet{Foley14} value measured by the high-resolution spectrum, which is also consistent with their color excess fits as well as the power-law model from \citet{Amanullah14}. Specifically, we take the value to be $1.8\pm0.9$ mag. 

\citet{Siverd15} find peak light on MJD 56690.62 using data from the Kilodegree Extremely Little Telescope North \citep{Pepper07, Siverd12}. 

\citet{Zheng14} demonstrate a two-component power-law fit to the early-time light curve rise, and \citet{Goobar15} also found that multiple components better fit the rising light curve. Thus, we categorize SN~2014J as a double SN Ia. Unfortunately, the extinction is so severe in the UV that SN~2014J is not comparable in the UV to other \Ia, thus we do not include it in plots and only include this discussion for completeness of the early-time SN Ia sample.

\subsection{SN~2015F}
Using MLCS2k2 \citep{Jha07}, \citet{Im15} find $E(B-V)_{hg}=0.04\pm0.03$ mag and $t_{B}^{max}$ on MJD 57105.98. \citet{Cartier17} estimate the $t_{B}^{max}$ to be MJD 57106.5. Using the optical and near-infrared colors, they calculate various $E(B-V)$ values, the weighted average of which they calculate to be $E(B-V)_{HG}=0.09\pm0.02$~mag. Their individual $E(B-V)$ calculations use bot the optical \citep{Phillips99} and near-infrared \citep{Krisciunas04a, Krisciunas04b} colors. The Milky Way extinction is $E(B-V)_{MW}=0.17$~mag \citep{Schlafly11}.

\citet{Im15} fit both single and double power laws to the early-time data of SN~2015F and find that the single-component power-law gives the best fit and that the double power law fit converges to the single power law result. Thus, we categorize SN~2015F as a single SN Ia, which is consistent with the categorization of \citet{Jiang18}. 

\subsection{SN~2015bq}
\citet{Li22} determined $t_{B}^{max}$ was on MJD 57084.1. By performing fits using the SALT2 modelling program \citep{Guy07}, \citet{Li22} calculate $E(B-V)_{HG}=0.15\pm0.07$ mag. The Milky Way extinction is $E(B-V)_{MW}=0.01$~mag \citep{Schlafly11}.

An early-time flux excess is claimed by \citet{Li22}. Their data show SN~2015bq has early-time flux which is greater than the normal \Ia\ they compare to. While the data is not early enough to be fit by power-law rises, we still opt to categorize SN~2015bq as an double SN Ia. Because SN~2015bq lacks \emph{Swift} detections in the critical $UVM2$ band, we do not include it in our analysis, but we mention it here for completeness.

\subsection{iPTF16abc}

\citet{Ferretti17} find a value of MJD 57498.8 for $t_{B}^{max}$ and $A_V=0.1\pm0.2$ mag from fitting the SED of SN~2011fe to the iPTF16abc data. Using SALT2, they find $A_V=-0.03\pm0.04$ mag. Despite finding deep \NaI\ D lines, \citet{Ferretti17} find weak reddening in the photometry of iPTF16abc. \citet{Dhawan18} fit iPTF16abc using \verb|SNooPy| and find $E(B-V)_{HG}=0.07\pm0.02$ mag and $R_V=3.1$. Finally, \citet{Miller18} adopt $E(B-V)_{HG} = 0.05$~mag from \citet{Ferretti17} and determine \timeB\ occurred on MJD 57499.5.
We adopt the values from \citet{Ferretti17}. Finally, the Milky Way extinction from \citet{Schlafly11} is $E(B-V)_{MW}=0.028$ mag.

\citet{Miller18} present early-rise fits and find that a $f\propto t^2$ model does not adequately describe the rising light curve of iPTF16abc whereas a nearly linear model fits the rise well. They did not fit a two-component power law, but given the single-component power law fits the rise well, we categorize iPTF16abc as a single SN Ia, which is different than the determination of \citet{Jiang18}.

\subsection{SN~2017cbv}
The five papers on SN~2017cbv agree on both $t_{B}^{max}$ and host-galaxy extinction. First, \citet{Hosseinzadeh17} find $t_{B}^{max}$ on MJD 57841.1 and negligible host-galaxy extinction due to the lack of \NaI\ D absorption in their spectra. Second, \citet{Ferretti17} find $E(B-V)=0.02\pm0.01$ mag from the \NaI\ D line using the method of \citet{Poznanski12}. Third, \citet{Wee18} determine $t_{B}^{max}$ to be on MJD 57840.4 and negligible host-galaxy extinction. Fourth, \citet{Wang20} again find no significant host-galaxy extinction and a $t_{B}^{max}$ value of MJD 57840.4. Finally, \citet{Burke22} find no host-galaxy extinction and MJD 57840.3 as $t_{B}^{max}$. For this work, we will use $t_{B}^{max}$ on MJD 57841.1 from \citet{Hosseinzadeh17} and $A_V=0.06\pm0.03$ mag from \citet{Ferretti17} (assuming $R_V=3.1$). The Milky Way extinction from \citet{Schlafly11} is $E(B-V)_{MW}=0.15$ mag. 

\citet{Hosseinzadeh17} and \citet{Burke22} show SN~2017cbv is fit better by multiple component models, and \citet{Jiang18} categorize SN~2017cbv as a SN Ia with an early excess. Given that the light curve for SN~2017cbv monotonically increases in all the optical and UV bands, we place SN~2017cbv in our double category rather than the bump category.

\subsection{SN~2017cfd}
\citet{Han20} estimate $A_V = 0.39\pm0.03$ mag from MLCS2k2 \citep{Jha07} fitting with an assumed $R_V=1.7$. This $R_V$ was presumably chosen by \citet{Han20} to yield a peak $B$-band absolute magnitude of $-19.2$ mag. They also derive $A_V=1.34\pm0.40$ mag from the \NaI\ D absorption line, however, they note that this large of an extinction value would make SN~2017cfd significantly brighter than most other \Ia\ at its $t_{B}^{max}$ on MJD 57843.4. The Milky Way extinction is $E(B-V)_{MW}=0.02$~mag \citep{Schlafly11}. Finally, \citet{Han20} show that SN~2017cfd is a normal SN Ia with a single power-law rise, thus we categorize SN~2017cfd as a single SN Ia. 

\subsection{SN~2017cyy}
\citet{Burke22} find $t_{B}^{max}$ on MJD 57870.1 and negligible host-galaxy extinction. Thus, we assume a nominal host-galaxy extinction of $A_V=0.01\pm0.01$ mag. The Milky Way extinction is $E(B-V)_{MW}=0.22$~mag \citep{Schlafly11}. \citet{Burke22} find no evidence of a two-component power-law rise. Thus, we categorize SN~2017cyy as a single SN Ia. 

\subsection{SN~2017erp}
\citet{Brown19} find MJD 57934.9 for $t_{B}^{max}$ as well as several values for the host-galaxy extinction which they list in their Table 3. The methods \citet{Brown19} use to estimate the host-galaxy extinction are the peak color, the \citet{Lira98} law, \NaI\ D fitting, and color excess from \verb|SNooPy| and MLCS2k2 fits. Consistently, \citet{Burke22} find $E(B-V)=0.10\pm0.01$ mag and $t_{B}^{max}$ on MJD 57934.4 from \verb|SNooPy| fits. We opt to combine the values from \citet{Brown19} using a weighted average for our final extinction of $A_V = 0.15\pm0.04$ mag. The Milky Way extinction is $E(B-V)_{MW}=0.09$~mag \citep{Schlafly11}.

We categorize SN~2017erp as a single SN Ia despite the claim of companion interaction from \citet{Burke22}. While they claim a two-component model fits the data better, the additional flux from the second component appears to only marginally improve the quality of the light curve fit.

\subsection{SN~2017fgc}
Three studies independently derive SN properties for SN~2017fgc. First, \citet{Zeng21} also fit SN~2017fgc with \verb|SNooPy| and find from the fits that $t_{B}^{max}$ was on MJD 57959.4 that the host-galaxy extinction is $E(B-V)=0.17\pm0.07$ mag. Second, \citet{Burgaz21} derive $t_{B}^{max}$ on MJD 57958.7 and adopt a host-galaxy extinction of $E(B-V)_{HG} = 0.29\pm0.02$ mag from the Lira law \citet{Phillips99}. Finally, \citet{Burke22} fit SN~2017 with \verb|SNooPY| and derived $t_{B}^{max}$ on MJD 57959.5 and a host-galaxy extinction of $E(B-V)=0.21\pm0.01$ mag. We adopt the values from \citet{Burke22} because their light curve is the most dense and their data is solely from Las Cumbres Observatory Global Telescope observations \citep{Brown13}, so $S$-corrections do not introduce an additional systematic uncertainty. The Milky Way extinction is $A_V = 0.094$ mag \citet{Schlafly11}. 

We categorize SN~2017fgc as a single SN Ia based upon the fits performed by \citet{Zeng21} and lack of companion signature from \citet{Burke22}.

\subsection{ASASSN-18bt (SN~2018oh)}
The three synoptic studies of ASASSN-18bt shortly after explosion (\citealt{Dimitriadis19,Li19,Shappee19}) all used the $t_{B}^{max}$ and $E(B-V)_{HG}$ derived by \citet{Li19}. \citet{Li19} performed fits using SALT2, \verb|SNooPy|, and MLCS2k2, which were all consistent. The values of their \verb|SNooPy| fits are $E(B-V)_{HG}=0.00\pm0.01$ and $t_{B}^{max}$ on MJD 58162.7. We adopt their values but alter their $E(B-V)_{HG}$ value to be $0.01\pm0.01$ such that we can extrapolate the small host-galaxy extinction into the UV. The Milky Way extinction is $E(B-V)_{MW}=0.04$~mag \citep{Schlafly11}.

\citet{Shappee19} showed the rising light curve of ASASSN-18bt was fit the best by a two-component power law, so we categorize ASASSN-18bt as a double SN Ia.  

\subsection{SN~2018gv}
Both \citet{Yang20} and \citet{Burke22} agree that there is negligible optical host-galaxy extinction. While \citet{Burke22} does not provide an estimate for the host galaxy-extinction, \citet{Yang20} uses the ``CMAGIC'' technique to estimate $E(B-V)_{HG}=0.03\pm0.03$ mag, which we adopt. The Milky Way extinction is $E(B-V)_{MW}=0.05$~mag \citep{Schlafly11}.

From fitting the light curves, \citet{Yang20} and \citet{Burke22} agree on $t_{B}^{max}$ as well, deriving $t_{B}^{max}$ on MJD 58149.7 and MJD 58149.6 respectively. We use the value from \citet{Yang20}. 

\citet{Yang20} found a single-component light curve rise consistent with $f\propto t^2$, and \citet{Burke22} also found a rise consistent with a fireball model. Thus, we categorize SN~2018gv as a single SN Ia. 

\subsection{SN~2018xx}
From \verb|SNooPy| fits, \citet{Burke22} determine $E(B-V)_{HG}=0.04\pm0.01$ mag and MJD 58183.9 for $t_{B}^{max}$, as well as that SN~2018xx does not show a two-component rise. Thus, we categorize SN~2018xx as a single SN Ia. The Milky Way extinction is $E(B-V)_{MW}=0.09$~mag \citep{Schlafly11}. 

\subsection{SN~2018yu}
From \verb|SNooPy| fits, \citet{Burke22} determine MJD 58183.3 for $t_{B}^{max}$. \citet{Burke22} claim negligible host-galaxy extinction, thus we adopt $A_V=0.01\pm0.01$ mag. The Milky Way extinction is $E(B-V)_{MW}=0.13$~mag \citep{Schlafly11}.

While \citet{Burke22} claim an early-excess flux from companion interaction in the rising light curve of SN~2018yu, the companion shock contribution to the early-time light curve rise is small, and neither \citet{Tucker20} nor \citet{Graham22} find evidence of H$\alpha$ in the nebular spectra. Additionally, the companion flux contribution appears small, so the improvement to the fit is marginal. Thus, we categorize SN~2018yu as a single SN Ia.

\subsection{SN~2018agk}
\citet{Wang21} determined \timeB\ was on MJD 58203.8 and from the \NaI\ D line relationship from \citet{Poznanski12} they estimate $E(B-V)_{HG}=0.11\pm0.05$ mag. The Milky Way extinction is $E(B-V)_{MW}=0.03$ mag \citep{Schlafly11}. 

The early-time light curve was fit best by a single power law by \citet{Wang21} using \emph{Kepler} data. Hence, we categorize SN~2018agk as a single SN Ia. SN~2018agk is distant ($d>100$~Mpc), so the UV data is faint -- there is only 1 detection in the $UVM2$ band and it is barely above 3$\sigma$. SN~2018agk is included in the tables for completion but is not factored into our analysis.

\subsection{SN~2018aoz}
\citet{Ni22} found $t_{B}^{max}$ on MJD 58222.2 via a \verb|SNooPy| fit and an upper limit for the host-galaxy extinction of $E(B-V)_{HG}<0.02$ mag via fitting the \NaI\ D line. 
\citet{Burke22} also performed \verb|SNooPy| fits and derived MJD 58222.1 for $t_{B}^{max}$ and negligible host-galaxy extinction. We adopt the values from \citet{Ni22}. The Milky Way extinction is $E(B-V)_{MW}=0.07$~mag \citep{Schlafly11}.

\citet{Burke22} find a single-component rise explains the early-time light curve of SN~2018aoz and \citet{Ni22} and \citet{Ni23} find that, while reddened, SN~2018aoz is consistent with a single power-law rise. Thus, we categorize SN~2018aoz as a single SN Ia.

\subsection{SN~2019np}
\citet{Burke22} used \verb|SNooPy| to fit SN~2019np and found MJD 58509.6 to be $t_{B}^{max}$. \citet{Sai22} fit a polynomial to the light  curve of SN~2019np and found $t_{B}^{max}$ on MJD 58510.2. From \verb|SNooPy| fits and the \NaI\ D absorption equivalent width, \citet{Sai22} determine $E(B-V)_{HG}=0.10\pm0.04$ mag. The Milky Way extinction is $E(B-V)_{MW}=0.02$~mag \citep{Schlafly11}.

\citet{Burke22} do not find signatures of companion interaction, which is consistent with the later work by \citet{Ni22} and \citet{Ni23}. However, \citet{Ni23} finds that, while reddened at early times \citep{Ni22}, SN~2019np has an early-excess flux consistent with a multiple-power-law rise. Thus, we categorize SN~2019np as a double SN Ia.  

\subsection{SN~2019ein}
\citet{Pellegrino20} find $t_{B}^{max}$ on MJD 58619.5 from SALT2 fits, and they assume the host-galaxy extinction from \citet{Kawabata20}. \citet{Kawabata20} determine $t_{B}^{max}$ on MJD 58618.2 from polynomial fit to data around maximum. We adopt the SALT2 fit result from \citet{Pellegrino20} as our $t_{B}^{max}$. The Milky Way extinction is $E(B-V)_{MW}=0.01$~mag \citep{Schlafly11}.

They estimate the host-galaxy extinction from \verb|SNooPy| fits, \citet{Phillips99} and \citet{Reindl05} $E(B-V)$ evolution relationships, and the \SiII\ $\lambda$ 6355 and peak intrinsic $E(B-V)$ color relationships from \citet{Foley11} and \citet{Blondin12}. Ultimately, they assume the \verb|SNooPy| value, which we will also adopt. This value is $E(B-V)_{HG}=0.09\pm0.06$ mag and $R_V=1.55$ which corresponds to $A_V=0.14\pm0.09$ mag.

SN~2019ein does not show an early excess \citet{Kawabata20}, so we label it as a single SN Ia.

\subsection{SN~2019yvq}
Both \citet{Miller20} and \citet{Burke21} are in agreement on when $t_{B}^{max}$ occurred, with respective values of MJD 58863.3 and 58863.1. We adopt the mean value of MJD 58863.2 for $t_{B}^{max}$.

With respect to host-galaxy extinction, \citet{Miller20} used \NaI\ D lines to find $E(B-V)_{HG}\approx0.032$ mag. \citet{Burke21} found that light curve fitting methods resulted in highly discordant estimates for the host-galaxy extinction. This is to be expected given how unique SN~2019yvq is. However, \citet{Burke21} fit the \NaI\ D line and found $E(B-V)_{HG} = 0.05^{+0.05}_{-0.03}$ mag. This is consistent with the derived host-galaxy extinction from \citet{Miller20}. We adopt the \citet{Burke21} host-galaxy extinction value. The Milky Way extinction is $E(B-V)_{MW}=0.02$ mag \citep{Schlafly11}.

\citet{Miller20}, \citet{Siebert20}, \citet{Burke21}, and \citet{Tucker21} all present analyses of the early-time bump of SN~2019yvq. We classify SN~2019yvq as a bump Ia. 

\subsection{SN~2020hvf}
\citet{Jiang21} suggest negligible host-galaxy extinction from a lack of \NaI\ D absorption in their spectra as well as a large distance from the center of the host galaxy and a $t_{B}^{max}$ of MJD 58979.3 from a polynomial fit to the data near peak. We adopt their values, assuming $A_V = 0.01\pm0.01$ mag. The Milky Way extinction is $E(B-V)_{MW}=0.04$ mag \citep{Schlafly11}. 

The data presented by \citet{Jiang21} show a clear early bump for SN~2020hvf, thus we categorize it as a bump SN Ia. 

\subsection{SN~2020nlb}
\citet{Sand21} fit a fourth-order polynomial to the data near peak and perform 1000 resamples based on the photometric uncertainties. They find negligible host-galaxy extinction based on the \citet{Phillips99} relationship and MJD 59041.8 as $t_{B}^{max}$. We adopt their value for $t_{B}^{max}$ and assume $A_V=0.01\pm0.01$ mag. The Milky Way extinction is $E(B-V)_{MW}=0.03$~mag \citep{Schlafly11}.

There are early-time data available for SN~2020nlb, however, \citet{Sand21} does not fit the early-time light curve. We perform an MCMC fit using \verb|emcee| \citep{Foreman-Mackey13} to the rising light curve and find it is consistent with a single-power-law rise. Thus, we categorize SN~2020nlb as a single SN Ia. 

\subsection{SN~2020tld}
\citet{Fausnaugh23} present TESS observations of SN~2020tld. The TESS data do not cover the light curve peak. Since their observations are only in one band and they have no spectra, the host-galaxy extinction is uncertain. However, SN~2020tld lies well outside its host, so we assume $A_V=0.01\pm0.01$~mag. Finally, the rising light curve fits from \citet{Fausnaugh23} disfavor companion interaction, so we categorize SN~2020tld as a single SN Ia. There are no pre-explosion images, nor post-explosion templates in \emph{Swift} for SN~2020tld, so we do not include it in our figures or analysis. We mention it here for completeness

\subsection{SN~2020udy}
\citet{Maguire23} found a $g$-band maximum light on MJD 59131.0, which we adopt as $t_{B}^{max}$. $t_{B}^{max}$ and $t_{g}^{max}$ are close enough that this assumption will not significantly impact any of our results. Since they do not observe any \NaI\ D absorption in their spectra, \citet{Maguire23} claim negligible host-galaxy extinction. Thus, we assume $A_V=0.01\pm0.01$ mag for SN~2020udy. The Milky Way extinction is $E(B-V)_{MW}=0.07$~mag \citep{Schlafly11}. Finally, \citet{Maguire23} fit a single power law to the early-time light curve rise, thus we categorize SN~2020udy as a single SN Ia. Unfortunately, SN~2020udy lacks pre-explosion imaging, so we exclude it from our analysis because the photometry would not be host-subtracted.

\subsection{SN~2021fxy}
\citet{DerKacy23} fit the light curves of SN~2021fxy with \verb|SNooPy|. From these fits, they find $t_{B}^{max}$ on MJD 59305.1 and a small host-galaxy extinction of $E(B-V)_{HG}=0.02\pm0.06$ mag. We adopt these values, along with a Milky Way extinction of $E(B-V)_{MW}=0.08$ mag \citep{Schlafly11}. The light curves presented by \citet{DerKacy23} are consistent with a single-component rise, so we categorize SN~2021fxy as a single SN Ia.

\subsection{SN~2021hpr}
\citet{Zhang22} fit the light curve peak with a 2nd order polynomial and determine $t_{B}^{max}$ was on MJD 59321.9. They posit negligible host-galaxy extinction. 
\citet{Ward22} fit SN~2021hpr using the \textsc{BayeSN} fitting program \citep{Thorp21,Mandel22}. They find $t_{B}^{max}$ on MJD 59321.4. Their Bayesian framework enables fitting with a fixed $R_V$ and with a uniform prior between 1 and 6. With $R_V$ fixed at $2.66$, they find $A_V=0.27\pm0.04$ mag; with $R_V$ drawn from a uniform prior distribution $\mathcal{U}(1,6)$, they find $A_V=0.28\pm0.07$ mag.
Finally, \citet{Lim23} use the \citet{Phillips99} relationship to estimate the host-galaxy extinction, which yields an estimate of $E(B-V)_{HG}=0.08\pm0.04$ mag. For $t_{B}^{max}$, they find that the light curve peaked on MJD 59321.7. We adopt the values from \citet{Ward22}. The Milky Way extinction is $E(B-V)_{MW}=0.02$~mag \citep{Schlafly11}.

\citet{Lim23} found a two-component best fit SN~2021hpr, thus we categorize it as an double SN Ia. 

\subsection{SN~2021zny}
\citet{Dimitriadis23} found that SN~2021zny peaked on MJD 59498.4 by fitting the light curve near peak with a polynomial. Using the \NaI\ D line and the \citet{Poznanski12} relationship, they derive $E(B-V)_{HG}=0.10\pm0.07$ mag. We adopt both of their values for this work. The Milky Way extinction is $E(B-V)_{MW}=0.04$ mag \citep{Schlafly11}.

TESS observed SN~2021zny, and the rising TESS lightcurve shows a small bump \citep{Dimitriadis23, Fausnaugh23}. However, \citet{Fausnaugh23} express some concerns over the veracity of this bump. They note similar light curve variations at the light curve peak. Furthermore, they note that SN~2021zny is on a CCD strap, which should produce noise similar to the observed peak variations. Their light curve fits disfavor companion interaction, so this is an argument against an astrophysical origin for the bump. We categorize SN~2021zny as a bump Ia with the caveats noted by \citet{Fausnaugh23}. 

\subsection{SN~2021aefx}
\citet{Ashall22} find $t_{B}^{max}$ on MJD 59547.2 and do not provide a host-galaxy extinction value. \citet{Hosseinzadeh22} find MJD 59546.5 as $t_{B}^{max}$ using a polynomial fit to the data near peak. They use their high-resolution spectrum to determine host-galaxy extinction from the \NaI\ D line using the method of \citet{Poznanski12}. They find  $E(B-V)_{HG}=0.10$ mag. Using the \citet{Phillips99} relationship, they find $E(B-V)_{HG}=0.04$~mag, which is consistent with their \NaI\ D-derived value. We adopt the values from \citet{Hosseinzadeh22} for both $t_{B}^{max}$ and $A_V$, specifically, $A_V=0.31$ mag from the \NaI\ D line. The Milky Way extinction is $E(B-V)_{MW}=0.01$ mag \citep{Schlafly11}.

Both \citet{Ashall22} and \citet{Hosseinzadeh22} find a two-component light curve fit to SN~2021aefx. Thus, we categorize SN~2021aefx as a double SN Ia.

\subsection{SN~2022eyw}
\citet{Fausnaugh23} present TESS observations of SN~2022eyw, however, the TESS sector only covers the rising light curve and not the peak. Since their observations are only in one band, we adopt an extinction of $A_V=0.02\pm0.05$~mag, motivated by the lack of \NaI\ D in the TNS spectra, but the position is in the host galaxy, so there may be some extinction. The Milky Way extinction is $E(B-V)_{MW}=0.01$~mag \citep{Schlafly11}.

The rising light curve fits from \citet{Fausnaugh23} disfavor companion interaction, so we categorize SN~2022eyw as a single SN Ia. There are no pre-explosion images, nor post-explosion templates in \emph{Swift} for SN~2022eyw, so we do not include it in our figures or analysis. We mention it here for completeness.

\subsection{SN~2022ilv}
\citet{Srivastav23a} searched for a host-galaxy, however, they did not satisfactorily locate a clear host-galaxy candidate. Because there is no sign of a host galaxy or \NaI\ D lines, we conclude that there is negligible host-galaxy extinction. In line with other \Ia\ in our sample with negligible host-galaxy extinction, we assume $A_V=0.01\pm0.01$ mag. While they do not clearly state the $t_{B}^{max}$ they derive, it can be inferred from other statements in the paper to be on MJD $\sim$59707.5. The Milky Way extinction is $E(B-V)_{MW}=0.10$~mag \citep{Schlafly11}.

SN~2022ilv shows an early-time bump based on an ATLAS non-detection. \citet{Srivastav23a} evaluate this non-detection and determine it is valid. We agree with their determination and thus, we categorize SN~2022ilv as a bump SN Ia.

\subsection{SN~2023bee}
Both \citet{Wang23} and \citet{Hosseinzadeh23} agree that host-galaxy extinction is negligible, thus we adopt $A_V=0.01\pm0.01$ mag. For $t_{B}^{max}$, we adopt MJD 59992.5 as our value, which is the average between the \citet{Wang23} and \citet{Hosseinzadeh23} values of MJD 59992.6 and MJD 59992.4, respectively. The Milky Way extinction is $E(B-V)_{MW}=0.01$~mag \citep{Schlafly11}.

SN~2023bee is the most recent SN with early-time light curve coverage. Both \citet{Wang23} and \citet{Hosseinzadeh23} report early-excess flux and two-component light curves. We categorize SN~2023bee as a double SN Ia.

\end{document}